\documentclass[graybox]{svmult}

\usepackage{mathptmx}       
\usepackage{helvet}         
\usepackage{courier}        
\usepackage{type1cm}        
\usepackage{makeidx}         
\usepackage{graphicx}        
\graphicspath{{Figures/}}
\usepackage{multicol}        
\usepackage[bottom]{footmisc}

\usepackage{amssymb}
\usepackage{amsmath}
\usepackage{hyperref}
\usepackage{subcaption}
\usepackage{soul}

\newcommand{\kms}{km\,s$^{-1}$}
\newcommand{\msun}{M$_{\odot}$}

\makeindex             

\begin{document}

\title*{Neutron stars formation and Core Collapse Supernovae}
\author{P. Cerda-Duran and N. Elias-Rosa}
\institute{Pablo Cerda-Duran \at Departamento de Astronom\'ia y Astrof\'isica, Universitat de Val\`encia, Dr. Moliner, 50, Burjassot, Valencia E-46100, Spain, \email{pablo.cerda@uv.es}
\and Nancy Elias-Rosa \at INAF - Osservatorio Astronomico di Padova, vicolo dell'Osservatorio 5, Padova I-35122, Italy \email{nancy.elias@oapd.inaf.it}}
%
%
\maketitle

\abstract{
In the last decade there has been a remarkable increase in our knowledge about core-collapse supernovae (CC-SNe), and the birthplace of neutron stars, from both the observational and the theoretical point of view.
Since the 1930's, with the first systematic supernova search, the techniques for discovering and studying extragalactic SNe have improved. Many SNe have been observed, and some of them, have been followed through efficiently and with detail. Furthermore, there has been a significant progress in the theoretical modelling  of the scenario, boosted by the arrival of new generations of supercomputers that have allowed to perform multidimensional numerical simulations with unprecedented detail and realism. The joint work of observational and theoretical studies of individual SNe over the whole range of the electromagnetic spectrum has allowed to derive physical parameters, which constrain the nature of the progenitor, and the composition and structure of the star's envelope at the time of the explosion. The observed properties of a CC-SN are an imprint of the physical parameters of the explosion such as mass of the ejecta, kinetic energy of the explosion, the mass loss rate, or the structure of the star before the explosion. 
In this chapter, we review the current status of SNe observations and theoretical modelling, the connection with their progenitor stars, and the properties of the neutron stars left behind.
}

\section{Introduction}
\label{sec_1}

\subsection{Core-collapse supernovae and their importance}
\label{SN_impo}

Neutron stars (NS) appear at the end point of the evolution of massive stars ($M>8~M_{\odot}$). At this stage, most of these stars have grown an iron core that cannot be supported by hydrostatic pressure and collapses.
In most of the cases it produces a supernova (SN), releasing a gravitational energy of about $\sim 10^{53}$ ergs (mainly in form of neutrino radiation), and leading to the complete destruction of the progenitor star (e.g. see \cite{Janka:2012}).
SNe leave behind compact remnants, such as neutron stars (NS) or black holes (BH) (e.g. see \cite{Heger:2003}). The first interpretation of a SN as a transition between massive stars and neutron stars was introduced by W. Baade \& F. Zwicky in the 1930s \cite{baade34} in order to explain the extraordinary amount of energy liberated. It still holds nowadays (see Section \ref{sec:th:ccsne} for a detailed description about how NS are born). 
\\

SNe are studied from multiple perspectives. They are crucial for a complete understanding of stellar evolution, being associated with NS and black holes (BH), as well as with other extreme events such as gamma-ray bursts (GRBs; e.g. \cite{galama98} or \cite{woosley06}) and X-ray flashes (XRFs; \cite{pian06}), mainly connected to stripped-envelope core-collapse SNe. SNe are also among the most influential events in the Universe regarding their energetic and chemical contribution to the interstellar medium in galaxies \cite{thielemann96}. In fact, SNe are the major ``factories" of heavy elements synthesized along the progenitors life and in the SN explosion itself, as well as sources of gravitational waves, neutrinos, and cosmic rays (e.g. \cite{Andersson:2013,hirata87,koyama95}). SNe also produce dust (e.g. \cite{todini01}), and induce star formation (e.g., \cite{krebs83}) since the shock waves generated in their explosion heat and compress interstellar molecular clouds. Finally, they can be used as cosmological distance indicators, and to set constraints on the equation of state of Dark Energy (e.g. \cite{riess98,perlmutter99,hamuy02}).\\

\subsection{Brief history of supernova observations}
\label{SN_history}

Temporary stars, comets and novae, as well as occasional supernovae, were fairly frequently recorded in East Asian history \cite{stephenson05}. Possibly, the first supernova for which written reports exist is SN~185, which took place in the year AD 185 in the Milky Way Galaxy (precisely it occurred in the direction of Alpha Centauri). 

The birth of modern SN astronomy occurred in 1885, when the first extragalactic SN (S Andromedae or SN~1885A) was detected with a telescope \cite{hartwig85}. In the 1920's scientists begun to realise that there was a particular class of very bright novae, and a decade later, W. Baade \& F. Zwicky \cite{baade34} named this kind of event ``supernova". The distinction between novae and supernovae was at first based on twelve objects discovered between 1895 and 1930, plus the galactic SN observed by Tycho in 1572. Between 1936 and 1941, the first systematic supernova search was started by W. Baade and F. Zwicky using the Palomar 18-inch Schmidt telescope and led to the discovery of 19 SNe. In the same years, SNe were first classified as Type I or Type II based on the lack or presence of hydrogen spectral lines, respectively \cite{minkowski41}. This subdivision is still the basis of the SN classification used today (Section \ref{sec_2_CCSN_class} and \cite{turatto03,galyam16}).

In the following years, the improvement of the instrumentation, the construction of new telescopes and also the numerous progresses in the understanding of the stellar evolution, stimulated the research and cataloguing of the SNe. By the Eighties, with the introduction of the CCD and the construction of larger diameter telescopes, not only did the number of SNe observed grow but it was possible also to obtain spectra with better resolution and to study their luminosity evolution for long times. SNe classification became consequently more complex due mainly to more careful comparison among the SNe, and the previous types of SNe were further sub-categorised attending to the presence/absence of chemical elements other than hydrogen. 

Computer controlled search programs of SNe were initiated in the following decade. Thanks to past and ongoing surveys (e.g. the All-Sky Automated Survey for SuperNovae -- ASAS-SN; \cite{shappee14}, the ESA Gaia transient survey -- \cite{hodgkin13}, or the Panoramic Survey Telescope \& Rapid Response System -- Pan-STARRRS; \cite{kaiser02}, among numerous others), and to the efforts of amateur astronomers, the rate of SN discoveries has dramatically increased, going from less than 20 SNe at the beginning of the 20th century, to more than $\sim$ 200 SNe per year in the first decade of the 21st century, and finding today up to 1000s of events per year.  

These technological advances also allowed SN searches at z $>$ 0.2 (e.g. Dark Energy Survey -- DES\footnote{\url{http://www.darkenergysurvey.org/}}), and the multi-wavelength observation of SNe. The optical band has played a fundamental role in the knowledge and classification of the SNe, but with the technological progress, it has been possible to observe the SNe in the IR bands, and in the radio, ultraviolet (UV) and X-ray. In particular, with the explosion of the SN~1987A in the LMC, the closest extragalactic SN observed (50 kpc), it has been possible also to directly identify for the first time the SN progenitor in archival images and to detect the neutrino flux produced during the explosion (see e.g. \cite{arnett89,mccray93} for reviews).

To further complicate this scenario, these new generation of deep, and wide surveys, are discovering new types of transients with unprecedented observational characteristics, as we will describe in Section \ref{sec_3_newclass}. For example, it has been found extreme SNe types such as superluminous SNe, hundreds of times brighter than those found over the last fifty years (M$_V < -21$ mag), whose energy regime is not explained by the standard core-collapse and neutrino-driven mechanisms. Or ultra-faint SNe (M$_V > -14$ mag), which are characterized by a very low explosion energy ($\sim 10^{49}$ ergs) and small amount of $^{56}$Ni mass ($< 10^{-3}$ M$_{\odot}$).

The wide variety of classes of SNe has also shown that core-collapse supernovae (CC-SNe) present observational heterogeneity, consequence of the different properties of the progenitor star at the moment of the explosion, their energetic, angular momentum, and environment. Thus, in the last decades SNe have been studied in order to better establish the link to their progenitor stars and thereby to clarify the evolution of massive stars. This make the SNe a very valuable probe of mass loss, circumstellar structure, and star formation rates. For nearby CC- events it has been possible to directly image the precursor star in pre-explosion images (mainly from the archive of the Hubble Space Telescope, {\sl HST}) and to verify its disappearance years after the SN exploded (\cite{smartt09,smartt15} and Section \ref{sec_2_directdet}).\\

\subsection{The theoretical perspective}

In parallel to the observations, our theoretical understanding of the processes leading to the formation of a neutron star as a result of a supernova explosion has grow significantly in the last century. Although the basic scenario was set by W. Baade \& F. Zwicky in 1934 \cite{baade34}, it was not until the 1960s, with the appearance of the first modern computers, that a significative progress was achieved. It was established that the collapsing core should bounce when reaching nuclear matter density \cite{Colgate:1960,Colgate:1961} and suggested that the primary energy source in supernova explosions should be in form of neutrino radiation \cite{Colgate:1966}. The basics behind most advanced models nowadays was set by H.~A. Bethe and J.~R. Wilson \cite{Bethe:1985} in the so-called delayed neutrino-heating mechanism. They realised that, when included the most sophisticated microphysics, a prompt explosion after bounce was not possible, but the energy deposited by neutrino radiation could revive the shock and power the explosion. We know now that multidimensional effects play a primary role in the explosion, and that a comprehensive numerical modelling of the scenario including all relevant physical effects is a computational challenge (see e.g. \cite{Janka:2007,Janka:2012,Burrows:2013,Mueller:2016}).
Therefore, our current understanding of the supernova mechanism relies heavily in the results of numerical modelling and the progress in the field has tracked closely the improvements of modern supercomputers and the development of high performance computing.

However, in order to understand the plethora of observations it is not sufficient to be able to model numerically the collapse of massive stars. One also has to be able to link these explosions with their progenitor stars, make predictions of how common or uncommon each type of event is, and what are the consequences for the galactic environment in which they live. The complete picture can only be achieved with the help of the complementary discipline of stellar evolution (see e.g. \cite{Woosley:2002}).
Even if more than 2500 CC-SNe have been discovered so far, it is still missing a complete picture of this progenitor-explosion link. In order to better understand the way in which these stars end their evolution, it is necessary to get stronger constrains of the physical characteristics of the progenitor stars alongside the explosion parameters through the observation of SNe, and thus test and constrain the models. At the same time, studying compact remnants like neutron stars, we can determinate explosion conditions and understand associated phenomena such as mass loss, r-process nucleosynthesis, gravitational waves and neutrino emission.

\subsection{Aim}

The aim of this work is to review the current knowledge about supernovae as the place of birth of neutron stars from both an observational and a theoretical perspective. In Section \ref{sec_2} we review the status of observations of core-collapse supernovae and our current knowledge about the scenario in which these explosions are produced. In Section \ref{sec_3} we discuss the present challenges and future perspectives in the field.

\section{Current status of supernova observations and modelling}
\label{sec_2}

\subsection{Traditional supernova types}
\label{sec_2_CCSN_class}

As we see in Section \ref{SN_history}, the first fundamental classification was given by \cite{minkowski41} who distinguished the SNe in two different classes based on the lack (SNe I) or presence (SNe II) of hydrogen lines such as H$\alpha$ 6563 \AA ~and H$\beta$ 4861 \AA~in their early spectra. Since SNe can be very different one from another as to spectral features (i.e. chemical composition, physical conditions), photometry, overall SED (spectral energy distribution), time evolution, radio and X-ray properties, in the 1980s sub-classes were introduced such as Ib, Ic, II-P, II-L, IIn, IIb which are related to characteristics of their spectra (small letter) or light curves (capital letter).

\begin{figure}[h]
\includegraphics[width=1.\textwidth]{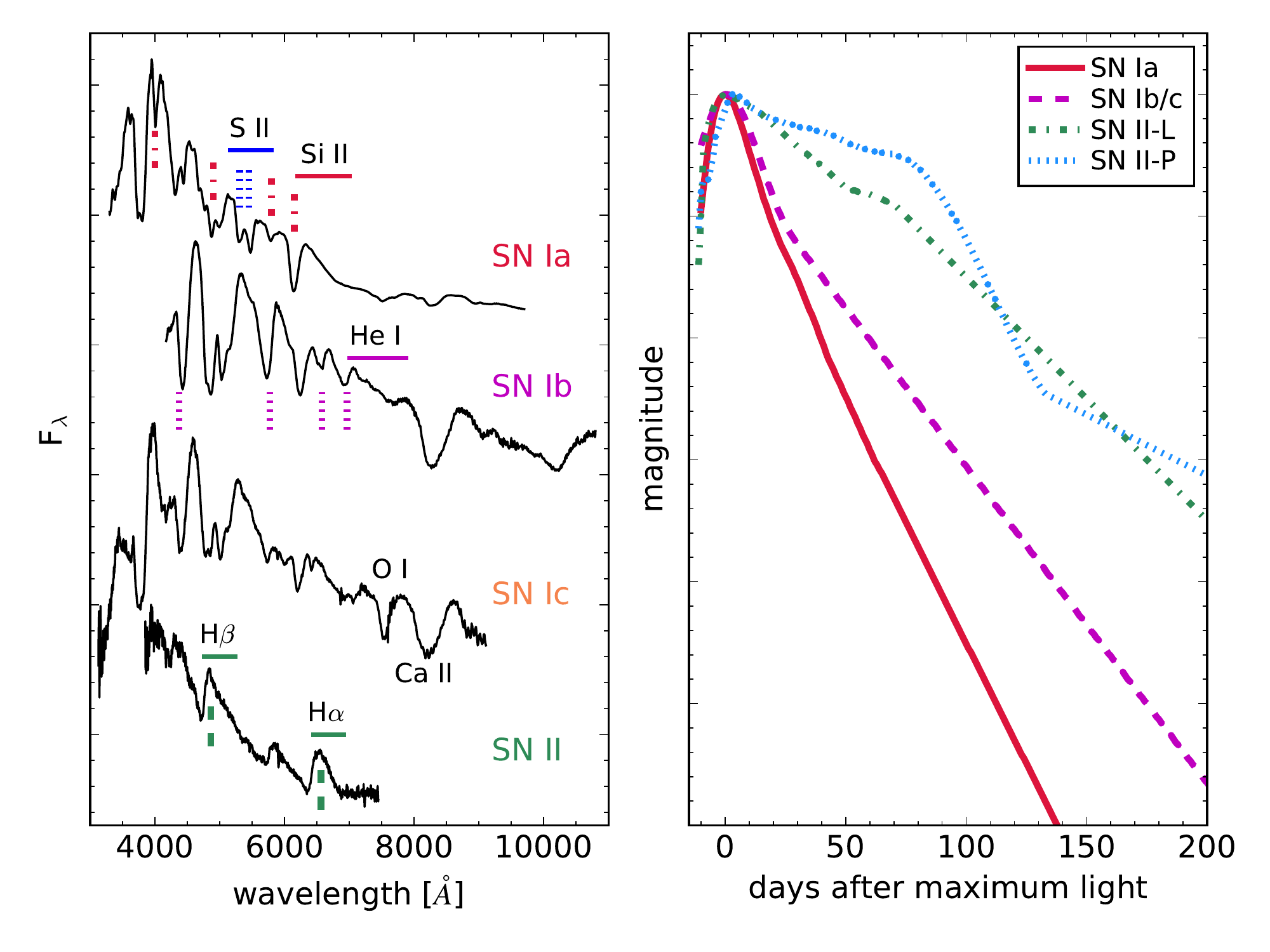}
\caption{{\it Left:} Representative spectra near maximum light of the main SN types (SN~Ia~2011fe \cite{pereira13}, Type~Ib~iPTF13bvn \cite{cao13}, Type~Ic~SN~2007gr \cite{valenti08}, and Type~II-P~SN~1999em \cite{leonard02}). The most prominent spectral features are indicated. The spectra are available in the public WISeREP repository \cite{yaron12}. {\it Right:} Average light curves of the main SN types \cite{li11}.
}
\label{fig_class}
\end{figure}

The SNe of type I are subdivided in three subclasses, depending basically on the presence or absence of Si\,{\sc ii} and He\,{\sc i} in the spectra (see left panel of Fig.~\ref{fig_class}). Type Ia SNe spectra present the line of Si\,{\sc ii} (rest wavelength 6347, 6371, and 6150 \AA) in absorption, while the spectra of the SNe Ib do not have these features but are characterized by pronounced lines of He\,{\sc i}, such as those at 5876, 6678 and 7065 \AA. Finally, the SNe Ic do not show Si\,{\sc ii} nor He\,{\sc i} lines (or He\,{\sc i} is very weak). Within this last subclass, we can distinguish events with fast-expanding ejecta (v $\sim$ 20000 \kms), named broad-line SNe Ic (Ic-BL) or hypernovae. These transients are sometimes related to long gamma-ray burst or X-ray flashes (SN~1998bw is the prototypical SN Ic-BL, discovered at the same time and location as GRB~980425; \cite{galama98}). Currently, it is known that type Ia SNe arise from the thermonuclear runaway of an accreting white dwarf (WD), whereas Type Ib and Ic originate from the core collapse of massive stars that had lost their hydrogen, or hydrogen plus helium envelopes before explosion, respectively.

The SNe II are also believed to be CC-SNe whose progenitors have retained (at least part of) their outer envelopes before exploding, reason for which their spectra are dominated by hydrogen lines all the time. As shown in the right panel of Fig.~\ref{fig_class}, SNe II are sub-classified as SNe II-P and SNe II-L, based on their light curve shape. The former exhibit light curves that decrease slowly after maximum for about three months displaying a ``plateau". The light curve of SNe II-L shows instead, a linear decline starting shortly past maximum. Recently, it has been argued that these two subclasses are at the extremes of a continuous distribution of SNe with different light-curve slopes (e.g. \cite{anderson14}). 

This classification, based on early phase spectra, is normally used when a new SN candidate is confirmed, but it is not always accurate, as it can also be that the appearance of a SN can change in time due to the characteristics of the progenitor or to those of the circumstellar material (CSM).

During the last part of the 1980s and the 1990s, the family of CC-SNe begun to grow with the identification of various sub-types. The SNe IIn (named after the detection of narrow emission lines in their spectra) present blue continuum spectra, with Balmer emission lines formed by several components that evolve in the time in various ways \cite{schlegel90}. The narrow component, with inferred full-width-at-half-maximum (FWHM) velocities from a few tens to a few hundreds \kms, are believed to arise from photoionised, slowly expanding gas which recombines and emits photons. This gas, most likely expelled by the progenitor star during the last phases of its evolutions, is located in the outer CSM and is not perturbed by the SN ejecta, at least during the early phases of the SN evolution. At later phases, intermediate-velocity line components (FWHM of a few thousands \kms) may be generated when the high velocity SN ejecta (a few $10^{-4}$) collides with the dense pre-existing CSM. Although the interaction may mask the innermost ejecta (as well as the explosion mechanism; \cite{chevalier94}), in some cases, particular geometric configurations may also favour the detection of high-velocity components (a few $10^{-4}$ \kms) arising from the photoionised SN ejecta. SNe IIn light curves are instead quite heterogeneous, showing both slow and fast declining SNe, as well as faint (M$_R \gtrsim -16$ mag) and very bright (M$_R \lesssim -19$ mag) objects (e.g. \cite{kiewe12} for a sample of SNe IIn). Note that recently it has been discussed that SNe IIn are not really a SN type, but an external phenomena where any type of SN (due to thermonuclear or core-collapse explosion) or not terminal outburst with fast ejecta and sufficient energy, interact with a slower and denser CSM. It produces a phenomena which appears or mimic what we know as a SN IIn. Therefore this sub-class is more commonly named ``interacting SNe" (e.g. see \cite{smith16}).

Finally, the spectrum of the SNe IIb is similar to that of the SNe II-P and II-L during maximum light, i.e. it has strong lines of H, but in the following week it metamorphoses to that of SNe Ib. This points out a physical link between these two classes (SN~1993J represents the prototypical object of this subclass; \cite{richmond94}), suggesting that SNe II and SNe Ib/c share a common origin, i.e., the CC of a massive star, but with just different amounts of stripping on the progenitors' outer layers.\\

Together the Type II-P and II-L represent the majority of CC-SNe (considering a volume-limited rate in the local Universe; \cite{li11}). Almost 9$\%$ is formed by interacting SNe. H-poor SNe (SNe IIb, Ib, Ic) constitute instead the remaining $\sim$ 37 $\%$. Recently, \cite{cappellaro15} find similar rates for CC-SNe groups considering a redshift range 0.15 $<$ z $<$ 0.35. Rare events like SN~1987A-like objects are estimated to form $\sim$ 3 $\%$ of the CC-SN population \cite{pastorello12}. 
\subsection{Observational constraints on the progenitors of core-collapse supernovae}
\label{sec_2_prog_obs}

In the last decades SNe have been studied in order to better establish the link to their progenitor stars and thereby to clarify the evolution of massive stars. 
 
\subsubsection{Observables of core-collapse supernovae}
\label{sec_2_indirectdet}

{\it Indirect clues} about the SN origin can be derived from the interaction of the material dismissed during the explosion with dense circumstellar medium lost by the progenitor during its turbulent life, or as said before, from their light curve and spectral evolution. 

\paragraph{Light curves}
The first observable of electromagnetic radiation from a SN is the shock-breakout. It is a short time scale (minutes for SNe with compact progenitors) in which the shock wave produced during the collapse of the stellar core,  reaches the stellar surface. In this instant a flash of soft X-rays and ultra-violet (UV) photons are released. The shock-breakout was directly observed in X-rays for the type Ib SN~2008D \cite{soderberg08}, whereas the fast cooling tail (due to the adiabatic expansion of the ejecta) after the shock-breakout has been observed in optical and UV bands for a few other objects (e.g. SNe~1993J and 2013df \cite{morales14}). This post shock-breakout phase mainly depends on the progenitor radius (e.g. \cite{chevalier92}).

Successively, the photons gradually leak out of the photosphere in a diffusion time scale ($t_{diff} \propto \rho \kappa R^2$, where $\rho$ is the density, $\kappa$ the opacity, and $R$ the SN radius) shorter than the expansion time scale  ($t_{exp} = R/V$, where $V$ is the expansion velocity). During this period (tens to days) the thermal shock energy decreases as the ejecta expands, and the radioactive decay of $^{56}$Ni (produced in the explosion, which has a lifetime of 8.8 days) and $^{56}$Co (lifetime of 111.3 days) becomes important, reaching a peak of luminosity. After this maximum the SN ejecta continue to expand and cool, eventually arriving to the hydrogen recombination temperature of the ejecta. While the recombination wave moves inward through the ejecta, the temperature remains practically constant. Hence, depending on the mass of the hydrogen envelope and the radius of the SN progenitor star, the SN luminosity could show a constant or plateau phase (e.g. the case of the SNe Type II-P), or a steep decline after maximum light (e.g. the SNe Type II-L). 

Once all the hydrogen has recombined, i.e. at late time (t $>$ 100 days), the light curve declines at the rate of the decay of $^{56}$Co $\rightarrow$ $^{56}$Fe (0.0098 mag d$^{-1}$). The late time light curve observations are useful to determine the mass of $^{56}$Co (and hence synthesized $^{56}$Ni). At times later than 1000 days past explosion, other radioactive elements with longer half-lives such as $^{56}$Co and $^{56}$Ti power the luminosity evolution of the CC-SNe.\\

Physical parameters from the SN explosion like the kinetic energy of the explosion, the total ejected mass, the mass of $^{56}$Ni synthesized and the radius of the progenitor, can be estimated through the fit of the SN bolometric light curves by semi-analytic functions (e.g. \cite{arnett82}) or more sophisticated hydrodynamical models (e.g. \cite{blinnikov00}). More details about the SN modelling can be found in Section \ref{sec:th:model}.

\paragraph{Spectral evolution}

SN spectral time evolution are a scan through the SN ejecta. They are important to study the chemical compositions of SNe and their progenitors, and the kinematic of the SN ejecta. A clear example is their utility is to classify SNe. \\

During the early phases or photospheric phase, the SN ejecta are not completely transparent and only the outer layers of the ejecta are observed. Consequently, spectra are characterized by showing a black-body continuum superposed by absorption and/or emission lines (e.g. see \cite{filippenko97} for a detailed review). These features are broad because the SN ejecta is expanding at high velocity. They often show P-Cygni profiles with blue-shifted absorption and red-shifted emission components.  Precisely via the absorption component of the P-Cygni profile it is possible to measured the velocity of the region at which the line predominantly forms (through the Doppler effect).

Late-time (nebular) SN spectra, taken several months after the explosion, are a unique way to peer into the very centre of the exploded stars. At those phases the opacity for optical photons has dropped substantially due to the expansion of the ejecta, so that the innermost parts, which were previously shielded by an effective photosphere, are now uncovered. The inner ejecta composition, unveiled by the nebular emission lines, is one of the most powerful tools to constrain the mechanism that gave rise to an explosion, since they all have a characteristic and widely unique nucleosynthesis. Late time spectra are characterized by strong emission lines on top of a faint continuum. Also the profile of a nebular emission line carries important information. The width of an emission line is a measure for the radial extent of the emitting species, and the detailed shape encodes asymmetries in its spatial distribution (e.g. \cite{taubenberger09}). 

When the SN ejecta interact with the CSM, the spectra present a blue continuum with superposed narrow emission lines, arisen from the CSM ionized by the shock interaction emission. If the CSM is thin, it is also possible to observe broader emission lines from the ionized ejecta. \\

The radiative-transport modelling of the SN late-time spectra can give us information about the kinematics of the ejecta and constrain the SN progenitor masses  (e.g. \cite{jerkstrand17}). For example, considering the sensitive dependency of the oxygen nucleosynthesis with the main-sequence mass of the star, the modelling of the [O\,{\sc i}] 6300, 6364 \AA~features helps us to constrain the progenitor mass (e.g. \cite{morales14}). \\

\paragraph{Supernova remnants}

Studying the observed properties of young SN remnants (SNR) is also an indirect path to connect with the SN progenitors (e.g. see the reviews \cite{chevalier05,patnaude17}). The remnant phase begins when, after the light from a SN fades away, the ejecta in expansion, cools down, and strongly interacts with the surrounding material, either the interstellar medium (ISM) or a more or less extended CSM modified by the SN progenitor. Thus SNR provide detailed information on the chemical composition of the ejecta, the explosion dynamics, and the progenitor star mass loss distribution. The most notable cases are the Galactic SNR Cassiopeia A, and the youngest know remnant, SN 1987A, in the Large Magellanic Cloud. The advantage of the study of Galactic SNR is that they can be resolved in fine detail. However, their link with CC-SN is complex given the large diversity of the CC-SN explosions and circumstellar environment, and the large mass range of the progenitors.

\subsubsection{Searching for SN progenitor stars}
\label{sec_2_directdet}

\begin{figure}[h]
\includegraphics[width=1\textwidth]{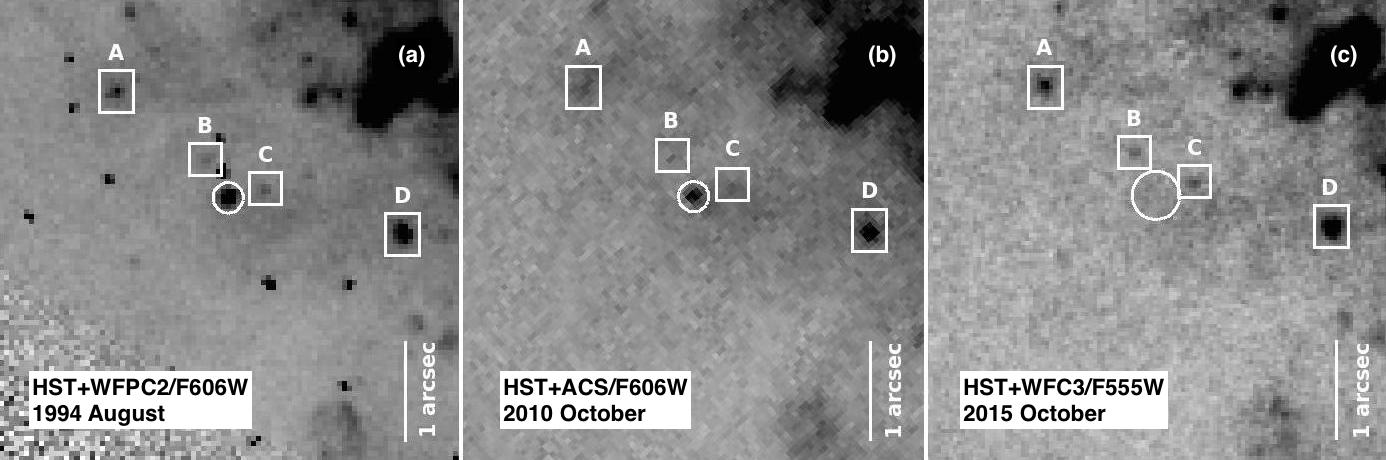}
\caption{Subsections of the pre-explosion ({\it panel a}), post-explosion ({\it panel b}), and late-time ({\it panel c}) {\sl HST} image of the SN~2010bt site. SN~2010bt is likely a Type IIn SN and its progenitor was a massive star that experienced a powerful outburst. The positions of the SN candidate progenitor and SN are indicated by {\it circles} each with radius of 3 pixels (between 0.08 and 0.15 arcsec), except for {\it panel c}, for which the radius is 6 pixels ($\sim$ 0.23 arcsec). The positions of three neighbouring sources of SN 2010bt, ``A", ``B", ``C", and ``D" are also indicated.
 }
\label{fig_prog}
\end{figure}

Still, the killer case is made with the {\it direct identification} of the star prior to explosion. Until the early 90s, only nearby events such SNe~1987A ($\sim$ 50 kpc; \cite{white87}) and 1993J ($\sim$ 3.6 Mpc; \cite{aldering94}) have allowed the direct progenitor identification in pre-explosion images. In recent years over a dozen CC-SN progenitors have been identified based on the inspection of archival, pre-explosion images. The identification relies on the positional coincidence between the candidate precursor and the SN transient. This requires high spatial resolution and very accurate astrometry because, at the typical distance of the targets ($>$ 30 Mpc\footnote{This distance limit is based on practical experience. \cite{smartt09} and \cite{eldridge13} set to 28 Mpc the distance limit for a feasible search of SN progenitors, although there are exceptions such as the massive progenitor of SN~2005gl at 60 Mpc \cite{galyam07}.}), source confusion becomes an issue. In practice, only {\sl HST} or 8-m ground-based telescopes mainly provided by adaptive optics images can be used to accurately pin-point (with a typical uncertainty of a few tens mas) the progenitor candidate. Even so, there is always the chance of mis-identification with foreground sources or associated companion stars. Thus, a final approach is visiting the SN field when the SN has weakened: if the candidate star has disappeared, then it was indeed the progenitor, otherwise it was a mis-identification (e.g. see \cite{maund09},\cite{vandyk13}, and Fig.~\ref{fig_prog}). 

Once a progenitor star candidate is identified, its initial mass and evolutionary state before the CC can be estimated by comparing its brightness and colour (if multi-colour imaging is available) measured with stellar evolution models. These models are chosen taking into account the metallicity in the SN environment, and the distance and the extinction to the star, derived from detailed light curves and spectral evolution of the SN with ground-based data (see Section \ref{sec:th:progenitor} for a discussion about stellar evolution models).

Taken together, the availability and depth of archive images of nearby galaxies is a determining factor that delimits the rate of SN progenitor stars identified. There is an approximate probability of about 25$\%$ to find an image of the host galaxy of a nearby SNe in the HST archive \cite{smartt09}.\\

Following the above or similar steps, direct detections or upper mass limits have been established for progenitors of some types of SNe: 

\noindent
$\bullet$ {\underline {Type II-P:}} Based on the statistics of around 15 SNe II-P, it appears that all of these progenitors exploded in the red supergiant phase from stars with initial mass range of 8-18 M$_{\odot}$, as we would theoretically expect (see Section \ref{sec:th:mass}). However there has been no detection of a higher mass stars in the range 20-40 M$_{\odot}$, which should be the most luminous and brightest stars in these galaxies. This has led to the intriguing possibility that higher mass stars undergo core-collapse, but form black-holes which prevents much of the stellar mass escaping the explosion \cite{reynolds15}. Theoretically, such quenched, low energy explosions have been proposed. Our lack of detection of high mass progenitors could be evidence for this missing population \cite{Kochanek:2008,smartt09}. But exceptions exist as the case of SN~1987A, considered a peculiar SN Type II because in spite of exhibiting prevalent hydrogen P-Cygni profiles in the early spectra, it had slow rise to maximum, faint, and broad light curves. Its progenitor was instead found to be a blue supergiant of around 14-20 M$_{\odot}$, with a hydrogen envelope of around 10 M$_{\odot}$.

\noindent
$\bullet$ {\underline {Type II-L:}} Only in a couple of cases, e.g. SNe~2009hd \cite{eliasrosa11} and 2009kr \cite{eliasrosa10,fraser10}, the progenitor has been identified and related with more yellow stars than expected by theoretical stellar evolution models, with estimated initial mass of between 18-25 M$_{\odot}$. However, a recent work argues that the progenitor of SN~2009kr is most probably a small compact cluster \cite{maund15}. It should be noted that it is no clear the separation between the historically defined SNe II-P and SNe II-L.

\noindent
$\bullet$ {\underline {Type IIb:}} These SNe likely originate from binary stripped progenitors with masses in the range 12-18 M$_{\odot}$. The precursor of the prototype SN~1993J (exploded in M81) was a K-type supergiant with $\sim$ 15 M$_{\odot}$ with a B-type star as companion (e.g. \cite{maund09}). In addition, it has been identified the progenitor of other four SNe IIb. SN 2011dh was found to have a yellow supergiant precursor \cite{maund11,vandyk11,bersten12}, likely part of a binary system with a compact companion \cite{benvenuto13}, and with relatively low initial mass ($\sim$ 13 M$_{\odot}$). This result was confirmed by the disappearance of the bright yellow star \cite{vandyk13}, and the detection of a UV source at the SN position \cite{folatelli14}. The progenitors of SNe~2013df \cite{vandyk14} and 2016gkg \cite{tartaglia17} have been also associated with moderate mass yellow supergiant stars of 13-17 M$_{\odot}$. Only in the case of SN~2008ax, the colours of the progenitor candidate might be those of a young Wolf-Rayet (WR) star (single massive stars that have lost their hydrogen envelopes and in some cases also helium layers), which had retained a thin, and low-mass shell of H, or with a system which has contaminated flux from an associated cluster or nearby stars \cite{crockett08}. 

\noindent
$\bullet$ {\underline {Type Ib/c:}} It has been proposed that these SNe arise from either massive stars with initial mass $>$ 25-30 M$_{\odot}$ (which have lost their outer hydrogen envelopes during a WR phase through radiatively driven winds or intense mass loss) or lower mass stars 8 $<$ M $<$ 30 M$_{\odot}$ in a binary system (which have been stripped of mass by their companion stars). Deep, high resolution images of pre-explosion sites of about 14 type Ib/c SNe have been analysed so far \cite{eldridge13,eliasrosa13}, but with not detections of progenitors, or progenitors systems in any of these. The probability of not finding a progenitor if these SNe originate from WR stars with initial masses $>$ 25 M$_{\odot}$ similar to those seen in the Milky Way and Magellanic Clouds is only 12 $\%$. Although some authors argue that the parameters derived for the observed WR population do not correspond to those of WR stars at the point of collapse \cite{yoon12,groh13}. In short, it is still not possible to set range of masses, or discriminate unambiguously between single WR stars or interacting massive stars in binary systems as the progenitors of Ib/c. Only for the case of iPTF13bvn \cite{cao13} a possible WR was detected at the position of the SN. However, further detailed stellar evolution modelling indicates that the progenitor candidate is a better match to a low mass progenitor in a binary system \cite{bersten14,eldridge15}. The recent analysis of post-explosion observations of the iPTF13bvn site finds the disappearance of the progenitor of iPTF13bvn, which is most likely a helium giant of 10-12 M$_{\odot}$ rather than a WR star \cite{eldridge16}.

\noindent
$\bullet$ {\underline {Interacting SNe:}} Some successful detection of progenitors of interacting SNe related these with either high luminosity or high-mass progenitors. A clear progenitor detection is the case of SN~IIn~2005gl \cite{galyam07,galyam09} which exploded at the same location as a luminous blue variable (LBV) star. For SN~2010jl \cite{smith11} it was also found blue flux coincident with the SN, likely from a very young cluster or a single massive star. \cite{smith11} argues that the initial mass of the progenitor should be, in any case, $>$ 30 \msun. In addition, the SN~Ibn~2006jc precursor was observed as a carbon-oxygen WR star embedded within a helium circumstellar medium with a mass between 60 and 100 \msun \cite{pastorello07}. It has been found pre-explosion outbursts for some of these objects. In fact, \cite{ofek14} allege that this activity is common in SNe IIn. As we will discuss in more detail in Section \ref{sec_3_newclass}, some interacting SNe show pre-SN photometric variability of the precursor stars which are invaluable to characterise the final stages of the progenitors. The most famous example is SN~2009ip for which were observed regular, but temporally sporadic, series of stellar eruptions in 2009 and 2012, before a final outburst in late 2012 (e.g. \cite{pastorello13,fraser13,mauerhan13a,margutti14}). A luminous progenitor for this transient (M$_V$ = -10) was detected in {\sl HST} archival images taken in 1999 \cite{smith10,foley11}. The nature of this transient or those of similar objects remains debated: some are undoubtedly genuine core-collapse SNe, while others may be giant non-terminal outbursts from LBV-like stars. The LBV progenitor scenario for SNe IIn is currently the favoured one, but given the large variety of this SN class it is natural to think about the existence of multiple precursor channels. For example, it has been suggested alternative scenarios such as a merger burst event in a close binary system (e.g. \cite{soker13}), or the collision of the SN ejecta of a red supergiant with the surrounded photoionization-confined shell formed through repeated mass loss events \cite{mackey14}. \\
 
\begin{figure}[h]
\includegraphics[width=1.\textwidth]{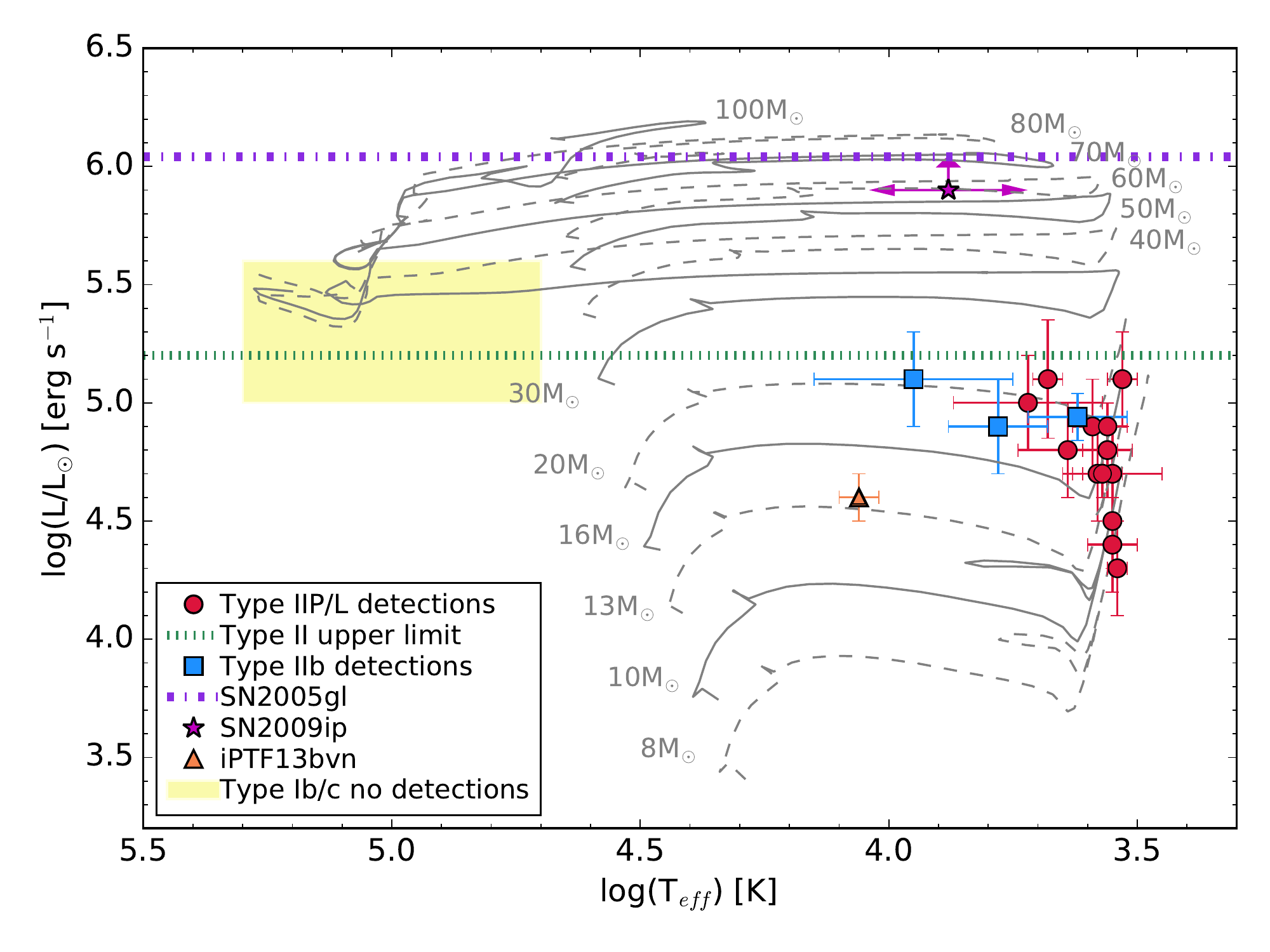} 
\caption{Hertzsprung Russell diagram showing the temperature and luminosity of the detected progenitor stars and upper limits of the main type of SNe presented in Section \ref{sec_2_directdet} and \cite{smartt15}. For comparison, model stellar evolutionary tracks from \cite{eldridge04} are also illustrated (figure reproduced from \cite{smartt15}).
}
\label{fig_hrd}
\end{figure}

Fig.~\ref{fig_hrd} shows a Hertzsprung-Russell diagram with a summary of the detected progenitor stars and estimated upper limits of the main SN types \cite{smartt15}. As we can see, although it is possible to delimit observationally SNe with their stellar progenitors, there are still many unknowns. Each case we study provides new clues.

\subsection{Theory of core collapse  supernovae}
\label{sec:th:ccsne}

Massive stars, with masses between  $\sim8$ and $\sim100~M_\odot$, end their lives collapsing under their 
own gravity. However, the most common kind of progenitors
for nearby supernovae and the likely origin of most observed neutron stars consist of 
isolated (or non-interacting) massive stars of a mass between $9$ and $30~M_\odot$ and solar metallicity.
We focus next on the case of a typical star of this kind while other kind of progenitors are explored in the next
sections.

\subsubsection{The collapse of a typical star}
\label{sec:th:typical}

In a typical star, the original material from which it was built, mainly hydrogen, has experienced a 
series of thermonuclear reactions leading to a stratified onion-like shell structure result of the intermediate ashes of
fusion chain reactions. The innermost region, the core, is composed by elements of the iron family, which posses the highest nuclear 
binding energy per nucleon and hence are unable to fusion further regardless of the density and temperature.
The iron core is fed by the surrounding silicon burning shell and it grows
until it reaches a mass of $1.2-2~M_\odot$, which is just below its Chandrasekhar mass \cite{Chandrasekhar:1938}. Typical conditions in the 
iron core, which determine its Chandrasekhar mass, are an electron fraction of $Y_{\rm e} \sim 0.42$, a specific entropy 
of $s\sim 1$~$k_{\rm B}$ per baryon ($k_{\rm B}$ being the Boltzmann constant), 
a temperature of $T\sim10^{10}$~K (about $1$~MeV) and a central density of $\rho\sim10^{10}$~g~cm$^{-3}$.
The Fermi energy of this gas of partially degenerate electrons is $\epsilon_{\rm F} \sim 8$~MeV, therefore, finite 
temperature corrections are important for the estimation of the ``effective" Chandrasekhar mass \cite{Woosley:2002}, which reads
\begin{equation}
\frac{M_{\rm Ch}}{M_\odot} = 5.83 Y_e^2  \left[ 1 + \left ( \frac{\pi k_{\rm B} T}{\epsilon_{\rm F}} \right )^2\right] \approx
1.03 \left ( \frac{Y_e}{0.42} \right)^2 \left [
1 + 0.15 \left ( \frac{kT}{1 {\rm MeV}} \right)^2 
\right ]
\end{equation}

\begin{figure}[h]
\centering
\includegraphics[width=0.49\textwidth]{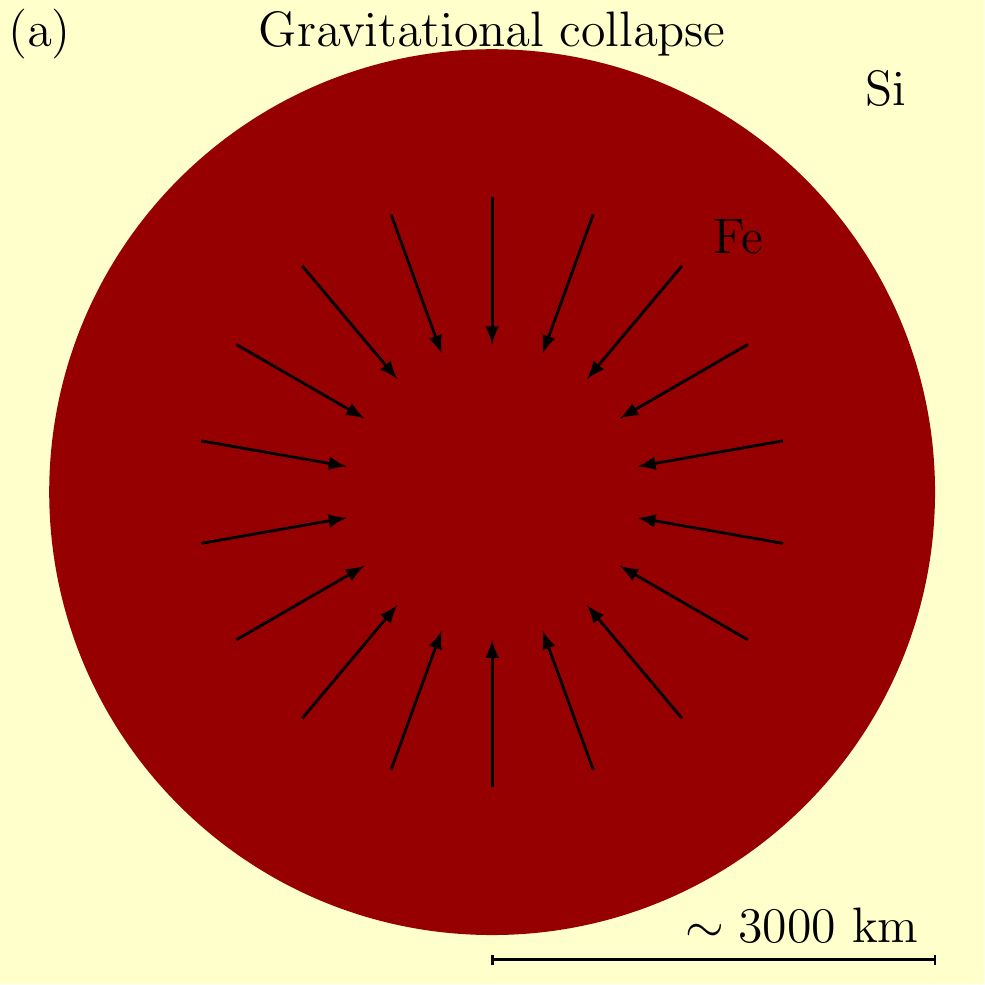}
\includegraphics[width=0.49\textwidth]{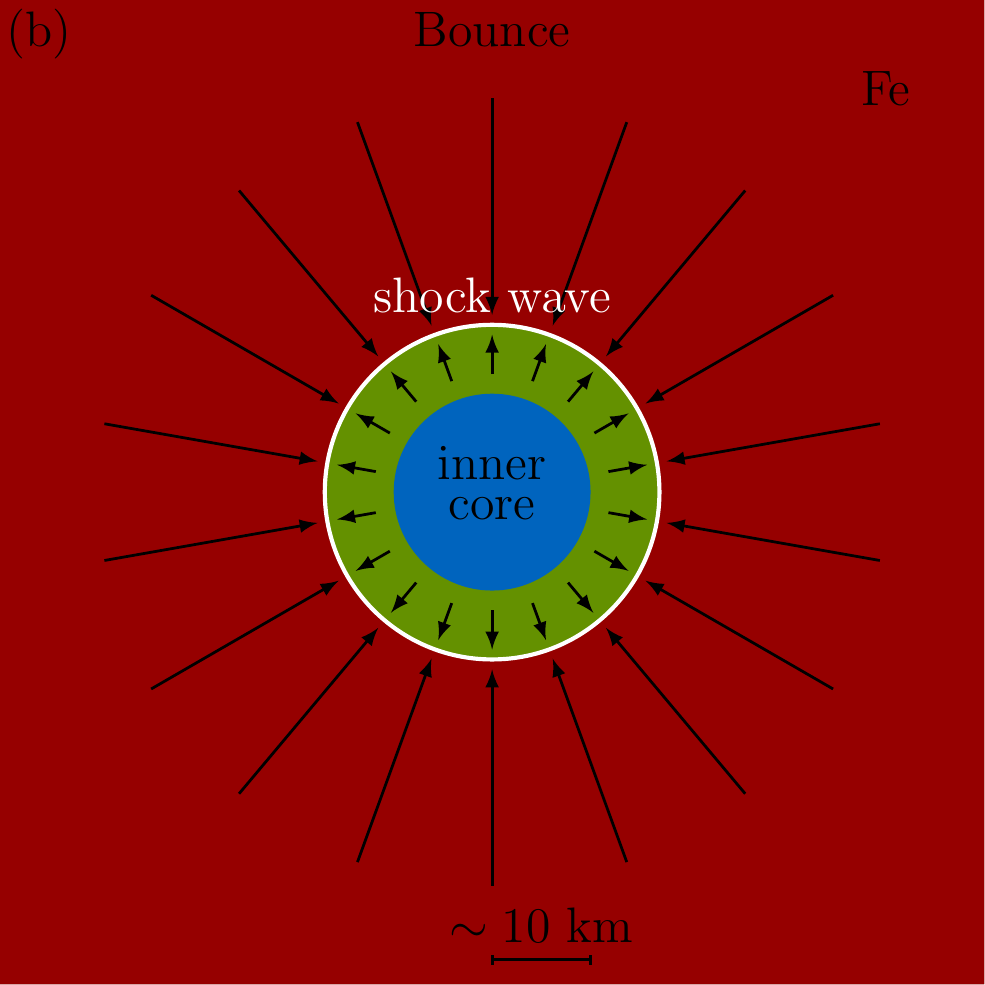} 
\includegraphics[width=0.49\textwidth]{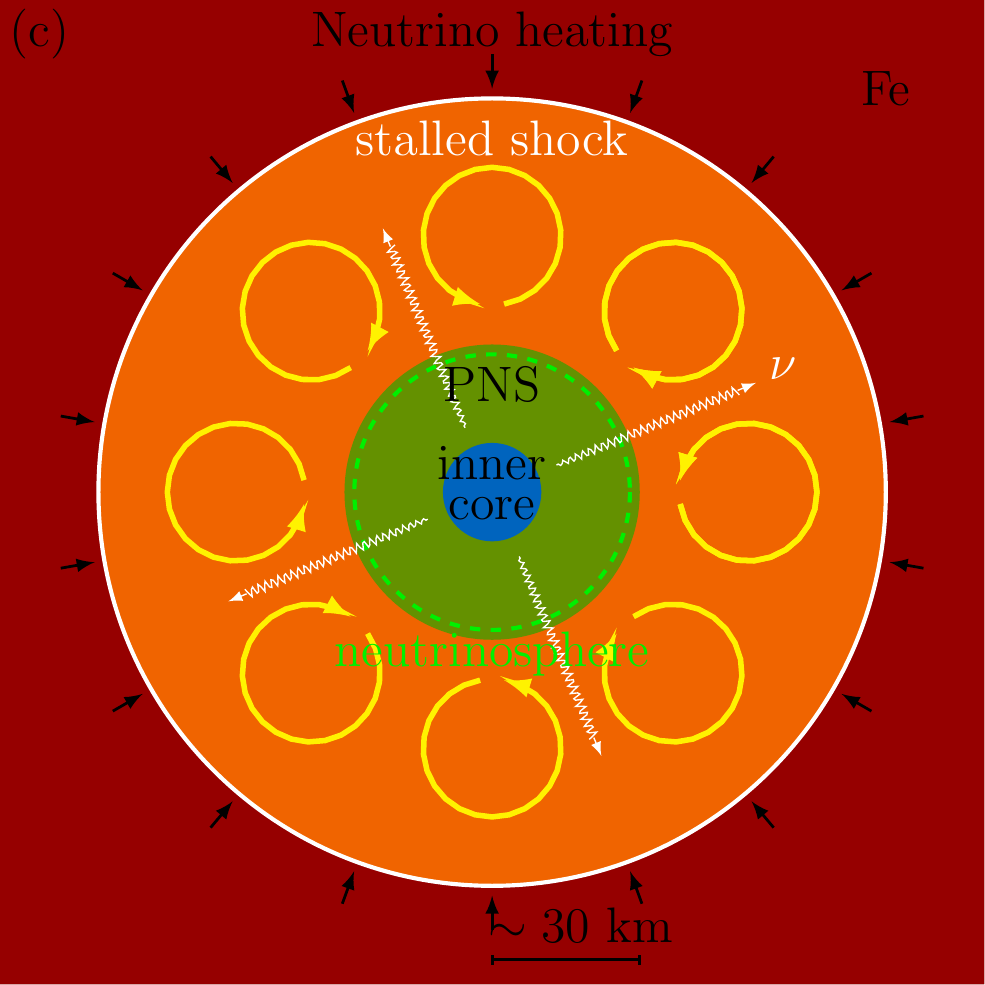} 
\includegraphics[width=0.49\textwidth]{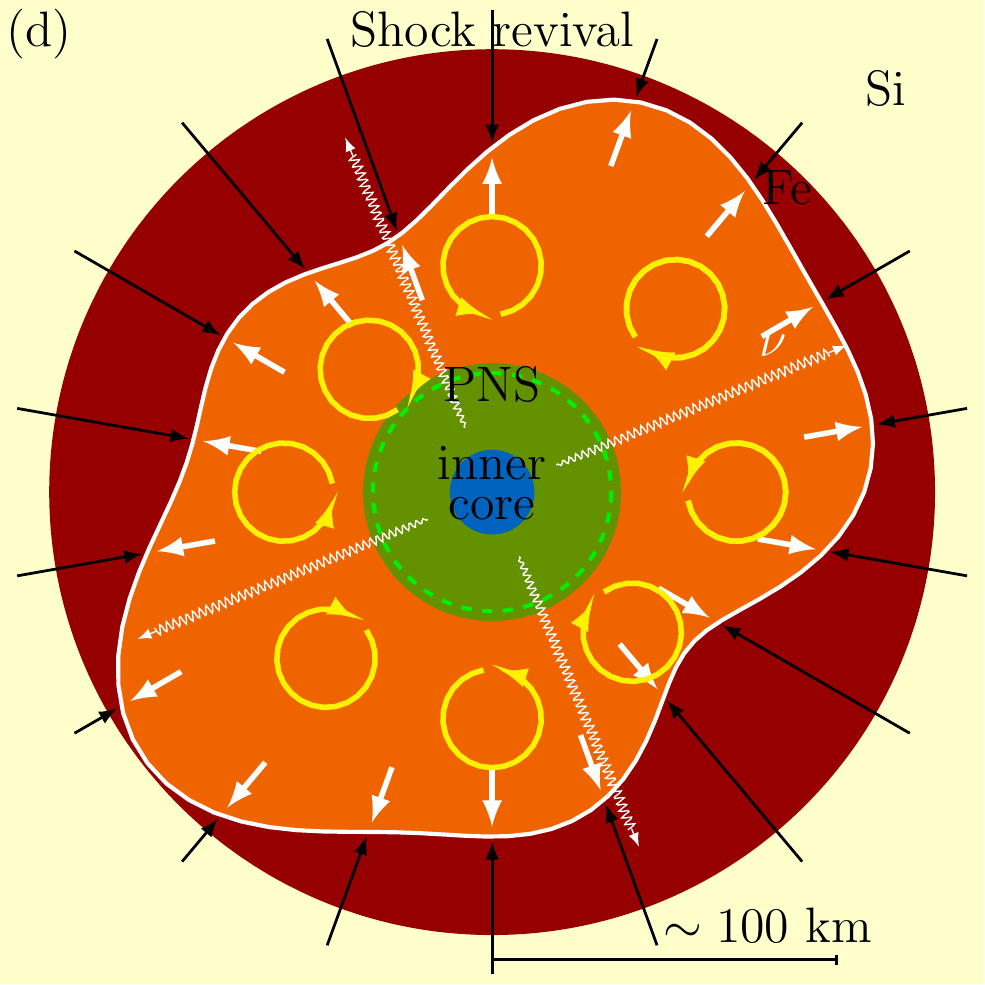} 
\caption{ Stages in neutrino-driven core collapse explosion:
gravitational collapse (a), core bounce (b), shock stagnation and neutrino heating (c) and shock revival (d).
Figure approximately at scale.
}
\label{fig:SNth}
\end{figure}

At this point the iron core starts collapsing due to two processes: 
first, electron captures by nuclei reduce the electron pressure  while the neutrinos produced carry away energy from the core.
Second, at densities above $10^{10}$~g~cm$^{-3}$, photo-disintegration 
of iron nuclei into alpha particles cools down the core. Both processes reduce the value of the Chandrasekhar mass until 
the pressure is not able to counteract the self gravity of the object and the collapse becomes inevitable (Fig.~\ref{fig:SNth}a).
The detailed structure of the progenitor, in particular the iron core, determines largely the outcome of 
the collapse and deserves a detailed discussion (see  section \ref{sec:th:progenitor}). 

In this case, the collapse progresses in dynamical timescales of several $100$~ms, enhanced by 
deleptonization and cooling processes.
Beyond densities of $\rho$ $\sim10^{12}$~g~cm$^{-3}$ neutrinos become trapped preventing further cooling and deleptonization. 
From this stage on, the collapse proceeds adiabatically inside the neutrinosphere, where matter reaches rapidly beta equilibrium.
At $\rho\sim2\times10^{14}$~g~cm$^{-3}$ nuclear interactions between nucleons become the fundamental source of pressure. This stops the collapse abruptly as the 
equation of state stiffens, i.e. as matter becomes less compressible. The supersonically infalling matter forms a shock at about $10$~km from 
the centre that propagates outwards (Fig.~\ref{fig:SNth}b). At the time of bounce, the shock encloses $\sim0.5 M_\odot$ of nuclear matter
the seed that will form a neutron star, which at this stage is known as proto-neutron star (PNS). The initial mass 
of the PNS is largely fixed by the deleptonization and cooling processes driving the collapse, which determine the 
local sound speed of the fluid and this, in turn, the location of the shock formation. As a result, massive stars always end their 
collapse with a bounce and the formation of a gravitationally supported object, a PNS, and never with a prompt 
collapse to a black hole\footnote{Note however that pop III stars have very massive ($M>100 M_\odot$) 
iron cores with high entropy ($s\sim 8 k_{\rm B}$ per baryon) and may form black holes promptly 
\cite{Sekiguchi:2011}.}.

As the shock propagates against the infalling material of the remaining iron core, nuclei are photo-dissociated into free
nucleons. About $10^{51}$~ergs per $0.1 M_\odot$ of infalling material is used in this process. This weakens the shock 
that stalls at about $100$~km from the centre in a time of a few $10$~ms (Fig.~\ref{fig:SNth}c).
The standing shock can survive for timescales of several $100$~ms and is effectively a gigantic transducer;  during 
the collapse of the iron core, a gravitational binding energy of 
\begin{equation}
E_{\rm bind} \sim \frac{G M_{\rm PNS}^2}{R_{\rm PNS}}  \sim 
10^{53} \left ( \frac{M_{\rm PNS}}{1 M_\odot}\right )^2
\left (\frac{30 {\rm km}}{R_{\rm PNS}} \right ){\rm erg}
\end{equation}
is released in form of kinetic energy of the infalling material, which is then transformed into thermal
energy at the shock. This is the main source of energy of all the events that will develop next.  At this time the PNS is growing rapidly 
in mass and has a size of about $30$~km. 
It consists of an cold inner core of $\sim10$~km radius, composed of the unshocked
material, and a thick hot envelope. The electron neutrinosphere is located right below its surface, preventing a rapid 
cooling of the PNS, which is now in thermal and beta equilibrium with neutrinos and stores about  $10^{53}$~erg of thermal energy.
Most of this energy is released as thermal neutrinos and antineutrinos streaming out in diffusion timescales of a few $\sim10$~s. 
However, this neutrino cooling process will not stop at that point but will continue for the next $10^6$~yr, acting as the main
cooling process in neutron stars. Under the appropriate conditions, a small fraction of the neutrino energy, of about $10^{51}$~erg, 
is deposited behind the stalled shock, which is revived disrupting the rest of the star and forming a supernova explosion
(Fig.~\ref{fig:SNth}d). 
This is the so-called delayed neutrino-heating mechanism \cite{Bethe:1985}.
If this mechanism fails, the PNS will grow in size until its maximum mass is reached and will collapse to a black hole
(see section~\ref{sec:th:mass}). 

In the region between the PNS surface and the shock, the temperature radial profile is such that neutrino emission rates,
responsible for the cooling, have a steeper radial decline than neutrino absorption rates, responsible for the heating.
This leads to the formation of a 
gain radius, above which there is a net absorption of energy by the fluid. This region is called the gain layer.
Neutrino heating is a self-regulated process because fluid elements being heated in the gain layer expand away, which in turn  
decreases the ability of this fluid to absorb energy. This process limits the amount of energy being released 
in a supernova explosion to $\sim10^{51}$~ergs and makes the outcome of the events invoking only the delayed neutrino-heating 
mechanism rather homogeneous (supernovae with additional sources of energy are discussed in the next sections).
As simple as it appears to be, heating proceeds in a rather complicated way due to multidimensional processes. 
Ultimately, what sets the amount of energy transferred to the shock is determined by the so-called dwell time, which is the
average time that takes for a fluid element to cross the gain layer \cite{Janka:2008,Buras:2006,Marek:2009}. Two processes
increase the residence time of matter in the gain layer and are crucial for the development of a successful explosion: convection and the 
standing accretion shock instability (SASI). On the one hand, convection overturn produces plumes of low-entropy matter that fall 
rapidly towards the PNS while hot bubbles of high entropy, which would otherwise fall into the PNS, rise constantly up into the gain layer.
On the other hand, the SASI \cite{Blondin:2003} produces a sloshing motion of the shock due to an unstable acoustic-advective cycle 
that couples the PNS surface with the shock \cite{Foglizzo:2000}. This instability enhances non-radial motions and helps to expand
the shock, which increases its energy deposition efficiency \cite{Marek:2009}. 
Which of both processes is the dominant effect leading to the supernova explosion is still matter of intense debate in the 
supernova modelling community. The analysis of \cite{Yamasaki:2007} suggests that SASI would be more favourable in cases
with lower neutrino luminosity while convection in cases with higher luminosities.
Regardless of this details, the explosion begins whenever the dwell time of matter in the gain layer is longer that the time 
required to transfer the amount of energy necessary to unbind this matter from the star. At this stage, the shock expands 
rapidly at the same time as it is continuously being heated by neutrinos. In this runaway situation, the globally expanding 
shock will gain energy for the next few seconds until reaching the final explosion energy. The shape of the expanding shock 
is highly asymmetric at this stage, dominated by either the sloshing motions produced by the SASI or by rising hot bubbles
produced by convection.

\subsubsection{Numerical modelling}
\label{sec:th:model}

While we start having a clear picture of the scenario in which neutron stars are formed, the exact conditions under 
which successful supernova explosions are produced are still not completely well understood. 
The modelling of core collapse supernova requires a wide variety of physical ingredients, including  among others
a nuclear physics motivated finite temperature equation of state, a detailed description of neutrino interactions  and general relativity
(see e.g. \cite{Janka:2007,Janka:2012,Burrows:2013,Mueller:2016}, for recent reviews).
The numerical modelling of this scenario is computationally challenging and even today, with the use of the largest 
scientific supercomputing facilities available, we cannot afford solving the full set of 6-dimensional Boltzmann equations, 
necessary for the neutrino transport,  or to use sufficiently high numerical resolution to resolve small-scale three-dimensional
turbulence that may affect the dynamics of the system. Additionally, the problem at hand, results in a set of non-linear equations
associated with the evolution of a fluid interacting with neutrino radiation, which, even if we could manage to solve them accurately, 
lead to a complex dynamics that may behave in a stochastic and chaotic way, very sensitive to small changes in the initial
conditions. Nevertheless, a huge progress have been made in the last four decades. We focus here on the current status of numerical
modelling and mostly in the developments of the last decade, which we hope will help the reader understanding the current uncertainties
in the theoretical knowledge of the scenario in which neutron stars are born. A more complete and historical perspective of the 
numerical modelling of supernova can be found e.g.  in \cite{Janka:2012}.

Most of the efforts have been focused in developing an accurate description for the neutrino transport in simulations, since this
is a crucial ingredient for the revival of the supernova explosion. So far, state-of-the-art multi-energy group solvers for 
three-flavour neutrino transport, including energy-bin coupling terms and velocity-dependent corrections, have only been 
possible in spherically symmetric (1D) simulations \cite{Yamada:1999,Liebendoerfer:2004}. Under such restrictive
symmetry conditions, the energy deposited by neutrinos is not sufficient to revive 
the shock and simulations fail to produce successful supernova explosions. 
The exception are stars with O-Ne-Mg cores (see section \ref{sec:th:progenitor}). 
In order to tackle the multidimensional case, one has to consider approximations
of the Boltzman transport equations. The two-momentum approximation is a popular choice, 
in which only equations for the neutrino number, energy and momentum are solved, supplemented
by an Eddington factor closure. This approach has been used in 1D simulations \cite{Burrows:2000,Mueller:2010}
and extended to multidimensional (2D and 3D) simulations in the so-called ray-by-ray-plus (RbR+) scheme \cite{Rampp:2002,Buras:2006}.
In this scheme a set of radial 1D problems is solved  for each angle in a spherical polar coordinates grid. Angular couplings appear
through neutrino pressure gradients and the advection of neutrinos trapped with the fluid in the optically thick regions. 
Truly multidimensional, multi-energy schemes have been only possible in 2D. The most sophisticated version of those 
are multi-group flux-limited (MGFD) schemes  \cite{Burrows:2006,Burrows:2007a,Burrows:2007b,Swesty:2009}, specially 
schemes including multi-angle treatment \cite{Ott:2008,Brandt:2011}.

As an alternative to those computationally expensive Boltzmann solvers, a series of additional approximations have appeared in the 
literature: In the M1 scheme \cite{Obergaulinger:2014}, a local algebraic closure was used in the two-momentum transport equations
yielding a set of hyperbolic equations. In the isotropic diffusion source approximation (IDSA) \cite{Liebendoerfer:2009}, neutrino 
distribution is decomposed into a trapped and a free streaming component. The fast multi-group transport (FMT) method  
\cite{Mueller:2015a,Mueller:2015b} solves the stationary neutrino transport problem in a ray-by-ray fashion. The so-called 
neutrino leakage scheme \cite{Oconnor:2010} is a simple prescription to describe cooling and heating of energy-averaged (grey)
neutrinos, which can be easily applied in multidimensional scenarios with a ray-by-ray approach \cite{Ott:2012}. The leakage
scheme has also been extended to the multi-energy group case \cite{Perego:2016}.
Although the results of the latter series of approximations cannot be truly compared to RbR+ or MGFLD methods, they are useful
to explore a large space of physical parameter or to study in details certain processes taking place during the explosion 
(e.g. instabilities, 3D effects and magnetic fields).

Also crucial is an accurate and complete treatment of neutrino interactions with matter. These set the emission 
and absorption rates, which are essential to determine neutrino luminosity and energy deposition rates at the shock
driving the supernova explosion. The most relevant interactions included in numerical simulations
(see e.g. \cite{Buras:2006,Lentz:2012}) are: $\beta$~processes, including neutrino and antineutrino absorption
and emission processes by nucleons \cite{Burrows:1998} and nuclei \cite{Langanke:2003}; 
scattering of neutrinos with nucleons \cite{Burrows:1998}, 
nuclei \cite{Horowitz:1997}, electrons and positrons \cite{Mezzacappa:1993};
nucleon-nucleon bremsstrahlung \cite{Hannestad:1998}; and neutrino-antineutrino annihilation to produce 
either electron-positron pairs \cite{Bruenn:1985,Pons:1998} or different flavour neutrino-antineutrino pairs
\cite{Buras:2003}. Special care has to be taken in the computation of these interactions to include
inelastic terms in the scattering that produce an interchange of neutrino energy with their targets,
Pauli blocking factors, high-density nucleon-nucleon correlations and weak magnetism corrections,
among others (for a more complete description see \cite{Janka:2012} and Chapter 9; 
for a particular implementation description see  \cite{Rampp:2000}).
 The use of incomplete 
sets of interactions with different corrections implemented has been one of the sources of disagreement 
in the results of numerical simulations among different supernova groups. This tendency has changed 
in the last few years as the main groups have adopted a complete and accurate set of interactions,
which has facilitated direct comparisons (see e.g. \cite{Burrows:2016}).

General relativity (GR)  also plays an important role in the dynamics of the supernova explosion \cite{Mueller:2012}.
In comparison to Newtonian gravity, GR deepens the gravitational potential well leading to 
more compact PNSs. This increases the amount of gravitational binding energy released during the collapse,
which adds up to the energy budget of the PNS and has a positive impact in triggering the explosion.
Full GR simulations of the core collapse scenario (e.g. \cite{Ott:2007a,Ott:2013, Moesta:2014,Abdikamalov:2015})
have been performed in the BSSN formulation  \cite{Shibata:1995,Baumgarte:1999}.
To avoid the use of computationally expensive full GR codes, several approximations have been developed. 
The conformally flat condition (CFC) approximation \cite{Isenberg:2008,Wilson:1996}
is a waveless approximation to GR, which is exact in spherical symmetry, and has been used 
in core collapse simulations codes (see e.g. \cite{Dimmelmeier:2002,Mueller:2010,Cerda-Duran:2013}).
Direct comparisons of the CFC approach with full GR simulations have shown that differences in the 
dynamics are minute \cite{Shibata:2004,Ott:2007a,Ott:2007b}. The CFC approximation has been 
reformulated (XCFC, \cite{Cordero-Carrion:2009}) to overcome uniqueness problems; with this 
improvement it is possible to study the formation of  black holes \cite{Cerda-Duran:2013} and, with the use of excision techniques
\cite{Cordero-Carrion:2014}, its posterior evolution. Second post-Newtonian corrections to the CFC metric 
(CFC+, \cite{Cerda-Duran:2005}) showed only quantitatively small differences ($<1\%$) in the dynamics.
Another popular approximation is the use of an effective pseudo-Newtonian
potential to mimic GR effects \cite{Rampp:2002,Marek:2006}. This approximation is widely used by the 
core-collapse community (see e.g. \cite{Buras:2006,Kitaura:2006,Obergaulinger:2006,Scheidegger:2008,
Wongwathanarat:2013,Hanke:2013,Bruenn:2013,Oconnor:2015,Bruenn:2016,Summa:2016,Burrows:2016}) 
and it produces an excellent agreement with GR in spherical symmetry. However, it agrees only qualitatively in 
the presence of fast rotation, for which either CFC of full GR are better suited. Regarding the computation of
gravitational waves, the approximate quadrupole formula is 
used almost exclusively in all simulations, regardless of the gravity treatment. This approximations has been shown 
to be accurate for the mildly relativistic gravity of  these system even when compared with more sophisticated waveform
extraction techniques \cite{Reisswig:2011}.

Additionally, simulations need realistic equations of state (EOS) able to handle a wide range of conditions
present in the core collapse scenario. These general purpose EOS, also known as Supernova EOS, typically
can handle densities from $\sim 10^8$ to $10^{15}$~g~cm$^{-3}$ and finite temperature dependence
to cover all steps in the collapse from the iron core to the final neutron star. The composition
is parametrised by the value of the electron fraction because nuclear statistical equilibrium (NSE) is considered.
This is possible since, for the typical conditions inside the iron core and in the PNS, nuclear reactions occur in 
timescales much shorter than the dynamical evolution timescales. These EOSs consider the fluid 
composed by a mixture of heavy nuclei, alpha particles, photons, neutrons and protons.
Several families of EOS are available including all these conditions: LS \cite{Lattimer:1991}, 
STOS \cite{HShen:1998a,HShen:1998b,HShen:2011}, FYSS \cite{Furusawa:2011,Furusawa:2013},
HS \cite{Hempel:2010}, SFH \cite{Steiner:2013} and SHO/SHT \cite{GShen:2011a,GShen:2011b}.
Additional details of those EOS and their variants can be found in the recent review by \cite{Oertel:2017}
and on Chapter 6.
The main uncertainties in EOS calculations are the properties of nuclear matter, parametrised in terms of constants 
such as the incompressibility modulus $K$ and the symmetry energy $J$, among others. Their values are currently constrained by 
nuclear physics experiments and neutron stars observations \cite{Oertel:2017}.
The impact of these EOS uncertainties  in the dynamics of the collapse is only moderate. 
It mainly affects three aspects:
first, below nuclear density, most supernova EOS use one average nucleus to describe all nuclei. This has an impact 
on electron capture rates, which in turn affects the shock formation location and the size of the inner core
\cite{Lattimer:2000,Langanke:2003,Sumiyoshi:2005,Suwa:2013, Steiner:2013}.
Secondly, the properties of matter at supranuclear densities influence the structure of the PNS. For a ``soft"
EOS, i.e. that producing more compressible matter at typical post-bounce densities, more compact PNSs are formed.  
This enhances neutrino emission and favours supernova explosions \cite{Marek:2009, Hempel:2012,Suwa:2013}.
Lastly, the EOS has an impact on the maximum mass that a neutron star can support, which is relevant for the possible
formation of a black hole (see section \ref{sec:th:mass}).

For the treatment of the matter outside the iron core, NSE is no longer a good approximation and nuclear burning
happens in longer timescales than dynamics ones. The EOS at densities below $\sim10^8$~g~cm$^{-3}$ depends on the full 
composition of the fluid in terms of the mass fractions of 
different isotopes, which have to be advected with the fluid during the evolution. Therefore, at low densities, the supernova EOS
described above is matched to a full composition EOS. A popular choice is the Timmes EOS \cite{Timmes_Arnett:1999,Timmes:2000}.
To avoid computationally intensive calculations of nuclear reaction networks (see e.g. \cite{Timmes:1999}), 
phenomenological ``flashing" prescriptions have been considered (see e.g. \cite{Rampp:2002,Buras:2006}) 
or simply NSE as a crude  approach (see e.g. \cite{Oconnor:2010}).

The most advanced core collapse simulations include all the physics described above using state-of-the-art numerical methods
and the best available approximations in each case. In axial symmetry (2D) recent simulations including full sets of neutrino 
interactions, three-flavour neutrinos and general relativistic corrections to the gravitational potential, have resulted in successful 
explosions for a wide range of progenitors ($12$-$25~M_\odot$) using both the RbR+ approach
\cite{Mueller:2014,Summa:2016,Bruenn:2016} and 2D-MGFLD transport \cite{Burrows:2016}. Comparison among 
different groups shows a qualitative agreement in the results but still differences in the time of the onset of the explosion 
and its energy \cite{Burrows:2016}. Analogous simulations with similar physical content have been performed in 3D by
 \cite{Lentz:2015,Melson:2015b} using a ray-by-ray approach for the neutrino transport. In both cases only low mass
 progenitors were considered ($15$ and $9.6~M_\odot$ respectively) and although both cases produce successful explosions, 
 3D effects weakened or enhanced the explosion with respect to 2D, depending on the work. Similar work for a $20~M_\odot$
 progenitor did not produced successful explosions unless strange-quark contributions to neutrino-nucleon scattering
 were considered. 2D and 3D simulations with less sophisticated neutrino transport and incomplete neutrino physics 
 \cite{Takiwaki:2016} (3D Newtonian, IDSA transport), \cite{Roberts:2016} (3D GR, M1 transport),
 \cite{Dolence:2015} (2D Newtonian, MGFLD transport), \cite{Nagakura:2017} (2D Newtonian, Boltzmann transport),
 \cite{Oconnor:2015} (2D GR, M1 transport) and \cite{Suwa:2016, Pan:2016} (2D Newtonian, IDSA),
 have also shown successful explosions in some cases but not in all models.

Overall, results from numerical 2D simulations show that, when an appropriate and complete description of the physics
involved is considered, it is possible to produce consistently successful supernova explosions. However, the more realistic 
3D case still proves to be a challenge. This is an indication that either the approximations used in the simulations 
are not sufficiently accurate (specially regarding neutrino transport) or that some important physical ingredient is missing.
Unresolved hydrodynamic turbulence could be also be a problem in 3D simulations
\cite{Melson:2015a, Abdikamalov:2015,Couch:2015} and should be investigated in more detail in the future.

\subsection{Progenitor dependence}
\label{sec:th:progenitor}

In section \ref{sec:th:ccsne} we have described the scenario in which most neutron stars are born, 
what is thought to be the typical  progenitor star of most supernova explosions. Here we discuss 
how this scenario changes when different progenitor are considered. There are several factors in 
stellar evolution that lead to differences in the progenitors of core collapse supernovae. The main one
is its mass at birth, the so-called zero-age main-sequence (ZAMS) mass. Additionally, there are variations
in the metallicity of the environment in which the star was born, its rotation rate, and the presence of a 
companion star.

\subsubsection{O-Ne-Mg cores}
\label{sec:th:onemg}

\begin{figure}[h]
\includegraphics[width=0.8\textwidth]{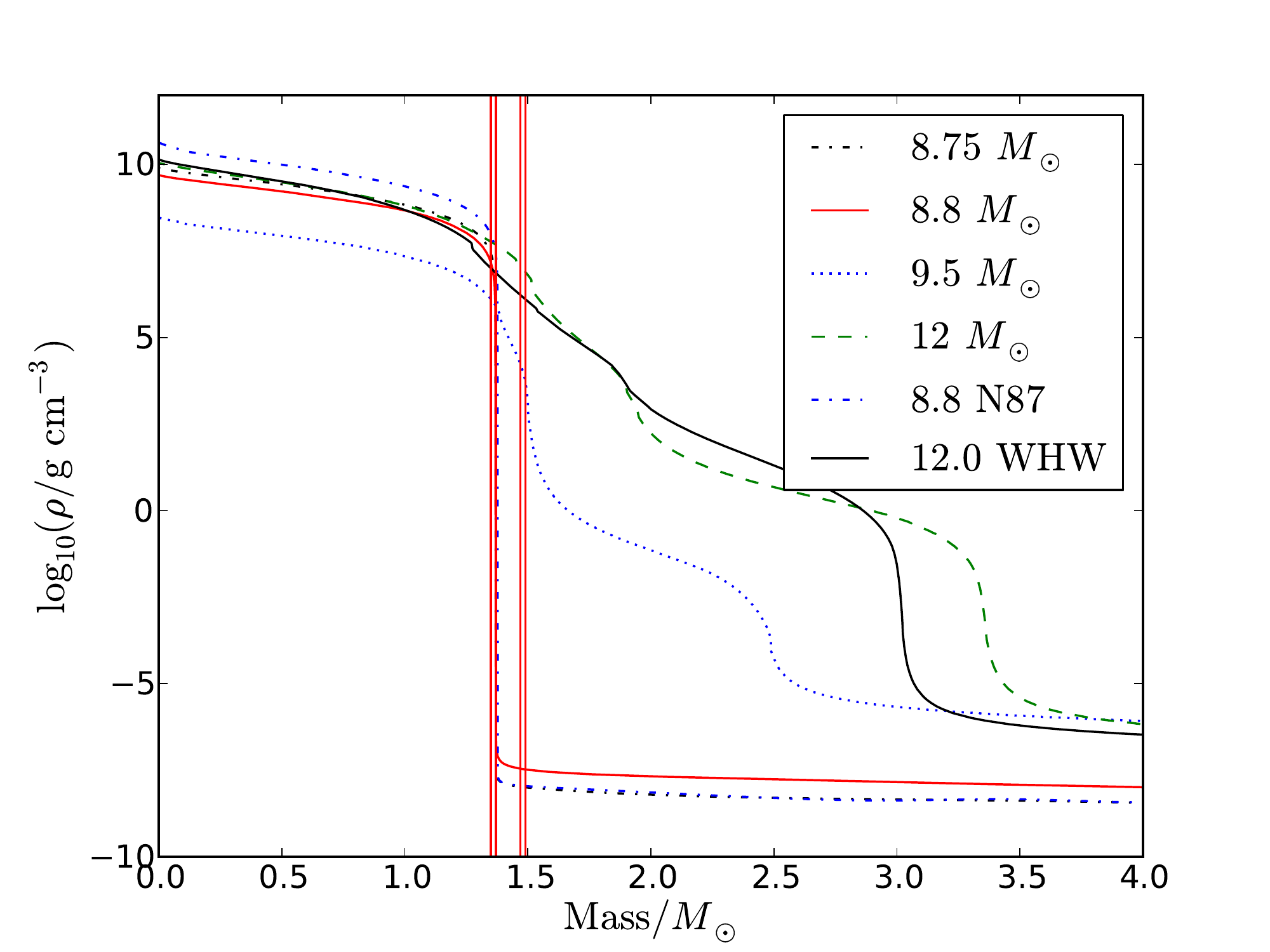}
\caption{Density profile as a function of mass coordinate for $8.75 - 12~M_\odot$ 
pre-supernova models of \cite{Jones:2013}, the $8.8~M_\odot$ model of \cite{Nomoto:1987} (N87), and  
a $12~M_\odot$ model from \cite{Woosley:2002}(WHW). From these models, those with a mass of $9.5$
or $12~M_\odot$ have formed an iron core and the rest an O-Ne-Mag core.
Vertical red lines show derived pre-collapse masses for the two peaks in the observed neutron star distribution of 
\cite{Schwab:2010} (Figure reproduced from \cite{Jones:2013}).
}
\label{fig:ECSN}
\end{figure}

Stars with ZAMS mass of $8 - 10~M_\odot$, can burn carbon to produce O-Ne-Mg cores. However, for such low mass
stars, temperatures are not sufficiently high to ignite Ne. Instead, the core, which is close to electron degeneracy,
grows to a mass of about $\sim 1.34~M_\odot$, close to the Chandrasekhar mass. At this point, electron captures, the dominant process under
this conditions, take over and trigger the collapse of the core. The resulting SNe are the so-called electron capture
supernovae (ECSNe) \cite{Nomoto:1984,Nomoto:1987}. The main feature of these progenitors is that they have a very 
steep density gradient outside the core (see Fig.~\ref{fig:ECSN}), 
with very thin carbon and helium layers and a rapid transition to the hydrogen envelope.
Due to this structure, once the core bounces, only a small fraction of the neutrino energy is necessary to power the explosion, 
shortly after bounce. Numerical simulations of this scenario show that it is possible to obtain successful explosions even in 1D
models \cite{Kitaura:2006,Fischer:2010}. This has been confirmed by multidimensional simulations \cite{Janka:2008}. 
The resulting supernova explosions are weak ($\sim 10^{50}$~erg) and Ni-poor. They are possibly associated with 
some subluminous type II-P supernovae e.g. see \cite{Smartt:2009,botticella09}. The Crab remnant associated with SN 1054 is likely the result of such an 
explosion \cite{Hillebrandt:1982,Nomoto:1982}.

For solar metallicity stars, the range in which O-Ne-Mg cores are formed is limited to a 
small interval ($\sim 0.2~M_{\odot}$) close to $9~M_\odot$ \cite{Poelarends:2008,Jones:2013}.
Depending on metallicity and the interaction with a close companion the range of masses may shift and 
widen \cite{Podsiadlowski:2004,Pumo:2009}.
As a consequence, the ECSNe represent a contribution of $\sim4\%$ to all supernovae 
in the local universe \cite{Poelarends:2008}. Comparable results are obtained when comparing
abundances of r-process elements in ECSNe simulations with observations of stars in the Galactic halo 
($\sim 4\%$ of all core-collapse SNe, \cite{Wanajo:2011}). This results are consistent with an estimated rate of low-luminosity 
type II SNe of $4-5\%$ \cite{Pastorello:2004}.

Neutron stars formed in this scenario are expected to have a lower mass than typical neutron stars formed from
progenitors with iron cores and appear in a very narrow range of masses fixed by the Chandrasekhar mass of
the O-Ne-Mg core \cite{Nomoto:1987, Podsiadlowski:2004}. 
Velocity kicks are also expected to be low \cite{Podsiadlowski:2004}, since there are neither large asymmetries involved
or a large amount of ejected mass (see section~\ref{sec:th:kicks}). This theoretical expectation matches some evidence
of a bimodal distribution in the observed masses of neutron stars that have not undergone accretion
 \cite{vandenHeuvel:2004,Schwab:2010}. The interpretation is that the lower and higher 
mass populations would correspond to a ECSNe and iron-core SNe, respectively. However, this bimodal distribution
has not been found in a more recent analysis of the observations \cite{Oezel:2012}.
\cite{Schwab:2010} also indicated that the observed orbital eccentricity of neutron stars in binaries is lower in cases with lower 
NS masses, which is consistent with the low velocity kicks resulting of ECSNe.

In the particular case of double neutron stars (DNS), population synthesis calculations favour systems with 
an ECSNe \cite{Andrews:2015}. The reason is that ECSNe produce weaker explosions with lower mass ejecta than 
iron-core SNe. This prevents the binary to unbind, leading to a low eccentricity binary neutron star. These results 
are qualitatively compatible with the narrow observed distribution of NS masses in DNS systems, with a mean mass of  $1.33~M_\odot$
and a dispersion $<0.1~M_\odot$ \cite{Oezel:2012,Oezel:2016}. Interestingly, these
are precisely the candidates for neutron star mergers. Therefore, neutron stars formed from O-Ne-Mg cores may be crucial 
for understanding phenomena such as short GRBs and to estimate merger rates relevant for GW detectors. 

\subsubsection{Iron cores in solar metallicity stars: mass dependence and black hole formation}
 \label{sec:th:mass}

\begin{figure}[h]
\includegraphics[width=0.47\textwidth]{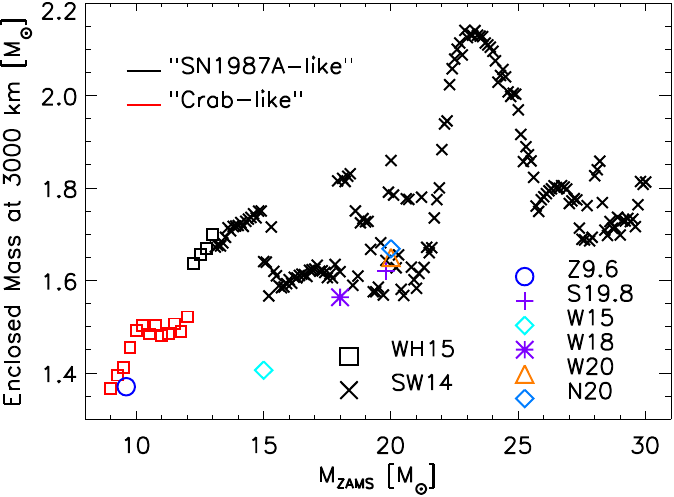}
\raisebox{-0.01\height}{\includegraphics[width=0.52\textwidth]{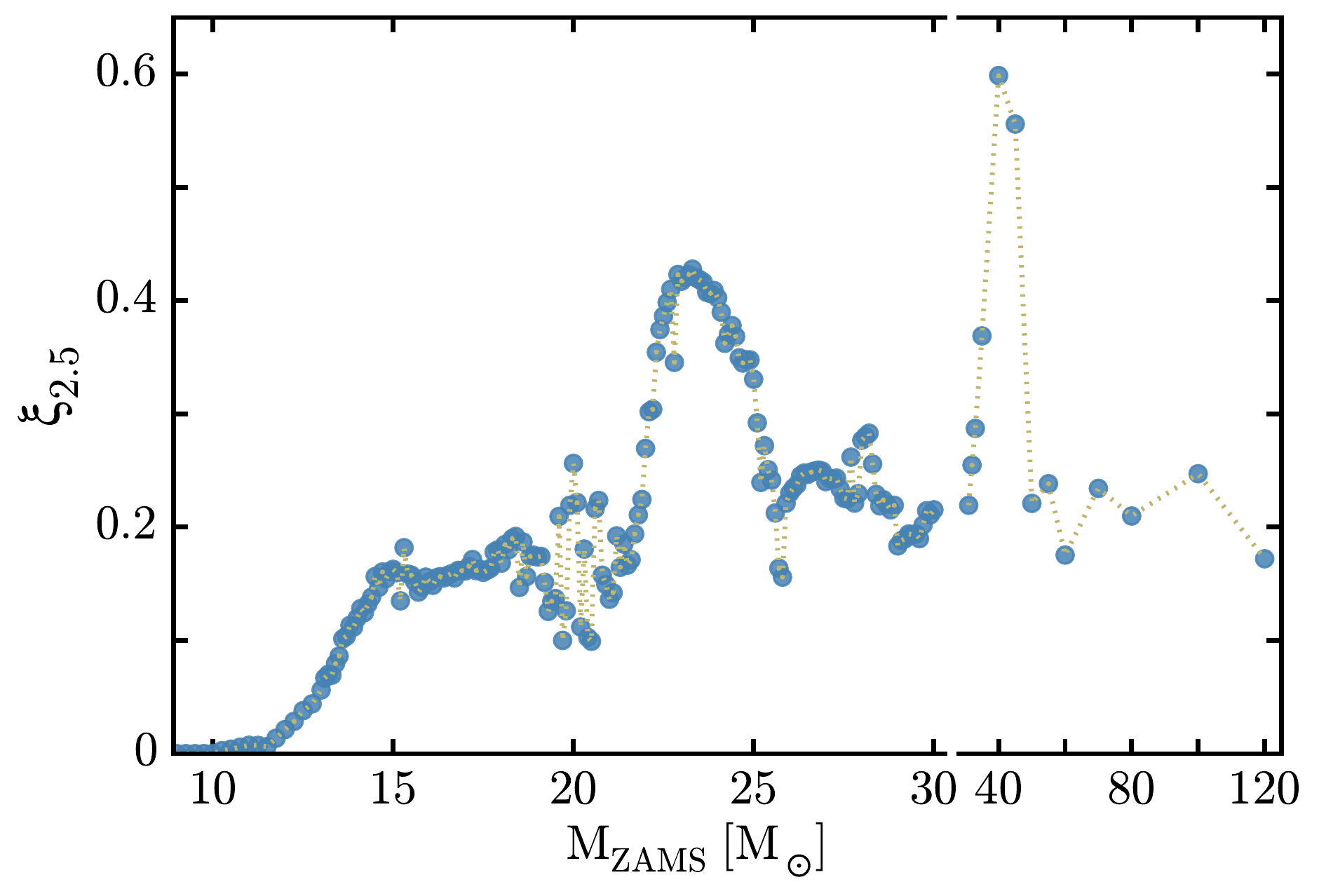}}
\caption{(Left) Mass inside a radius of 3000\,km for progenitors up to
  30 $M_\odot$, at the time when the central density reaches the same
  value of $3\times 10^{10}$\,g\,cm$^{-3}$, as a function of ZAMS mass. Black and red symbols denote models
  forming an iron core (``SN1987-like'') or an O-Ne-Mg core (``Crab-like''), respectively.
  Progenitor models are the results from simulations by \cite{Woosley:2015} (WH15, squares) and
  \cite{Sukhbold:2014}(SW14, crosses). The rest of the symbols are calibration models to SN~1987A 
  (see \cite{Sukhbold:2016}, for details).
  (Right) Compactness parameter $\xi_{2.5}$ (see Eq.~(\ref{eq:compactness})) as a function of
  ZAMS mass for the 200 models between 9.0 and 120 $M_\odot$ of \cite{Sukhbold:2014,Woosley:2015}
  used in \cite{Sukhbold:2016}. Note the scale break above $32~M_\odot$.
(Both figures reproduced from \cite{Sukhbold:2016})
}
\label{fig:progenitors}
\end{figure}

We consider next mass dependence in solar metallicity stars forming an iron core, i.e. those with a mass
above $\sim9 M_\odot$. The evolution of massive stars with solar metallicity to their pre-supernovae stage has been studied 
by a number of authors \cite{Woosley:2002,Limongi:2006,Nomoto:2006,Woosley:2007, Chieffi:2013,Sukhbold:2014,Woosley:2015}.
These models are the result of 1D simulations in which multidimensional effects  (convection, rotation, magnetic fields, mass transfer
in binaries) are included in a phenomenological way (see e.g. \cite{Woosley:2002}).
 Generally speaking, more massive stars reach higher temperatures at the core
and thus have higher specific entropies. The thermal contribution increases the Chandrasekhar mass, and, 
as a consequence, more massive stars can host larger iron cores \cite{Woosley:2002}.
This can be seen in the left panel of Fig.~\ref{fig:progenitors}, which shows the dependence of the pre-collapse 
iron core mass with the ZAMS mass for solar metallicity progenitors (red and black symbols). The relation is however
non monotonic due different processes taking place at different mass ranges.

Stars with masses in the range  $9-12~M_\odot$ form iron cores. However, their stellar structure is somewhat similar to the O-Ne-Mg cores 
described in section~\ref{sec:th:onemg}, with a steep decline in density outside the core (see $9.5~M\odot$ model in 
Fig.~\ref{fig:ECSN}). Numerical simulations
of the collapse of these low mass iron cores in 1D \cite{Mueller:2012b}, 2D \cite{Mueller:2013} and 3D \cite{Melson:2015a}, 
show that the resulting SN explosions are weak, resembling to ECSNe. Although 1D simulations produce successful
explosions, multidimensional effects play here a more important role than in ECSNe, increasing the explosion energy  
in about a factor $5$ \cite{Melson:2015a} with respect to spherical symmetry.

In the range $12-20~M_\odot$ the stars are highly convective during their carbon burning phase.
Convection enhances neutrino cooling, which drives the core towards degeneracy. This in turn 
produces light and compact iron cores. In the range $20-30~M_\odot$ radiative (non-convective) 
carbon burning occurs, which hinders neutrino cooling leading to large and hot iron cores. Around $20~M_\odot$,
in the limit between convective and radiative carbon burning, there is a high variability in the resulting 
iron core masses and compactness \cite{Sukhbold:2014}. 
However, not all features can be explained in terms of carbon burning alone. Details in burning of O, Ne and Si, as well as 
shell burning introduce a non-trivial dependence of the structure of the final iron core as a function of the 
initial mass (see \cite{Woosley:2002}, for a review on the topic). At the pre-supernova stage, these stars appear
as blue or red supergiants, depending on the details of hydrogen shell burning and mixing.

In the range $30-100~M_\odot$ density profiles outside the iron core become shallower, due to the presence of thick 
shells of heavy elements. These stars experience an important mass loss due to winds and as a result their
pre-supernova masses decrease for increasing ZAMS mass, reaching pre-supernova values of $6-8~M_\odot$. 
Above $35~M_\odot$ these stars appear as
blue variables or WR stars. If winds are able to strip away the hydrogen envelope they may explode 
as type Ib/Ic SNe. Rotation may also play an important role in the evolution and explosion of these stars, by 
adding an additional source of energy (see section \ref{sec:th:rot}). In that case they may be the origin
of long GRBs.

Multidimensional core-collapse simulations with the most advanced microphysics and transport have focused
mainly in progenitors in the mass range $12-30~M_\odot$ \cite{Mueller:2014,Summa:2016,Bruenn:2016, Burrows:2016,
Lentz:2012,Melson:2015b}. The lower side of this range of masses ($12-18~M_\odot$) is the main contributor to all core collapse supernovae observed in the local universe (see \cite{smartt15} and Section~\ref{sec_3_progquestions}). SN~1987A is a peculiar example of this class, with a blue supergiant progenitor that has been estimated to 
have a mass of $14-20~M_\odot$ and solar metallicity \cite{Shigeyama:1990,Woosley:1988,Woosley:2002}. 
In this mass range, simulations show that stars with masses below $20~M_{\odot}$ result ``easier" to explode 
(shorter time to explosion, higher energies), 
while  above this limit they become ``harder" to explode (longer time to explosion, lower energies).
By performing a wide set of 1D simulations using a simplified neutrino leakage scheme, \cite{OConnor:2011} identified 
that the relevant parameter determining the ``explodability" of a progenitor in the compactness of the core, $\xi_{2.5}$ 
defined as
\begin{equation}
\xi_M = \frac{M/M_\odot}{R(M) / 1000~km}.
\label{eq:compactness}
\end{equation}
A small value of $\xi_{2.5}$ indicates that the layers surrounding the core are extended and that there is a steep
density profile outside the core (as e.g. in O-N-Mg cores). On the contrary, a large value of $\xi_{2.5}$ indicates 
that the core is surrounded by a thick envelope, where the density profile is shallow. As a result,
cores with high values of $\xi_{2.5}$ are harder to explode than those with lower values\footnote{The 
reader should not confuse the parameter $\xi_{2.5}$ with the compactness of a neutron star 
($M_{\rm NS}/R_{\rm NS}$) often appearing in the literature, despite of its similarity in name and definition.
While more compact neutron stars have higher values of $M_{\rm NS}/R_{\rm NS}$, a higher value of $\xi_{2.5}$
indicates that the core of the star is more extended (less compact).}.
Right panel of Fig.~\ref{fig:progenitors} shows the compactness of the core as a function of the progenitor ZAMS mass.
The compactness of the core is highly correlated with the mass of the iron core (compare both panels of 
Fig~\ref{fig:progenitors}, below $30~M_\odot$).  Above $40~M_\odot$, compactness declines to a nearly constant value,
due to the significant mass loss of these stars. The compactness of the iron core is thus relevant to determine if 
a progenitor star will produce a neutron star or a black hole as a result of the collapse. 

Traditionally, the onset of black hole formation has been thought to occur for progenitors with masses above
certain threshold of $\sim30~M_\odot$. But recent numerical simulations suggest that this simple picture is likely 
wrong and that there are interleaved intervals of masses forming either neutron stars or black holes  
A number of works have addressed this issue by exploring systematically the dependence of the supernova properties 
with the initial progenitor mass \cite{Zhang:2008,Ugliano:2012,Pejcha:2015,Ertl:2016,Sukhbold:2016}.
To be able to cover a wide range of progenitor masses and to evolve the supernova explosion to very late times,
including the fallback into the neutron star, these simulations were performed in spherical symmetry (1D) and 
with a very simplified parametrised explosion engine. Despite of these simplifications important 
information can be extracted from these simulations. We review the main results next.

{\it Explodability:}  Below $15~M_\odot$ and in the range $25-28~M_\odot$, coinciding with low compactness
progenitors (see right panel of Fig.~\ref{fig:progenitors}) the outcome is preferentially a neutron star accompanied by a 
supernova explosion. Similarly, for masses of about $15-16~M_\odot$ and $22-25~M_\odot$ 
the outcome is preferentially a black hole. In the range $16-22~M_\odot$ and above $\sim 28 M_\odot$ 
both neutron stars and black holes can appear.
Although the $\xi_{2.5}$ provides a rough guide to explain this behaviour, it is clearly insufficient to explain 
all features observed \cite{Ertl:2016,Sukhbold:2016}; for example a star in the range $25-28~M_\odot$ has a 
larger value of  $\xi_{2.5}$ than one of about $15~M_\odot$ and nevertheless, the latter is more likely to collapse to a black 
hole than the former. The introduction of a second parameter has proven to be sufficient to explain the phenomenology
of all the 1D simulations \cite{Ertl:2016}. 

{\it Black hole production rate:} When these results are weighted with the initial mass function (IMF),
the resulting fraction of massive stars forming black holes is about $20-30\%$ 
\cite{Ugliano:2012,Sukhbold:2016}, increasing significantly previous estimates of $\sim10\%$ based 
on a single mass threshold model \cite{Woosley:2002}. These results partially solve the so-called
``SN rate problem" \cite{Horiuchi:2011}, namely that the SN rate predicted from the star formation rate is higher than the 
SN rate measured by SN surveys.
It is currently unclear what would be the observational signature
of BH forming events.
Even if no supernova explosion is produced there are suggestions that some
weak transient could be associated with these events \cite{Nadezhin:1980,Lovegrove:2013,Piro:2013}
rather than an unnovae \cite{Kochanek:2008} (also know as failed SNe or dark SNe; see Section~\ref{sec_3_progquestions}).
A non-negligible fraction of
the dim core-collapse SNe in the very local Universe (10 Mpc), could in fact be related to BH forming 
events \cite{Horiuchi:2011}. Additionally, there are observational indications of a paucity of potential SN progenitors (red supergiants) 
in the mass range $16.5-25~M_\odot$ \cite{Kochanek:2008}, which could be associated with BH forming
progenitors. Direct observations of SNe progenitors are also consistent with most of the stars above $\sim18~M_\odot$
forming BHs (see \cite{smartt15} and Section~\ref{sec_3_progquestions}).
Fast rotation has a strong influence in the behaviour of these systems and is treated separately in the next sections. 

{\it Explosion energy:} One could expect that, since $\xi_{2.5}$ is related to the explodability of the star, 
stars with lower value of this parameter, and thus ``easier" to explode, would produce more energetic events. 
However, all stars producing supernova explosions with initial masses above $\sim10~M_\odot$, have
very similar explosion energies of $\sim10^{51}$~erg \cite{Ugliano:2012,Pejcha:2015,Sukhbold:2016}, due to 
the self-regulation of the delayed neutrino-heating mechanism (see section~\ref{sec:th:typical}). Below $10~M_\odot$, 
regardless of whether they have an iron core or an O-Ne-Mg core, the resulting explosions are weaker, with energies 
of $\sim10^{50}$~erg in the lowest end of the progenitor masses. The exception are progenitors very close to the 
onset of black hole formation, which are very hard to explode and produce explosion a factor of a few weaker 
than typical explosions. In general, there is a positive weak correlation between Nickel mass 
production and explosion energy \cite{Zhang:2008,Pejcha:2015,Sukhbold:2016}.

\begin{figure}[t]
\centering
\includegraphics[width=0.7\textwidth]{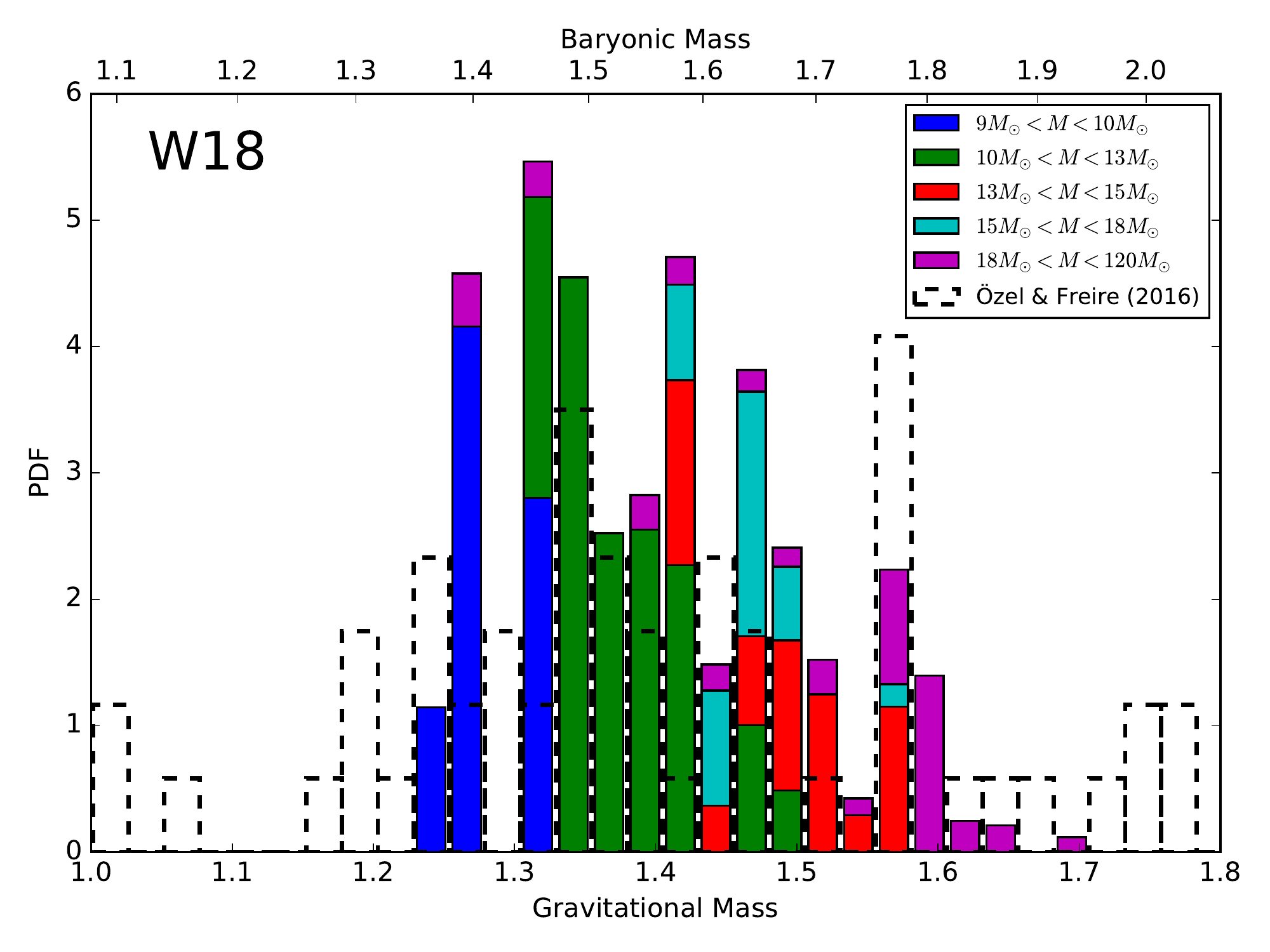}
\caption{
Distributions of neutron star masses for the explosions
calculated by \cite{Sukhbold:2016} (colour coded), plotted against  the observational 
data from \cite{Oezel:2016}. Different colour indicate the ZAMS mass of the progenitor star.
The calibration W18 was used in these calculations (see 
\cite{Sukhbold:2016} for additional details). 
(Figure reproduced from \cite{Sukhbold:2016})
}
\label{fig:nsdist}
\end{figure}

{\it Remnant mass:} All simulations show that progenitors with initial masses larger than $\sim 10~M_\odot$ form
neutron stars with an average mass of $M_{\rm NS}\sim1.4 M_\odot$, while, below this threshold, lighter neutron stars
are formed with ($M_{\rm NS}\sim1.2 M_\odot$) \cite{Pejcha:2015,Sukhbold:2016}. When these results are weighted
with the IMF \cite{Sukhbold:2016}, the resulting mass distribution (see Fig.~\ref{fig:nsdist})
is weakly bimodal, resembling 
the observed distribution of NS masses \cite{vandenHeuvel:2004,Schwab:2010} (see discussion in section~\ref{sec:th:onemg}). 
In this distribution, most neutron stars  
with masses above/below $1.3~M_\odot$ belong to progenitors with masses above/below $10~M_\odot$.
The higher end of neutron star masses, above $1.6~M_\odot$, belong to progenitors mostly above 
$\sim20~M_\odot$\cite{Pejcha:2015,Sukhbold:2016}, very close to the onset of black hole formation, and that produce
weak explosions.
As a general result it is fair to say that, according to current numerical modelling, the most common type of neutron 
stars was formed with
a mass of $\sim 1.4~M_\odot$ in supernova explosions of progenitors in the range $10-20~M_\odot$.
However, these numerical simulations still lack of a complete modelling of the neutrino-heating mechanism 
and of important multidimensional effects, so the predicted NS mass distribution functions have to be taken with care.
Regarding systems forming black holes,
the work of \cite{Ugliano:2012,Pejcha:2015,Sukhbold:2016} predicts masses above 
$\sim5 M_\odot$, producing a gap in the range $\sim 2-5~M_\odot$ in the remnant mass distribution.
  
{\it Fallback:} 
Even if the explosion is sufficiently energetic to disrupt the star, some of the material of the star may fall back 
onto the neutron star, increasing its mass and potentially forming a black hole \cite{Colgate:1971,Chevalier:1989}. 
This process occurs mainly as the supernova shock crosses composition interfaces, e.g. that between the He shell 
and the H envelope, with steep density declines, and a reverse shock is formed. The formation of such a reverse 
shock has been observed in multidimensional simulations of the supernova shock propagation through the star in 
2D \cite{Kifonidis:2003,Scheck:2006} and 3D \cite{Hammer:2010,Joggerst:2010,Wongwathanarat:2015}.
However, estimating the total amount of mass accreted into the NS by this process is difficult, since it implies 
following the matter falling back for timescales of hours, which is numerically challenging.
This limits fallback mass estimates to 1D numerical simulations \cite{Ugliano:2012,Ertl:2016,Sukhbold:2016}.
In general, the amount of fallback material is expected to be small, mostly in the range $10^{-4}-10^{-2}~M_\odot$, 
progenitors with lower initial mass experiencing a lower fallback accretion. In simulations by  \cite{Ertl:2016,Sukhbold:2016},
a few stars of $\sim 30~M_\odot$ formed black holes as a result of the fallback accretion, however this mechanism does 
not appear to be a significant channel for BH formation. 
There have been suggestions that, for weak explosions with strong fallback, it would produce an under-luminous
SN, that could explain some observed peculiar Type Ia SN \cite{Moriya:2010}.
Fallback has also been invoked to explain the low magnetic field
observed in central compact objects (CCOs), young neutron stars located near the centre  of SN remnants. In this 
``hidden magnetic-field" scenario \cite{Young:1995,Muslimov:1995,Geppert:1999,Shabaltas:2012}, the fallback 
material is able to bury the NS magnetic field explaining the observations of young NSs. In a longer timescale of $1-10^7$~kyr, 
the magnetic field is able to re-emerge explaining magnetic fields in older objects \cite{Young:1995,Muslimov:1995,Geppert:1999}.
Numerical simulations of the accretion of matter onto magnetised material have shown that this mechanism is
indeed feasible \cite{Payne:2004,Payne:2007,Bernal:2010,Mukherjee:2013a,Bernal:2013,Mukherjee:2013b}. It has been
estimated that it is sufficient an accreted mass of $10^{-3}-10^{-2}~M_\odot$ to bury typical NSs magnetic fields
\cite{Torres-Forne:2016}. As a consequence, it is expected that some neutron stars are born with hidden magnetic fields (CCOs)
while other appear as regular pulsars (e.g. Crab). This scenario could also explain the lack of NS detected in SN~1987A.

\subsubsection{Supernova kicks and and binary disruption}
\label{sec:th:kicks}

Considering that the most of stars live in binaries \cite{Sana:2012}, 
if the supernova explosion is able to impart some kick velocity on the newly 
formed neutron star, then this could be ejected or at least introduce some eccentricity
in the resulting system. Kicks are genuinely multidimensional effects that cannot  
be accounted for in the 1D numerical simulations reviewed in the previous section, so we deal with them 
separately here. There are two mechanism that produce kicks in neutron stars: asymmetries in the 
supernova explosion and the ejection of large amounts of mass by the explosion.
These kicks have been observed in young neutron stars that show typical velocities of several $100$~km/s,
with some neutron stars moving at more than $1000$~km/s (see e.g. \cite{Hobbs:2005, Arzoumanian:2002}) .

As we show in section~\ref{sec:th:typical}, supernova explosions are likely to occur in a very asymmetric
way, tracing the multidimensional instabilities that helped to revive the stalled shock. Therefore, the expanding
shock it is likely to have some net linear momentum in a random direction. This produces a reaction on the 
neutron star that, being less massive than the envelope, can experience a sudden increase of its velocity, 
the so called kick, in the opposite direction. Since the kick is the result of an instability breaking the 
initial symmetry of the star, the resulting kick direction and velocity are highly stochastic.
Numerical simulations \cite{Janka:1994,Fryer:2007} have shown that this mechanism alone is not able to 
accelerate NSs to more than a few $100$~km/s, which is insufficient to explain observations.
However, this initial kick is not the final velocity of the neutron star. In longer time-scales
the gravitational interaction between the remnant and the slow moving massive ejecta accelerates 
further the neutron star in the so called  ``tug-boat" mechanism \cite{Nordhaus:2010,Nordhaus:2012}.
A series of 2D \cite{Scheck:2004,Scheck:2006,Nordhaus:2010,Nordhaus:2012} and 3D 
\cite{Wongwathanarat:2010,Wongwathanarat:2013} simulations have shown that this mechanism can explain
natal kick velocities of more than $1000$~km/s. The extensive study of \cite{Wongwathanarat:2013} showed 
that, not only the kick velocities are consistent with the observed velocity distribution of NSs, but 
also the same mechanism would impart a spin in the NS (see also \cite{Spruit:1998}). 
The resulting NS periods, in the range $100-8000$~ms, are similar to those encountered in pulsars.

The second mechanism is specific of binaries.
During a supernova explosion, the stars loses most of its mass in very short time, compared to the orbital period 
of the binary. Even if the explosion is perfectly spherically symmetric with respect to the remnant compact object, 
there is always a strong asymmetry with respect to the centre  of mass of the system. That leads inevitably to a 
recoil of the compact object, which acquires a kick velocity. This case can be studied analytically 
\cite{Blaauw:1961,Boersma:1961} and gives very interesting predictions. If more than half of the mass of the 
binary is ejected during the explosion, the binary is disrupted, and the compact remnant flies away 
(see e.g. \cite{Postnov:2014}). For lower ejected masses, the kick is not able to disrupt the system, but can 
introduce a significant eccentricity to the binary. Note that, unless the previous kick mechanism, this one
has low degree of stochasticity, and it can be predicted, which kind of binaries are likely to survive
a SN explosion and which are not. This systematic effect has important consequences in stellar evolution
that are beyond of the scope of this review (see e.g. \cite{Postnov:2014}, for more information).

\subsubsection{Metallicity}
\label{sec:th:rot}

Metallicity and rotation play an important role in stellar evolution. They may be responsible for the wide
variety  of observed properties in supernova explosions and neutron stars. There is often an interplay 
between both rotation and metallicity, so we start summarising the results for non-rotating stars, 
and we deal with the effect of metallicity in rotating stars in the next section.

The metallicity of the environment in which a star was born has a significant impact in processes of mass
loss. Metallicity increases the opacity of the envelope of the star and allows for the formation of a radiation driven wind.
As a consequence, stars with higher metallicity have have stronger winds and hence
a higher mass loss. In the previous sections we have dealt exclusively with solar metallicity stars,
which have a significant mass loss above $20~M_\odot$, and become WR stars (basically
bare He cores) above $\sim 30~M_\odot$. In low metallicity stars these mass limits are shifted upwards
and the typical iron cores formed are more massive, specially for the most massive stars
(see e.g. \cite{Woosley:2002}). A number of authors have performed 1D stellar evolution calculations 
of low metallicity stars
\cite{Woosley:1995,Marigo:2002,Heger:2002,eldridge04,Umeda:2005,Hirschi:2006,Tominaga:2007,Limongi:2012,Sukhbold:2014}.
The more extreme cases can be found in population III stars.
In very low metallicity environments, stars may form with masses above $100~M_\odot$. 
These stars produce a copious amount of $e^- -e^+$ pairs after central carbon burning, 
which cools down the star and leads to a gravitational pair instability. The outcome 
is either a black hole or a thermonuclear explosion (pair-instability SN) 
(see \cite{Woosley:2002,Heger:2003} for details), but no neutron stars are formed.

Below $\sim100~M_\odot$, simplified 1D core collapse simulations \cite{Zhang:2008,Pejcha:2015} show that, 
since the iron core masses are in average
larger for low metallicity, in stars with initial masses above $\sim 25~M_\odot$ the most common outcome
are black holes. In this mass range, those stars not forming a black hole, have larger masses than in the solar 
metallicity case \cite{Zhang:2008,Pejcha:2015}. In fact, many of these neutron stars experience a significant
fallback, which is favoured under low-metallicity conditions. For initial masses $M<25~M_\odot$, 
the properties of the supernova explosions and the resulting neutron stars are similar, regardless of 
metallicity, because these stars do not experience a significant mass during their lives.
Taking the IMF into account, the fraction of stars forming black holes could be as large as $50\%$ 
at low metallicities \cite{Pejcha:2015}. Multidimensional simulations of low metallicity stars have been
performed mainly for fast rotating progenitors, which are discussed in the next sections.

\subsubsection{Rotation and magnetic fields}
\label{sec:th:rot}

Main sequence stars rotate rapidly, with typical observed surface velocities 
of the order of $200$~km/s \cite{Fukuda:1982}. 
Such a high rotation rate can have important implications on the evolution of the star
and in the spin period of the remnant compact object after the supernova explosion. 
Depending on whether the core is able to retain its angular momentum during its evolution
or not, the resulting iron core will produce a neutron star with a period of about $1$~ms, 
very promising as a progenitor of GRBs and magnetars, or a slowly rotating neutron star,
compatible with measurements of spin periods in pulsars 
(see section~\ref{sec:th:kicks} for the case of a non-rotating core). It is thus clear that
there has to be important differences in the rate of 
angular momentum loss among different stars, depending on initial mass, rotation, and
metallicity, to produce the variety of events observed. 

Loss of angular momentum in massive stars occurs mainly during the red supergiant phase,
for stars with $M<30 M_\odot$, or WR phase, above this limit (see e.g. \cite{Woosley:2002}). 
As the hydrogen envelope expands it spins down, due to angular momentum conservation, and 
starts rotating differentially with respect to the core. The angular momentum of the envelope
can be extracted from the star if there is mass loss due to winds. This process is specially important
in WR stars that will lose the whole hydrogen envelope 
during this phase. If there is some process 
transporting angular momentum efficiently, coupling the core with the outer layers
of the star, the wind will 
extract  angular momentum from the core as well \cite{Langer:1998}. Therefore, the final total angular momentum of the core 
in the pre-supernova stage depends on both,  the efficiency of angular momentum
transport processes and the amount mass lost by winds. Mass loss depends mainly on metallicity and
has been discussed in the previous section. We focus next on angular momentum transport.

Multiple hydrodynamic instabilities contribute to the transport of angular momentum, both in convective
and radiative regions \cite{Heger:2000,Maeder:2000a,Maeder:2000b}. However, in the absence
of magnetic fields, this transport is rather inefficient and leads to rapidly rotating cores 
\cite{Heger:2000,Hirschi:2004,Chieffi:2013}.
The main mechanism responsible for the transport of angular momentum are magnetic fields
\cite{Spruit:1998,Spruit:1999}. Regardless of the initial magnetic field, instabilities in combination 
with rotation can  lead to the formation of a dynamo, able to support magnetic fields during 
the life of the star \cite{Spruit:2002}. A detailed understanding of magnetic field dynamos
in stars is a long standing problem, which is not even completely solved for the most studied 
star, the Sun (see e.g. \cite{Charbonneau:2013}). The incorporation of magnetic fields in stellar 
evolution codes  has only been attempted using simplified 1D models for the magnetic torques so far
\cite{Heger:2005}. In this case, for stars with $M < 30~M_\odot$, magnetic torques are able to enforce
rigid rotation in the star and spin down the core by a factor  $30-50$ with respect to the case in
which magnetic fields are not  considered, producing progenitors of pulsar-like objects. This  leads to the conclusion
that, for the most common type of progenitor of core-collapse supernovae, the iron core is likely to have
lost most of its angular momentum during its evolution, and the outcome of its collapse would be very similar to the case 
of non-rotating progenitors, as described in section~\ref{sec:th:ccsne}. The remaining 
spin of the iron core, in combination with the spin imparted by the SN explosion (see section~\ref{sec:th:kicks}), would 
explain the variety of periods observed in most pulsars. 

Furthermore, rotation accelerates burning by introducing additional mixing. This has consequences
in the mass of the core, the amount of mass loss, the colour of the star in the late evolution
(blue instead of red supergiant) and the type of supernova producing (Type Ib/c instead of Type II)
\cite{Hirschi:2004,Chieffi:2013}.

\subsubsection{Fast rotation: Hypernovae, long GRBs and magnetic field amplification}
\label{sec:th:fast}

Massive stars with very large initial rotation velocity ($\sim400$~km/s) may undergo a completely 
different evolutionary path to that described in the previous sections. 
Rotationally induced mixing produce stars that are chemically homogeneous and 
 that are able to burn efficiently all the hydrogen and 
skip the red supergiant phase \cite{Maeder:1987}. The result is a bare helium core without a 
hydrogen envelope, very similar to a WR star. Similar situation could be reached if the star
is stripped down to a helium core as the result of mass transfer in a binary system \cite{Woosley:2006}.
In solar metallicity environments, this helium core can still lose 
a significant amount of angular momentum due to winds. However, for low metallicities
winds cannot spin down the star and the core is able to retain its angular momentum until the 
pre-supernova stage \cite{Yoon:2005,Woosley:2006}. These stars can potentially form
neutron stars with periods of $\sim 1$~ms or fast spinning black holes, and are potential candidates
for long GRB progenitors. Additionally, the absence of a hydrogen envelope would explain the 
observed association of long GRBs with Type Ib/c supernovae.

Although it is not well known the exact evolutionary path that leads to the formation of high spinning cores,
it is clear that they are necessary to clarify the phenomenology associated with long GRBs and 
hypernovae. For those cases, the neutrino heating mechanism is clearly insufficient to explain the 
presence of a highly collimated jet or the energetics of the explosion in hypernovae. The current 
understanding is that, in the presence of magnetic fields, the energy stored in the rotation of the star can 
be extracted and used to drive a powerful magnetorotational explosion. In some cases a relativistic jet
may be produced by black hole accretion (collapsar model, \cite{MacFadyen:1999})
or by the presence of a millisecond magnetar \cite{Usov:1994,Wheeler:2000}. 
Fast rotation could also be an explanation for the recently discovered class of superluminous SNe (see \cite{nicholl13} and Section~\ref{sec_3_newclass})

Multidimensional numerical simulations of
the collapse of rapidly rotating magnetised cores have been performed both in 2D 
\cite{Shibata:2006,Obergaulinger:2006,Ott:2006,Burrows:2007,CerdaDuran:2008,Takiwaki:2011, 
Sawai:2013,Sawai:2016,Obergaulinger:2017} and 3D
\cite{Mikami:2008, Kuroda:2010, Scheidegger:2010, Moesta:2014,Moesta:2015,Winteler:2012},
with different degree of sophistication in the treatment of the neutrinos.
For sufficiently high 
initial magnetic field in the progenitor, numerical simulations show that magneto-rotational effects
help in the explosion, producing more energetic events exploding at earlier times than the corresponding
non-magnetized models. Explosions are usually highly asymmetric and aligned with the rotation 
axis. In 2D simulations mildly relativistic outflows have been observed along the axis, but 3D simulations
have shown that this effect is likely exaggerated by the axisymmetry imposed to the system \cite{Moesta:2014}. 
Recent 2D simulations have shown that, the asymmetries in the magneto-rotational explosion
make it possible to keep the accretion onto the neutron star, even after the shock expands \cite{Obergaulinger:2017}.
This would allow for the formation of a black hole after the supernova explosion, which could in turn 
serve as central engine for a long GRB. This would solve the problem that, if a black hole is formed too
early, no supernova explosion would be formed. 

All the simulations assume a large scale initial magnetic field in the iron core in the range $10^9-10^{12}$~G, 
consistent with stellar evolution results by \cite{Heger:2005}. However, this magnetic field is expected to be
amplified once the proto-neutron star has formed by the action of the magneto-rotational instability (MRI) 
\cite{Akiyama:2003}. The MRI \cite{Velikhov:1959,Chandrasekhar:1960} is an instability that appears in differentially
rotating magnetised fluids and was proposed as the main mechanism driving accretion in discs 
\cite{Balbus:1991}. The MRI is able to amplify the magnetic field and generate turbulence, so it  has been invoked
in numerous times in the literature as a justification to start numerical simulations with an artificially  enhanced
magnetic field strength, using as an argument that the MRI would be responsible for this effect.
However, the direct simulation of the MRI in core collapse simulations is challenging, because it develops 
in very small length-scales that require an amount of numerical resolution not feasible with present day supercomputers
(see discussion in \cite{Rembiasz:2016b}). Nevertheless, a few attempts have been carried out using artificially 
enlarged magnetic fields \cite{CerdaDuran:2008,Sawai:2013,Sawai:2016,Moesta:2015}; by increasing 
the magnetic field the MRI length-scale grows being easier to resolve numerically. However, it is unclear 
how close to reality are these simulations. Increasing artificially the global magnetic field may not 
be equivalent to the turbulent state that would be expected form the development of MRI.

To explore the amplification of the magnetic field due to the MRI, local and semi-local 
simulations have been performed with the conditions present in proto-neutron stars
\cite{Obergaulinger:2009,Guilet:2015a,Rembiasz:2016a, Rembiasz:2016b}. The result of the 
simulations show that the amplification of the magnetic field by MRI is severely limited 
by the presence of parasitic instabilities of the Kelving-Helmholtz type. These instabilities can limit 
the magnetic field amplification to a 
factor of about $\sim10$ \cite{Rembiasz:2016b}. Additionally, the development of the MRI
could be severely affected by the presence of neutrinos inside the neutrinosphere \cite{Guilet:2015b}.
Nevertheless, MRI is able to create and sustain
turbulence with non-zero kinetic helicity. In presence of differential rotation, this turbulence could 
create a large scale dynamo (see e.g. \cite{Brandenburg:2005}) capable of  generating a
large scale field similar to what is needed to power a hypernova or a GRB. The results by 
\cite{Moesta:2015} may be indicative that this is the case although more detailed simulations 
will be required in the future to explore this possibility.

\section{Challenges and future prospects}
\label{sec_3}

\subsection{New supernova types}
\label{sec_3_newclass}

As discussed previously, thanks to the current sky surveys, astronomers are collecting unprecedented samples of known classes of objects, as well as discovering novel types of stellar transients that cannot be explained with traditional explosion channels. Some examples are described below:

\noindent
$\bullet$ {\underline {Superluminous SNe (SLSNe):}} 
Their absolute magnitude exceed $-$20 mag, and may occasionally reach $-$22 mag, which is one order of magnitude brighter than Type Ia SNe. They have a slow photometric evolution, and are preferentially hosted in very faint, likely metal-poor, dwarf galaxies \cite{quimby11}. Most of them are H-poor \cite{pastorello10}, though some others show evidence of H spectral lines originated in the stellar CSM  \cite{benetti14}. In addition, due to their intrinsic luminosity, some SLSNe have been proposed as standardisable candle candidates to redshifts z $\approx$ 3-4 \cite{inserra14}. However, the sample is still small (only eight to ten objects currently have enough data to test this method) and there are still some concerns about the unknown progenitor systems or the explosion physics. SLSNe seems to be incompatible with models powered only by the radioactive decay chain $^{56}$Ni~$\rightarrow$~$^{56}$Co~$\rightarrow$~$^{56}$Fe and to favour alternative explosion scenarios, in which the additional energy is provided by the spin-down of a rapidly rotating young magnetar, pair-instability and/or strong interaction of SN ejecta with a very massive opaque CSM (e.g. see \cite{nicholl13} and Section \ref{sec:th:progenitor} for more details). No models yet provide an excellent and unique fit with observations.

\noindent
$\bullet$ {\underline {Type Ibn SNe:}}
There is wide heterogeneity in SNe showing strong interaction with their CSM. They can be generated in a H-rich environment as we have seen before (Type IIn), or also by stars exploding in a He-rich medium, being labelled as Type Ibn SNe \cite{pastorello08}. These Ibn SNe can be luminous objects (peaking at $\sim -19$ mag), usually followed by quickly declines ($\sim$ 0.1 mag day$^{-1}$), and show a spectroscopically heterogeneity \cite{pastorello16}. Since the discovery of its prototypical object, SN~2006jc \cite{pastorello07,foley07}, other $\sim$ 30 Type Ibn SNe has been observed at this time. 

\noindent
$\bullet$ {\underline {Faint stripped-envelope SNe:}} These SNe show faint luminosity peaks (M $< -15$ mag), fast-evolving light curves and weird spectroscopic properties (narrow features, very weak Si and S lines, sometimes evidence of He, and prominent [Ca {\sc II}] in the nebular phases; e.g. \cite{valenti09}). The possible explanation of these weird observables are exotic explosions involving white dwarfs below the Chandrasekhar mass limit \cite{perets10}, failed SN explosions, or unusual CC-SNe (e.g. fallback SNe from very massive stars or electron-capture SNe from super-AGB stars; \cite{pumo10} and Section \ref{sec:th:progenitor}). 

\noindent
$\bullet$ {\underline {SN impostors:}}
Sometimes there are sneak cases of powerful eruptions of LBVs that mimic the true appearance of an interacting SN. The latter are commonly known as ``SN impostors" (e.g. \cite{vandyk_rev12}). This misclassification used to happen because the spectra of these giant eruptions are characterized by incipient narrow hydrogen lines in emission, like those of type IIn SNe. However, their luminosities are considerably lower (fainter than $-14$ mag). Although SN impostors may not be SNe, they are included here because it is not clear yet their nature. A grown number of SN impostors have interestingly heralded the terminal SN explosion, from weeks to a few years after the outburst episode (e.g. SN~2009ip \cite{pastorello13}, LSQ13zm \cite{tartaglia16}, or SN~2015bh \cite{eliasrosa16}), but some others show repeated intermediate-luminosity without leading (so far) to a SN explosion (e.g. SN~2000ch \cite{pastorello10}, or SN~2007sv \cite{tartaglia15}). LBV stars are the most usual channel to explain the bursty activity of the SN impostors, however, these outburst have been also linked to lower mass stars, interacting massive binary system, or pre-SN nuclear burning instabilities. Nevertheless, the mechanisms triggering this class of outbursts are still very poorly understood.

\noindent
$\bullet$ {\underline {Other unusual SNe:}} 

\noindent
\begin{itemize}
\renewcommand\labelitemi{--}
\item There are a handful of events that have broad light curves with much {\it longer rise times} from estimated explosion to peak, like SN~2011bm \cite{valenti12}, a Type Ic SN with a rise time of 35 d (instead of 10-22 d for normal SNe of this class), or Type~II~SN~1987A with an exceeding long rise rime of $>$ 80 days \cite{arnett89}. The long rise times suggest much larger ejected masses. 

\item Other objects show {\it faster decline rates}, such as SN~2005ek \cite{drout13}. It is spectroscopically a Type Ic with a decline of $2.5-3$ mag in 15 days.

\item The wide range of magnitudes during the plateau phase of type II-P SNe is a well established fact \cite{hamuy02}. However, {\it very faint and also luminous Type II-P} exist (e.g. \cite{spiro14,inserra13}). These rare objects can allow us to better constrain the correlation between $^{56}$Ni and progenitor mass, and to test the reliability of other tight relations among physical parameters of Type II-P SNe, on the most and least energetic events. An examples is SN~1997D \cite{turatto98}, which showed a small amount of $^{56}$Ni and narrow P-Cygni profiles, suggesting a low explosion energy.

\end{itemize}

\subsection{Multi-messenger astronomy.}
\label{sec_3_GW}

The core-collapse of massive stars has been associated with SNe (and long GRBs, $>$ 2 sec). CC-SNe are canonical examples of multi-messenger astrophysical sources since the gravitational energy driven by the CC-SN at the time of explosion is released as $\sim$ 99\% of neutrinos, $\sim$ 1\% is converted to ejecta kinetic energy, $\sim$ 0.01 \% becomes photons, and an uncertain, though likely smaller fraction is carried away by gravitational waves (GW). 

So far, we have focused this work on electromagnetic observation of core-collapse supernovae. However, there are additional observational channels that will provide in the future complementary information about the scenario in which neutron stars are formed: cosmic rays, neutrinos and GWs. While  electromagnetic observations  mainly carry information about the matter ejected during the explosion, and on the central regions at late time, both neutrinos and GWs will allow us to infer directly the properties of the compact remnant, and provide information about the thermodynamics and dynamics of the SN engine. In the last decade a big effort has been done to improve current neutrino and GW observatories to reach sensitivities that would allow us to do multi-messenger astronomy of nearby SN explosions, within $\sim100$~kpc. 

\subsubsection{Cosmic rays}

The shockwaves of CC-SNe accelerate charged particles such as protons, some of which end up raining on Earth as {\it cosmic rays}. The importance in detecting cosmic rays is that they are the only particles detected on Earth which have traversed a considerable distance through the ISM and which were accelerated in events such as SN in a relatively recent past. In fact, data from Fermi Large Area Telescope \cite{atwood09} have been interpreted as a fraction of primary cosmic rays originate from a SN. However, the astrophysical source identification is complicated without any other messenger information.

\subsubsection{Neutrinos}

CC-SNe emit low-energy neutrinos, as confirmed by the observations of the only recent nearby supernova, SN~1987A (a few tens of MeV; e.g. see \cite{hirata87}). The searches for neutrinos are based on three kind of analysis: searches for significant neutrino excess with respect to the atmospheric neutrino background, searches for anisotropies on the sky to identify point sources, and multi-messenger searches in time and spatial coincidence with other messenger signals. Having information from neutrino observations, we may constrain the exact time of the CC bounce (with just EM information we have an uncertainty of many hours), and establish a tight correlation with GWs. Moreover,  with information from neutrinos, we will be able to set limits on neutrino mass, and if this signal is more copious (for example a SN in the Milky Way) we will be able to make distinctions between different theoretical models of CC-SNe explosion. Current neutrino detectors (e.g. SuperKamiokande \cite{SuperKamiokande:2007}, IceCube \cite{IceCube:2009} and Antares \cite{Antares:2012}) are capable of measuring neutrinos from a supernova as far as in the Large Magellanic Cloud.

\subsubsection{Gravitational waves}

The last multi-messenger channel that has been opened are gravitational waves, with the 
first detection of a binary black hole merger, GW150914 \cite{abbott16a}, by the Advanced
Laser Interferometer Gravitational-Wave Observatory (aLIGO). In order to be detectable by GW observatories, these waves should be produced by a very dynamic and compact relativistic system, associated to high energy processes. 
Sources of GWs can be classified according to their frequency (see e.g. \cite{Cutler:2002}, for a review on GW sources). 
High frequency sources ($10-2000$~Hz)  include binary mergers of compact objects of stellar origin (NS and BH), asymmetric core-collapse of massive stars, and rotating isolated neutron stars. Low frequency sources  ($10^{-9}-1$~Hz) include the inspiral phase of binary NS, BH and white dwarfs, as well as the merger of WD and supermassive BHs, at cosmological distances. Observations in the high frequency range are 
accessible to ground-based  laser interferometers such as aLIGO \cite{aLIGO:2015} and Virgo \cite{aVirgo:2015}, and the upcoming KAGRA \cite{KAGRA:2013} experiment, while low frequency observations will be performed by space-based laser interferometers such as LISA \cite{LISA:2017} in the range $10^{-4}-0.1$~Hz, and by the pulsar timing array (PTA) at the lowest frequencies, $10^{-9}-10^{-8}$~Hz \cite{PTA:2010}. 

Differently from the binary black hole merger case, for which reliable and accurate GW templates have been developed during the last decade, the GW signal from the core collapse case still presents many theoretical uncertainties related to the complexity in the numerical modelling of the scenario (see Section  \ref{sec:th:ccsne}). The signal can be divided in two parts. {\it The core bounce} is the part of the waveform which is best understood~\cite{Dimmelmeier:2002b}. Its frequency (at about 800 Hz) can be directly related to the rotational properties of the core~\cite{Dimmelmeier:2008, Abdikamalov:2014, Richers:2017}. However, fast-rotating progenitors are uncommon (see Section \ref{sec:th:progenitor}) and this bounce signal is not likely to be observed in non-rotating galactic events. More interesting is the signal related to the {\it post-bounce evolution} of the newly formed proto-neutron star, which is produced by convection and the excitation of highly damped modes in the PNS \cite{Murphy:2009,Mueller:2013,Cerda-Duran:2013,Kuroda:2016,Andresen:2017}. Typical duration of this signal is of $\sim0.5$~s in the case of successful SN explosions (see e.g. \cite{Mueller:2013}) but can last for seconds if the final outcome is a BH \cite{Cerda-Duran:2013}. Typical frequencies raise monotonically with time due to the contraction of the PNS, whose mass is steadily increasing. Characteristic frequencies of the PNS can be as low as $\sim100$~Hz, specially those related to g-modes, which make them a perfect target for ground-based interferometers with the highest sensitivity at those frequencies. In the future, it may be possible to infer the properties of PNS based on the identification of mode frequencies in their waveforms (see e.g. \cite{Sotani:2016}). Depending on their rotation rate, supernovae could be detected as far as $1$~Mpc for extreme events, but typical distances are more likely $1-100$~kpc \cite{Abbot:2016c}. The rate of CC of massive stars in the Milky Way is around 2 per century (e.g. see \cite{ott09}). 

One of the main challenges is the difficulty of detecting an electromagnetic counterpart inside the localisation error box of the GW signal. With only two GW detectors (LIGO interferometers in Hanford and Livingston), typical sky localisation is so far poor (hundreds of square degrees) \cite{SkyLoc:2016}. However, this will change in the near future by the addition of new detectors to the network (advanced Virgo, KAGRA and LIGO India), improving the sky localisation to an error box of a few $10$~deg$^2$.
The next challenge is then to detect the electromagnetic counterparts inside the error box of GW signals. Finding these counterparts is a proof of the veracity of the origin of the signal.

\subsection{Challenges and future prospects from an observational point of view}
\label{sec_3_futureprospects_obs}

\subsubsection{Progenitor hunting}
\label{sec_3_progquestions}

In previous sections we have reviewed the progenitor properties derived from the SN observations and the direct observations of the SN site before the explosion. This connection is crucial to test our understanding of stellar evolution. However, it has not been found all what expected.\\

There is information of about 30 SN progenitors between detections and limits. Still, none of these has an estimated luminosity above of log $L/L_{\odot} \backsimeq 5.1$, which corresponds to an initial evolutionary mass of about 18 \msun~(except for interacting SNe, whose progenitors seems to have high masses, but in some cases the nature of that transients is still debated). This estimate is significantly lower than the stellar evolution textbook value (about 25 \msun, and Section \ref{sec:th:progenitor}), or than the red supergiants found in the Local Universe, with luminosity up to log $L/L_{\odot} \backsimeq 5.5$ (approximately initial mass of 30 \msun; \cite{levesque09}). What is more, according a typical Salpeter initial mass function (of slope $\alpha = -2.35$), a $\sim$ 30$\%$ of stars between 8-100 \msun\ have masses $>$ 18 \msun. Thus one would expect to have found some SN progenitor stars with higher masses by now, which is not the case. 

A good review exploring possible bias and explanations for this deficit can be found in \cite{smartt15}. They discussed how the circumstellar dust or the systematic errors in the analysis can affect the luminosity and mass estimates. However, they also hold that these biases, although may affect our estimates, do not seem to explain the missing mass progenitor stars.

Furthermore, it is natural to think that depending on the stellar evolution code used for estimating the progenitor star  mass, more uncertainties could be added to the measure. Still, \cite{smartt15} compare the end points of three stellar evolution models (STARS models, \cite{eldridge04}, rotating Geneva models, \cite{Hirschi:2004}; and KEPLER models, \cite{Woosley:2007}) and show that while the use of one model or another could affect the lower mass limit to produce a CC-SN (masses between 7 and 10 \msun), the high mass upper limit appears be secure.  

It has been proposed as a possible explanation that massive stars above 18 \msun\ evolve into WR stars and evade the detection because they are too hot and faint at the point of core-collapse. In this case they would produce Ib/c SNe, although this disagrees with the growing evidence that the majority of Ib/c SNe comes from lower mass stars in interacting binaries.

Another plausible explanation could be that these missed massive stars may have collapsed without producing a detectable SN. Such events may happen when the shock wave dies out before reaching the stellar surface, or because the stellar mantle fallback onto the nucleus. These failed explosions generate a black hole and eject negligible amounts of radioactive isotopes (see also Section \ref{sec:th:mass}). While these events cannot be directly observed, it is possible to detect the disappearance of massive stars through the comparison (subtraction) of reference ``template" images \cite{Kochanek:2008,reynolds15,adams16}. 

Advances in this field could be achieved from the collection of deep, multi-wavelength, wide-field imaging of nearby galaxies for future SN progenitor characterization. This feat is plausible by using high resolution images from {\sl HST}, or from the next generation of ground-based and space telescopes. In the near future, key information will be obtain thanks to the wide field of the 8-m Large Synoptic Survey Telescope (expected to start operations on 2021-2022), the fully adaptive and diffraction-limited optics of the 39-m European Extremely Large Telescope (planned first light for 2024), or the infrared domain of the 6.5 m mirror of the James Webb Space Telescope (scheduled to launch in October 2018).

\subsubsection{Pre-SN outbursts in SN impostors/IIn SNe}
\label{sec_3_IInquestions}

In the last years there has been some success in detecting the progenitors of interacting transients, including Type IIn SNe (Section \ref{sec_2_prog_obs} and \ref{sec_3_newclass}). The nature of these transients remains debated: some are undoubtedly genuine core-collapse SNe, while others may be giant non-terminal outbursts from LBVs. Observational constraints on their progenitors can help light on their true nature. We note that classical LBVs are not expected to explode directly as CC-SNe, and so these results are challenging models of massive stellar evolution.

A lot of work still needs to be done to improve our knowledge on the mechanisms triggering luminous stellar outbursts and major eruptions. Observational time is needed to trace the photometric history of a these transients and to keep relaxed monitoring of SN impostors to pinpoint any possible re-brighting of the object. In short, only the discovery of a large number of similar transients, extensive follow-up campaigns in a wide wavelength range, and the availability of rich archives with deep images at different domains will allow us to give more robust conclusions on the variety of properties of SN impostors, and how this heterogeneity is connected to stellar parameters (mass, radius, chemical composition, rotation, binarity).

\subsubsection{Flash spectroscopy}
\label{sec_3_flashquestions}

The SNe require rapid and intense follow-up to characterize their explosions. Thanks to the current high-cadence optical surveys dedicated to the observations of transients, it has been possible to discover new SNe only few hours after explosion. This technique is called flash spectroscopy \cite{galyam14}. Some Type II SNe, like 2014G \cite{terreran16}, have shown narrow emission lines superimposed to a hot blue continuum, which disappears in few days after the explosion. These emissions arise from the photoionized surrounding CSM distributed around the exploding star, which was previously originated through winds or violent eruption. The analysis of these early spectra can provide unique information about the physical distribution of gas around each event, and therefore from the stellar evolution during the final phases before the explosion. 

Some current surveys such as ePESSTO (extension of the Public ESO Spectroscopic Survey of Transient Objects - PESSTO; \cite{smartt15_pessto}) are dedicating part of their observational time to the classification and intensive early monitoring of Type II SNe discovered after $2-3$ d (at most) from the explosion.

\subsection{Future prospects in numerical modelling}

\subsubsection{Advances in high performance computing}

The main challenge in the numerical modelling of core collapse supernovae is associated with the 
huge amount of computational resources needed for the completion of realistic simulations. Current 
three-dimensional state-of-the-art numerical simulations use millions of CPU hours distributed 
among a few  $1000$ to $10000$ cores and may take months to finish. This kind of simulations can 
only be performed in high performance computing (HPC) facilities, and CPU time has to be requested
in a similar fashion as observational time in observatories. 
The number of simulations and the numerical resolution affordable by dedicated groups  
is limited by the availability of CPU time and the size and number of such HPC facilities.
Best supercomputers have a few $100$ thousands of cores, a few petaflops ($10^{15}$ operations per second)
of computing power and are distributed all around the world in USA (e.g. Titan, Sequoia), Japan (e.g. K Computer),
China (e.g. Tiahne2) and Europe (e.g. FZJ, SuperMUC, Marconi). 

Next generation of supercomputers will be available
during the next decade and will reach the exaflop scale, improving by a factor $1000$ previous machines.
In these facilities, it will be  possible to perform numerical 3D simulations with unprecedented numerical resolution, which will 
allow us to study in detail the impact of turbulence in the explosion mechanism and its role in the amplification 
of the magnetic field due to dynamos. It will also help to increase the
realism in the neutrino transport by getting closer to solve the full 6-dimensional Boltzmann equations.
Additionally, these detailed simulations will allow to settle down the issue of whether our current understanding of the
core-collapse mechanism is sufficient to explain typical SN events without practically any numerical uncertainty, or if there
is some unknown physical ingredient missing.
Furthermore, it will allow us to explore the connection between progenitor stars and explosion properties
by performing parametric studies involving hundreds to thousands of 3D simulations.
The challenge here consist in developing numerical algorithms and codes that are capable to scale in parallel to 
a few $10^5$ cores. This will involve an even closer collaboration with applied mathematicians and computer scientists.

\subsubsection{Subgrid modelling}

For the high Reynolds number conditions present in PNSs (see e.g. \cite{Thompson:1993}), fluid motions can 
break up into smaller scales in a turbulent cascade that goes down to the dissipative scale.
Resolving turbulence numerically in global numerical simulations is a challenge, due to the smallness of the 
dissipative scale. In the presence of magnetic fields, the problem is even harder, because, under the right conditions,
small scale turbulence can develop an inverse cascade giving raise to large scale magnetic field
(see e.g. \cite{Brandenburg:2005}). Next generation of supercomputers,
will help us to understand better magneto-hydrodynamc (MHD) 
turbulence and dynamos by means of local simulations, but it may not allow us
to resolve MHD turbulence properly in global simulations of core-collapse supernovae. A possibility would be
to do subgrid modelling, i.e. incorporate the effect of the small scales (e.g. turbulence) in a phenomenological
way to the equations for the large scales (bulk fluid motions). Subgrid modelling of non-magnetized turbulence 
is common in other fields of science (e.g. meteorology) and engineering (e.g. aerodynamics). Although the case of
MHD turbulence is comparably harder to model, there have been some attempts along this line. The most well know 
cases are the alpha-viscosity model \cite{Shakura:1973} describing angular transport in disks by MRI-driven turbulence, 
and the mean field dynamo formalism \cite{Moffatt:1978}, popular in the study of solar dynamos (see e.g. \cite{Charbonneau:2013}) 
and accretion disks (e.g. \cite{Bugli:2014,Sadowski:2015}). There has been some recent progress in this respect
in the context of binary neutron stars \cite{Giacomazzo:2015,Shibata:2017}. However, applications in the core-collapse
scenario are yet to come. The challenge in subgrid modelling is to find the proper closure relations that link  small 
and large scales. In this respect, local simulations may be crucial to fix the parameters and dependencies appearing
in such closures.

\subsubsection{Multidimensional stellar evolution}

To a great extent, the link between supernova progenitors and explosion properties depends on the accuracy 
of the stellar evolution models. Nowadays, pre-supernova models  are the result of 1D calculations (see Section \ref{sec:th:progenitor}),
where multidimensional effects, such as convection and dynamos, are incorporated in a phenomenological way. 
However, the increase in computing power will allow in the future to perform 3D simulations of 
the evolution of stars, or at least of the relevant phases and regions, where multidimensional effects are most important
(see \cite{Baraffe:2011} and examples therein).
This has lead to an increasing interest in the numerical study of the last stages of the star before collapsing
(see e.g. \cite{Mueller:2016b} and references therein). In fact, the asymmetries at the pre-supernova stage, e.g. induced by convection, have been suggested to help the shock revival after bounce \cite{Couch:2013,Couch:2015b,Mueller:2015}.

Multidimensional effects also play a role in interacting binaries. They can induce mass transfer, go through common envelope
phases, or strip down the envelope of stars. This influences the evolution of stars and can lead to completely different 
evolutionary paths, e.g. to rapidly rotating cores (see Section \ref{sec:th:fast}). There has been some 
work in recent years trying to address the problem using multidimensional simulations of the scenario 
(see e.g. \cite{Motl:2002, Lajoie:2011,Lombardi:2011}) and in the future we foresee a great advances in the field.


\begin{acknowledgement}
This work is supported by the New Compstar COST action MP1304. 
N.E.R. acknowledges financial support by the 1994 PRIN-INAF 2014 (project `Transient Universe: unveiling new types of stellar explosions with PESSTO'). N.E.R. acknowledges the hospitality of the ``Institut de Ci\`encies de l'Espai" (CSIC), where part of this work has been done.
P.C.D. acknowledges the financial support from the Spanish Ministerio de Econom\'ia y Competitividad (grant AYA2015-66899-C2-1-P) and from the Generalitat Valenciana (grant PROMETEO-II-2014-069).
\end{acknowledgement}

\bibliographystyle{spphys}
\bibliography{references}

\begin{thebibliography}{100}
\providecommand{\url}[1]{{#1}}
\providecommand{\urlprefix}{URL }
\expandafter\ifx\csname urlstyle\endcsname\relax
  \providecommand{\doi}[1]{DOI \discretionary{}{}{}#1}\else
  \providecommand{\doi}{DOI \discretionary{}{}{}\begingroup
  \urlstyle{rm}\Url}\fi

\bibitem{Janka:2012}
H.T. {Janka}, Annual Review of Nuclear and Particle Science \textbf{62}, 407
  (2012).
\newblock \doi{10.1146/annurev-nucl-102711-094901}

\bibitem{Heger:2003}
A.~{Heger}, C.L. {Fryer}, S.E. {Woosley}, N.~{Langer}, D.H. {Hartmann}, \apj
  \textbf{591}, 288 (2003).
\newblock \doi{10.1086/375341}

\bibitem{baade34}
W.~{Baade}, F.~{Zwicky}, Physical Review \textbf{46}, 76 (1934).
\newblock \doi{10.1103/PhysRev.46.76.2}

\bibitem{galama98}
T.J. {Galama}, P.M. {Vreeswijk}, J.~{van Paradijs}, C.~{Kouveliotou},
  T.~{Augusteijn}, H.~{B{\"o}hnhardt}, J.P. {Brewer}, V.~{Doublier}, J.F.
  {Gonzalez}, B.~{Leibundgut}, C.~{Lidman}, O.R. {Hainaut}, F.~{Patat},
  J.~{Heise}, J.~{in't Zand}, K.~{Hurley}, P.J. {Groot}, R.G. {Strom}, P.A.
  {Mazzali}, K.~{Iwamoto}, K.~{Nomoto}, H.~{Umeda}, T.~{Nakamura}, T.R.
  {Young}, T.~{Suzuki}, T.~{Shigeyama}, T.~{Koshut}, M.~{Kippen},
  C.~{Robinson}, P.~{de Wildt}, R.A.M.J. {Wijers}, N.~{Tanvir}, J.~{Greiner},
  E.~{Pian}, E.~{Palazzi}, F.~{Frontera}, N.~{Masetti}, L.~{Nicastro},
  M.~{Feroci}, E.~{Costa}, L.~{Piro}, B.A. {Peterson}, C.~{Tinney}, B.~{Boyle},
  R.~{Cannon}, R.~{Stathakis}, E.~{Sadler}, M.C. {Begam}, P.~{Ianna}, \nat
  \textbf{395}, 670 (1998).
\newblock \doi{10.1038/27150}

\bibitem{woosley06}
S.E. {Woosley}, J.S. {Bloom}, \araa \textbf{44}, 507 (2006).
\newblock \doi{10.1146/annurev.astro.43.072103.150558}

\bibitem{pian06}
E.~{Pian}, P.A. {Mazzali}, N.~{Masetti}, P.~{Ferrero}, S.~{Klose},
  E.~{Palazzi}, E.~{Ramirez-Ruiz}, S.E. {Woosley}, C.~{Kouveliotou}, J.~{Deng},
  A.V. {Filippenko}, R.J. {Foley}, J.P.U. {Fynbo}, D.A. {Kann}, W.~{Li},
  J.~{Hjorth}, K.~{Nomoto}, F.~{Patat}, D.N. {Sauer}, J.~{Sollerman}, P.M.
  {Vreeswijk}, E.W. {Guenther}, A.~{Levan}, P.~{O'Brien}, N.R. {Tanvir},
  R.A.M.J. {Wijers}, C.~{Dumas}, O.~{Hainaut}, D.S. {Wong}, D.~{Baade},
  L.~{Wang}, L.~{Amati}, E.~{Cappellaro}, A.J. {Castro-Tirado}, S.~{Ellison},
  F.~{Frontera}, A.S. {Fruchter}, J.~{Greiner}, K.~{Kawabata}, C.~{Ledoux},
  K.~{Maeda}, P.~{M{\o}ller}, L.~{Nicastro}, E.~{Rol}, R.~{Starling}, \nat
  \textbf{442}, 1011 (2006).
\newblock \doi{10.1038/nature05082}

\bibitem{thielemann96}
F.K. {Thielemann}, K.~{Nomoto}, M.A. {Hashimoto}, \apj \textbf{460}, 408
  (1996).
\newblock \doi{10.1086/176980}

\bibitem{Andersson:2013}
N.~{Andersson}, J.~{Baker}, K.~{Belczynski}, S.~{Bernuzzi}, E.~{Berti},
  L.~{Cadonati}, P.~{Cerd{\'a}-Dur{\'a}n}, J.~{Clark}, M.~{Favata}, L.S.
  {Finn}, C.~{Fryer}, B.~{Giacomazzo}, J.A. {Gonz{\'a}lez}, M.~{Hendry}, I.S.
  {Heng}, S.~{Hild}, N.~{Johnson-McDaniel}, P.~{Kalmus}, S.~{Klimenko},
  S.~{Kobayashi}, K.~{Kokkotas}, P.~{Laguna}, L.~{Lehner}, J.~{Levin},
  S.~{Liebling}, A.~{MacFadyen}, I.~{Mandel}, S.~{Marka}, Z.~{Marka},
  D.~{Neilsen}, P.~{O'Brien}, R.~{Perna}, J.~{Read}, C.~{Reisswig},
  C.~{Rodriguez}, M.~{Ruffert}, E.~{Schnetter}, A.~{Searle}, P.~{Shawhan},
  D.~{Shoemaker}, A.~{Soderberg}, U.~{Sperhake}, P.~{Sutton}, N.~{Tanvir},
  M.~{Was}, S.~{Whitcomb}, Classical and Quantum Gravity \textbf{30}(19),
  193002 (2013).
\newblock \doi{10.1088/0264-9381/30/19/193002}

\bibitem{hirata87}
K.~{Hirata}, T.~{Kajita}, M.~{Koshiba}, M.~{Nakahata}, Y.~{Oyama}, Physical
  Review Letters \textbf{58}, 1490 (1987).
\newblock \doi{10.1103/PhysRevLett.58.1490}

\bibitem{koyama95}
K.~{Koyama}, R.~{Petre}, E.V. {Gotthelf}, U.~{Hwang}, M.~{Matsuura},
  M.~{Ozaki}, S.S. {Holt}, \nat \textbf{378}, 255 (1995).
\newblock \doi{10.1038/378255a0}

\bibitem{todini01}
P.~{Todini}, A.~{Ferrara}, \mnras \textbf{325}, 726 (2001).
\newblock \doi{10.1046/j.1365-8711.2001.04486.x}

\bibitem{krebs83}
J.~{Krebs}, W.~{Hillebrandt}, \aap \textbf{128}, 411 (1983)

\bibitem{riess98}
A.G. {Riess}, A.V. {Filippenko}, P.~{Challis}, A.~{Clocchiatti}, A.~{Diercks},
  P.M. {Garnavich}, R.L. {Gilliland}, C.J. {Hogan}, S.~{Jha}, R.P. {Kirshner},
  B.~{Leibundgut}, M.M. {Phillips}, D.~{Reiss}, B.P. {Schmidt}, R.A.
  {Schommer}, R.C. {Smith}, J.~{Spyromilio}, C.~{Stubbs}, N.B. {Suntzeff},
  J.~{Tonry}, \aj \textbf{116}, 1009 (1998).
\newblock \doi{10.1086/300499}

\bibitem{perlmutter99}
S.~{Perlmutter}, G.~{Aldering}, G.~{Goldhaber}, R.A. {Knop}, P.~{Nugent}, P.G.
  {Castro}, S.~{Deustua}, S.~{Fabbro}, A.~{Goobar}, D.E. {Groom}, I.M. {Hook},
  A.G. {Kim}, M.Y. {Kim}, J.C. {Lee}, N.J. {Nunes}, R.~{Pain}, C.R.
  {Pennypacker}, R.~{Quimby}, C.~{Lidman}, R.S. {Ellis}, M.~{Irwin}, R.G.
  {McMahon}, P.~{Ruiz-Lapuente}, N.~{Walton}, B.~{Schaefer}, B.J. {Boyle}, A.V.
  {Filippenko}, T.~{Matheson}, A.S. {Fruchter}, N.~{Panagia}, H.J.M. {Newberg},
  W.J. {Couch}, T.S.C. {Project}, \apj \textbf{517}, 565 (1999).
\newblock \doi{10.1086/307221}

\bibitem{hamuy02}
M.~{Hamuy}, P.A. {Pinto}, \apjl \textbf{566}, L63 (2002).
\newblock \doi{10.1086/339676}

\bibitem{stephenson05}
F.R. {Stephenson}, D.A. {Green}, in \emph{1604-2004: Supernovae as Cosmological
  Lighthouses}, \emph{Astronomical Society of the Pacific Conference Series},
  vol. 342, ed. by M.~{Turatto}, S.~{Benetti}, L.~{Zampieri}, W.~{Shea} (2005),
  \emph{Astronomical Society of the Pacific Conference Series}, vol. 342, pp.
  63--70

\bibitem{hartwig85}
E.~{Hartwig}, Astronomische Nachrichten \textbf{112} (1885)

\bibitem{minkowski41}
R.~{Minkowski}, \pasp \textbf{53}, 224 (1941).
\newblock \doi{10.1086/125315}

\bibitem{turatto03}
M.~{Turatto}, S.~{Benetti}, E.~{Cappellaro}, in \emph{From Twilight to
  Highlight: The Physics of Supernovae}, ed. by W.~{Hillebrandt},
  B.~{Leibundgut} (2003), p. 200.
\newblock \doi{10.1007/10828549_26}

\bibitem{galyam16}
A.~{Gal-Yam}, ArXiv e-prints  (2016)

\bibitem{shappee14}
B.J. {Shappee}, J.L. {Prieto}, D.~{Grupe}, C.S. {Kochanek}, K.Z. {Stanek},
  G.~{De Rosa}, S.~{Mathur}, Y.~{Zu}, B.M. {Peterson}, R.W. {Pogge},
  S.~{Komossa}, M.~{Im}, J.~{Jencson}, T.W.S. {Holoien}, U.~{Basu}, J.F.
  {Beacom}, D.M. {Szczygie{\l}}, J.~{Brimacombe}, S.~{Adams}, A.~{Campillay},
  C.~{Choi}, C.~{Contreras}, M.~{Dietrich}, M.~{Dubberley}, M.~{Elphick},
  S.~{Foale}, M.~{Giustini}, C.~{Gonzalez}, E.~{Hawkins}, D.A. {Howell}, E.Y.
  {Hsiao}, M.~{Koss}, K.M. {Leighly}, N.~{Morrell}, D.~{Mudd}, D.~{Mullins},
  J.M. {Nugent}, J.~{Parrent}, M.M. {Phillips}, G.~{Pojmanski}, W.~{Rosing},
  R.~{Ross}, D.~{Sand}, D.M. {Terndrup}, S.~{Valenti}, Z.~{Walker}, Y.~{Yoon},
  \apj \textbf{788}, 48 (2014).
\newblock \doi{10.1088/0004-637X/788/1/48}

\bibitem{hodgkin13}
S.T. {Hodgkin}, L.~{Wyrzykowski}, N.~{Blagorodnova}, S.~{Koposov},
  Philosophical Transactions of the Royal Society of London Series A
  \textbf{371}, 20120239 (2013).
\newblock \doi{10.1098/rsta.2012.0239}

\bibitem{kaiser02}
N.~{Kaiser}, H.~{Aussel}, B.E. {Burke}, H.~{Boesgaard}, K.~{Chambers}, M.R.
  {Chun}, J.N. {Heasley}, K.W. {Hodapp}, B.~{Hunt}, R.~{Jedicke}, D.~{Jewitt},
  R.~{Kudritzki}, G.A. {Luppino}, M.~{Maberry}, E.~{Magnier}, D.G. {Monet},
  P.M. {Onaka}, A.J. {Pickles}, P.H.H. {Rhoads}, T.~{Simon}, A.~{Szalay},
  I.~{Szapudi}, D.J. {Tholen}, J.L. {Tonry}, M.~{Waterson}, J.~{Wick}, in
  \emph{Survey and Other Telescope Technologies and Discoveries},
  \emph{\procspie}, vol. 4836, ed. by J.A. {Tyson}, S.~{Wolff} (2002),
  \emph{\procspie}, vol. 4836, pp. 154--164.
\newblock \doi{10.1117/12.457365}

\bibitem{arnett89}
W.D. {Arnett}, J.N. {Bahcall}, R.P. {Kirshner}, S.E. {Woosley}, \araa
  \textbf{27}, 629 (1989).
\newblock \doi{10.1146/annurev.aa.27.090189.003213}

\bibitem{mccray93}
R.~{McCray}, \araa \textbf{31}, 175 (1993).
\newblock \doi{10.1146/annurev.aa.31.090193.001135}

\bibitem{smartt09}
S.J. {Smartt}, J.J. {Eldridge}, R.M. {Crockett}, J.R. {Maund}, \mnras
  \textbf{395}, 1409 (2009).
\newblock \doi{10.1111/j.1365-2966.2009.14506.x}

\bibitem{smartt15}
S.J. {Smartt}, \pasa \textbf{32}, e016 (2015).
\newblock \doi{10.1017/pasa.2015.17}

\bibitem{Colgate:1960}
S.A. {Colgate}, M.H. {Johnson}, Physical Review Letters \textbf{5}, 235 (1960).
\newblock \doi{10.1103/PhysRevLett.5.235}

\bibitem{Colgate:1961}
S.A. {Colgate}, W.H. {Grasberger}, R.H. {White}, \aj \textbf{66}, 280 (1961).
\newblock \doi{10.1086/108573}

\bibitem{Colgate:1966}
S.A. {Colgate}, R.H. {White}, \apj \textbf{143}, 626 (1966).
\newblock \doi{10.1086/148549}

\bibitem{Bethe:1985}
H.A. {Bethe}, J.R. {Wilson}, \apj \textbf{295}, 14 (1985).
\newblock \doi{10.1086/163343}

\bibitem{Janka:2007}
H.T. {Janka}, K.~{Langanke}, A.~{Marek}, G.~{Mart{\'{\i}}nez-Pinedo},
  B.~{M{\"u}ller}, \physrep \textbf{442}, 38 (2007).
\newblock \doi{10.1016/j.physrep.2007.02.002}

\bibitem{Burrows:2013}
A.~{Burrows}, Reviews of Modern Physics \textbf{85}, 245 (2013).
\newblock \doi{10.1103/RevModPhys.85.245}

\bibitem{Mueller:2016}
B.~{M{\"u}ller}, \pasa \textbf{33}, e048 (2016).
\newblock \doi{10.1017/pasa.2016.40}

\bibitem{Woosley:2002}
S.E. {Woosley}, A.~{Heger}, T.A. {Weaver}, Reviews of Modern Physics
  \textbf{74}, 1015 (2002).
\newblock \doi{10.1103/RevModPhys.74.1015}

\bibitem{pereira13}
R.~{Pereira}, R.C. {Thomas}, G.~{Aldering}, P.~{Antilogus}, C.~{Baltay},
  S.~{Benitez-Herrera}, S.~{Bongard}, C.~{Buton}, A.~{Canto},
  F.~{Cellier-Holzem}, J.~{Chen}, M.~{Childress}, N.~{Chotard}, Y.~{Copin},
  H.K. {Fakhouri}, M.~{Fink}, D.~{Fouchez}, E.~{Gangler}, J.~{Guy},
  W.~{Hillebrandt}, E.Y. {Hsiao}, M.~{Kerschhaggl}, M.~{Kowalski}, M.~{Kromer},
  J.~{Nordin}, P.~{Nugent}, K.~{Paech}, R.~{Pain}, E.~{P{\'e}contal},
  S.~{Perlmutter}, D.~{Rabinowitz}, M.~{Rigault}, K.~{Runge}, C.~{Saunders},
  G.~{Smadja}, C.~{Tao}, S.~{Taubenberger}, A.~{Tilquin}, C.~{Wu}, \aap
  \textbf{554}, A27 (2013).
\newblock \doi{10.1051/0004-6361/201221008}

\bibitem{cao13}
Y.~{Cao}, M.M. {Kasliwal}, I.~{Arcavi}, A.~{Horesh}, P.~{Hancock},
  S.~{Valenti}, S.B. {Cenko}, S.R. {Kulkarni}, A.~{Gal-Yam}, E.~{Gorbikov},
  E.O. {Ofek}, D.~{Sand}, O.~{Yaron}, M.~{Graham}, J.M. {Silverman}, J.C.
  {Wheeler}, G.H. {Marion}, E.S. {Walker}, P.~{Mazzali}, D.A. {Howell}, K.L.
  {Li}, A.K.H. {Kong}, J.S. {Bloom}, P.E. {Nugent}, J.~{Surace}, F.~{Masci},
  J.~{Carpenter}, N.~{Degenaar}, C.R. {Gelino}, \apjl \textbf{775}, L7 (2013).
\newblock \doi{10.1088/2041-8205/775/1/L7}

\bibitem{valenti08}
S.~{Valenti}, N.~{Elias-Rosa}, S.~{Taubenberger}, V.~{Stanishev},
  I.~{Agnoletto}, D.~{Sauer}, E.~{Cappellaro}, A.~{Pastorello}, S.~{Benetti},
  A.~{Riffeser}, U.~{Hopp}, H.~{Navasardyan}, D.~{Tsvetkov}, V.~{Lorenzi},
  F.~{Patat}, M.~{Turatto}, R.~{Barbon}, S.~{Ciroi}, F.~{Di Mille},
  S.~{Frandsen}, J.P.U. {Fynbo}, P.~{Laursen}, P.A. {Mazzali}, \apjl
  \textbf{673}, L155 (2008).
\newblock \doi{10.1086/527672}

\bibitem{leonard02}
D.C. {Leonard}, A.V. {Filippenko}, E.L. {Gates}, W.~{Li}, R.G. {Eastman}, A.J.
  {Barth}, S.J. {Bus}, R.~{Chornock}, A.L. {Coil}, S.~{Frink}, C.A. {Grady},
  A.W. {Harris}, M.A. {Malkan}, T.~{Matheson}, A.~{Quirrenbach}, R.R.
  {Treffers}, \pasp \textbf{114}, 35 (2002).
\newblock \doi{10.1086/324785}

\bibitem{yaron12}
O.~{Yaron}, A.~{Gal-Yam}, \pasp \textbf{124}, 668 (2012).
\newblock \doi{10.1086/666656}

\bibitem{li11}
W.~{Li}, J.~{Leaman}, R.~{Chornock}, A.V. {Filippenko}, D.~{Poznanski},
  M.~{Ganeshalingam}, X.~{Wang}, M.~{Modjaz}, S.~{Jha}, R.J. {Foley},
  N.~{Smith}, \mnras \textbf{412}, 1441 (2011).
\newblock \doi{10.1111/j.1365-2966.2011.18160.x}

\bibitem{anderson14}
J.P. {Anderson}, S.~{Gonz{\'a}lez-Gait{\'a}n}, M.~{Hamuy}, C.P.
  {Guti{\'e}rrez}, M.D. {Stritzinger}, F.~{Olivares E.}, M.M. {Phillips},
  S.~{Schulze}, R.~{Antezana}, L.~{Bolt}, A.~{Campillay}, S.~{Castell{\'o}n},
  C.~{Contreras}, T.~{de Jaeger}, G.~{Folatelli}, F.~{F{\"o}rster}, W.L.
  {Freedman}, L.~{Gonz{\'a}lez}, E.~{Hsiao}, W.~{Krzemi{\'n}ski},
  K.~{Krisciunas}, J.~{Maza}, P.~{McCarthy}, N.I. {Morrell}, S.E. {Persson},
  M.~{Roth}, F.~{Salgado}, N.B. {Suntzeff}, J.~{Thomas-Osip}, \apj
  \textbf{786}, 67 (2014).
\newblock \doi{10.1088/0004-637X/786/1/67}

\bibitem{schlegel90}
E.M. {Schlegel}, \mnras \textbf{244}, 269 (1990)

\bibitem{chevalier94}
R.A. {Chevalier}, C.~{Fransson}, \apj \textbf{420}, 268 (1994).
\newblock \doi{10.1086/173557}

\bibitem{kiewe12}
M.~{Kiewe}, A.~{Gal-Yam}, I.~{Arcavi}, D.C. {Leonard}, J.~{Emilio Enriquez},
  S.B. {Cenko}, D.B. {Fox}, D.S. {Moon}, D.J. {Sand}, A.M. {Soderberg},
  T.~{CCCP}, \apj \textbf{744}, 10 (2012).
\newblock \doi{10.1088/0004-637X/744/1/10}

\bibitem{smith16}
N.~{Smith}, ArXiv e-prints  (2016)

\bibitem{richmond94}
M.W. {Richmond}, R.R. {Treffers}, A.V. {Filippenko}, Y.~{Paik},
  B.~{Leibundgut}, E.~{Schulman}, C.V. {Cox}, \aj \textbf{107}, 1022 (1994).
\newblock \doi{10.1086/116915}

\bibitem{cappellaro15}
E.~{Cappellaro}, M.T. {Botticella}, G.~{Pignata}, A.~{Grado}, L.~{Greggio},
  L.~{Limatola}, M.~{Vaccari}, A.~{Baruffolo}, S.~{Benetti}, F.~{Bufano},
  M.~{Capaccioli}, E.~{Cascone}, G.~{Covone}, D.~{De Cicco}, S.~{Falocco},
  M.~{Della Valle}, M.~{Jarvis}, L.~{Marchetti}, N.R. {Napolitano},
  M.~{Paolillo}, A.~{Pastorello}, M.~{Radovich}, P.~{Schipani}, S.~{Spiro},
  L.~{Tomasella}, M.~{Turatto}, \aap \textbf{584}, A62 (2015).
\newblock \doi{10.1051/0004-6361/201526712}

\bibitem{pastorello12}
A.~{Pastorello}, M.L. {Pumo}, H.~{Navasardyan}, L.~{Zampieri}, M.~{Turatto},
  J.~{Sollerman}, F.~{Taddia}, E.~{Kankare}, S.~{Mattila}, J.~{Nicolas},
  E.~{Prosperi}, A.~{San Segundo Delgado}, S.~{Taubenberger}, T.~{Boles},
  M.~{Bachini}, S.~{Benetti}, F.~{Bufano}, E.~{Cappellaro}, A.D. {Cason},
  G.~{Cetrulo}, M.~{Ergon}, L.~{Germany}, A.~{Harutyunyan}, S.~{Howerton}, G.M.
  {Hurst}, F.~{Patat}, M.~{Stritzinger}, L.G. {Strolger}, W.~{Wells}, \aap
  \textbf{537}, A141 (2012).
\newblock \doi{10.1051/0004-6361/201118112}

\bibitem{soderberg08}
A.M. {Soderberg}, E.~{Berger}, K.L. {Page}, P.~{Schady}, J.~{Parrent},
  D.~{Pooley}, X.Y. {Wang}, E.O. {Ofek}, A.~{Cucchiara}, A.~{Rau}, E.~{Waxman},
  J.D. {Simon}, D.C.J. {Bock}, P.A. {Milne}, M.J. {Page}, J.C. {Barentine},
  S.D. {Barthelmy}, A.P. {Beardmore}, M.F. {Bietenholz}, P.~{Brown},
  A.~{Burrows}, D.N. {Burrows}, G.~{Byrngelson}, S.B. {Cenko}, P.~{Chandra},
  J.R. {Cummings}, D.B. {Fox}, A.~{Gal-Yam}, N.~{Gehrels}, S.~{Immler},
  M.~{Kasliwal}, A.K.H. {Kong}, H.A. {Krimm}, S.R. {Kulkarni}, T.J.
  {Maccarone}, P.~{M{\'e}sz{\'a}ros}, E.~{Nakar}, P.T. {O'Brien}, R.A.
  {Overzier}, M.~{de Pasquale}, J.~{Racusin}, N.~{Rea}, D.G. {York}, \nat
  \textbf{453}, 469 (2008).
\newblock \doi{10.1038/nature06997}

\bibitem{morales14}
A.~{Morales-Garoffolo}, N.~{Elias-Rosa}, S.~{Benetti}, S.~{Taubenberger},
  E.~{Cappellaro}, A.~{Pastorello}, M.~{Klauser}, S.~{Valenti}, S.~{Howerton},
  P.~{Ochner}, N.~{Schramm}, A.~{Siviero}, L.~{Tartaglia}, L.~{Tomasella},
  \mnras \textbf{445}, 1647 (2014).
\newblock \doi{10.1093/mnras/stu1837}

\bibitem{chevalier92}
R.A. {Chevalier}, \apj \textbf{394}, 599 (1992).
\newblock \doi{10.1086/171612}

\bibitem{arnett82}
W.D. {Arnett}, \apj \textbf{253}, 785 (1982).
\newblock \doi{10.1086/159681}

\bibitem{blinnikov00}
S.~{Blinnikov}, P.~{Lundqvist}, O.~{Bartunov}, K.~{Nomoto}, K.~{Iwamoto}, \apj
  \textbf{532}, 1132 (2000).
\newblock \doi{10.1086/308588}

\bibitem{filippenko97}
A.V. {Filippenko}, ARA\&A \textbf{35}, 309 (1997).
\newblock \doi{10.1146/annurev.astro.35.1.309}

\bibitem{taubenberger09}
S.~{Taubenberger}, S.~{Valenti}, S.~{Benetti}, E.~{Cappellaro}, M.~{Della
  Valle}, N.~{Elias-Rosa}, S.~{Hachinger}, W.~{Hillebrandt}, K.~{Maeda}, P.A.
  {Mazzali}, A.~{Pastorello}, F.~{Patat}, S.A. {Sim}, M.~{Turatto}, \mnras
  \textbf{397}, 677 (2009).
\newblock \doi{10.1111/j.1365-2966.2009.15003.x}

\bibitem{jerkstrand17}
A.~{Jerkstrand}, ArXiv e-prints  (2017)

\bibitem{chevalier05}
R.A. {Chevalier}, \apj \textbf{619}, 839 (2005).
\newblock \doi{10.1086/426584}

\bibitem{patnaude17}
D.~{Patnaude}, C.~{Badenes}, ArXiv e-prints  (2017)

\bibitem{white87}
G.L. {White}, D.F. {Malin}, \nat \textbf{327}, 36 (1987).
\newblock \doi{10.1038/327036a0}

\bibitem{aldering94}
G.~{Aldering}, R.M. {Humphreys}, M.~{Richmond}, \aj \textbf{107}, 662 (1994).
\newblock \doi{10.1086/116886}

\bibitem{eldridge13}
J.J. {Eldridge}, M.~{Fraser}, S.J. {Smartt}, J.R. {Maund}, R.M. {Crockett},
  \mnras \textbf{436}, 774 (2013).
\newblock \doi{10.1093/mnras/stt1612}

\bibitem{galyam07}
A.~{Gal-Yam}, D.C. {Leonard}, D.B. {Fox}, S.B. {Cenko}, A.M. {Soderberg}, D.S.
  {Moon}, D.J. {Sand}, {Caltech Core Collapse Program}, W.~{Li}, A.V.
  {Filippenko}, G.~{Aldering}, Y.~{Copin}, \apj \textbf{656}, 372 (2007).
\newblock \doi{10.1086/510523}

\bibitem{maund09}
J.R. {Maund}, S.J. {Smartt}, Science \textbf{324}, 486 (2009).
\newblock \doi{10.1126/science.1170198}

\bibitem{vandyk13}
S.D. {Van Dyk}, W.~{Zheng}, K.I. {Clubb}, A.V. {Filippenko}, S.B. {Cenko},
  N.~{Smith}, O.D. {Fox}, P.L. {Kelly}, I.~{Shivvers}, M.~{Ganeshalingam},
  \apjl \textbf{772}, L32 (2013).
\newblock \doi{10.1088/2041-8205/772/2/L32}

\bibitem{reynolds15}
T.M. {Reynolds}, M.~{Fraser}, G.~{Gilmore}, \mnras \textbf{453}, 2885 (2015).
\newblock \doi{10.1093/mnras/stv1809}

\bibitem{Kochanek:2008}
C.S. {Kochanek}, J.F. {Beacom}, M.D. {Kistler}, J.L. {Prieto}, K.Z. {Stanek},
  T.A. {Thompson}, H.~{Y{\"u}ksel}, \apj \textbf{684}, 1336-1342 (2008).
\newblock \doi{10.1086/590053}

\bibitem{eliasrosa11}
N.~{Elias-Rosa}, S.D. {Van Dyk}, W.~{Li}, J.M. {Silverman}, R.J. {Foley},
  M.~{Ganeshalingam}, J.C. {Mauerhan}, E.~{Kankare}, S.~{Jha}, A.V.
  {Filippenko}, J.E. {Beckman}, E.~{Berger}, J.C. {Cuillandre}, N.~{Smith},
  \apj \textbf{742}, 6 (2011).
\newblock \doi{10.1088/0004-637X/742/1/6}

\bibitem{eliasrosa10}
N.~{Elias-Rosa}, S.D. {Van Dyk}, W.~{Li}, A.A. {Miller}, J.M. {Silverman},
  M.~{Ganeshalingam}, A.F. {Boden}, M.M. {Kasliwal}, J.~{Vink{\'o}}, J.C.
  {Cuillandre}, A.V. {Filippenko}, T.N. {Steele}, J.S. {Bloom}, C.V.
  {Griffith}, I.K.W. {Kleiser}, R.J. {Foley}, \apjl \textbf{714}, L254 (2010).
\newblock \doi{10.1088/2041-8205/714/2/L254}

\bibitem{fraser10}
M.~{Fraser}, K.~{Tak{\'a}ts}, A.~{Pastorello}, S.J. {Smartt}, S.~{Mattila},
  M.T. {Botticella}, S.~{Valenti}, M.~{Ergon}, J.~{Sollerman}, I.~{Arcavi},
  S.~{Benetti}, F.~{Bufano}, R.M. {Crockett}, I.J. {Danziger}, A.~{Gal-Yam},
  J.R. {Maund}, S.~{Taubenberger}, M.~{Turatto}, \apjl \textbf{714}, L280
  (2010).
\newblock \doi{10.1088/2041-8205/714/2/L280}

\bibitem{maund15}
J.R. {Maund}, M.~{Fraser}, E.~{Reilly}, M.~{Ergon}, S.~{Mattila}, \mnras
  \textbf{447}, 3207 (2015).
\newblock \doi{10.1093/mnras/stu2658}

\bibitem{maund11}
J.R. {Maund}, M.~{Fraser}, M.~{Ergon}, A.~{Pastorello}, S.J. {Smartt},
  J.~{Sollerman}, S.~{Benetti}, M.T. {Botticella}, F.~{Bufano}, I.J.
  {Danziger}, R.~{Kotak}, L.~{Magill}, A.W. {Stephens}, S.~{Valenti}, \apjl
  \textbf{739}, L37 (2011).
\newblock \doi{10.1088/2041-8205/739/2/L37}

\bibitem{vandyk11}
S.D. {Van Dyk}, W.~{Li}, S.B. {Cenko}, M.M. {Kasliwal}, A.~{Horesh}, E.O.
  {Ofek}, A.L. {Kraus}, J.M. {Silverman}, I.~{Arcavi}, A.V. {Filippenko},
  A.~{Gal-Yam}, R.M. {Quimby}, S.R. {Kulkarni}, O.~{Yaron}, D.~{Polishook},
  \apjl \textbf{741}, L28 (2011).
\newblock \doi{10.1088/2041-8205/741/2/L28}

\bibitem{bersten12}
M.C. {Bersten}, O.G. {Benvenuto}, K.~{Nomoto}, M.~{Ergon}, G.~{Folatelli},
  J.~{Sollerman}, S.~{Benetti}, M.T. {Botticella}, M.~{Fraser}, R.~{Kotak},
  K.~{Maeda}, P.~{Ochner}, L.~{Tomasella}, \apj \textbf{757}, 31 (2012).
\newblock \doi{10.1088/0004-637X/757/1/31}

\bibitem{benvenuto13}
O.G. {Benvenuto}, M.C. {Bersten}, K.~{Nomoto}, \apj \textbf{762}, 74 (2013).
\newblock \doi{10.1088/0004-637X/762/2/74}

\bibitem{folatelli14}
G.~{Folatelli}, M.C. {Bersten}, O.G. {Benvenuto}, S.D. {Van Dyk},
  H.~{Kuncarayakti}, K.~{Maeda}, T.~{Nozawa}, K.~{Nomoto}, M.~{Hamuy}, R.M.
  {Quimby}, \apjl \textbf{793}, L22 (2014).
\newblock \doi{10.1088/2041-8205/793/2/L22}

\bibitem{vandyk14}
S.D. {Van Dyk}, W.~{Zheng}, O.D. {Fox}, S.B. {Cenko}, K.I. {Clubb}, A.V.
  {Filippenko}, R.J. {Foley}, A.A. {Miller}, N.~{Smith}, P.L. {Kelly}, W.H.
  {Lee}, S.~{Ben-Ami}, A.~{Gal-Yam}, \aj \textbf{147}, 37 (2014).
\newblock \doi{10.1088/0004-6256/147/2/37}

\bibitem{tartaglia17}
L.~{Tartaglia}, M.~{Fraser}, D.J. {Sand}, S.~{Valenti}, S.J. {Smartt},
  C.~{McCully}, J.P. {Anderson}, I.~{Arcavi}, N.~{Elias-Rosa}, L.~{Galbany},
  A.~{Gal-Yam}, J.B. {Haislip}, G.~{Hosseinzadeh}, D.A. {Howell}, C.~{Inserra},
  S.W. {Jha}, E.~{Kankare}, P.~{Lundqvist}, K.~{Maguire}, S.~{Mattila},
  D.~{Reichart}, K.W. {Smith}, M.~{Smith}, M.~{Stritzinger}, M.~{Sullivan},
  F.~{Taddia}, L.~{Tomasella}, \apjl \textbf{836}, L12 (2017).
\newblock \doi{10.3847/2041-8213/aa5c7f}

\bibitem{crockett08}
R.M. {Crockett}, J.J. {Eldridge}, S.J. {Smartt}, A.~{Pastorello}, A.~{Gal-Yam},
  D.B. {Fox}, D.C. {Leonard}, M.M. {Kasliwal}, S.~{Mattila}, J.R. {Maund}, A.W.
  {Stephens}, I.J. {Danziger}, \mnras \textbf{391}, L5 (2008).
\newblock \doi{10.1111/j.1745-3933.2008.00540.x}

\bibitem{eliasrosa13}
N.~{Elias-Rosa}, A.~{Pastorello}, J.R. {Maund}, K.~{Tak{\'a}ts}, M.~{Fraser},
  S.J. {Smartt}, S.~{Benetti}, G.~{Pignata}, D.~{Sand}, S.~{Valenti}, \mnras
  \textbf{436}, L109 (2013).
\newblock \doi{10.1093/mnrasl/slt124}

\bibitem{yoon12}
S.C. {Yoon}, G.~{Gr{\"a}fener}, J.S. {Vink}, A.~{Kozyreva}, R.G. {Izzard}, \aap
  \textbf{544}, L11 (2012).
\newblock \doi{10.1051/0004-6361/201219790}

\bibitem{groh13}
J.H. {Groh}, G.~{Meynet}, C.~{Georgy}, S.~{Ekstr{\"o}m}, \aap \textbf{558},
  A131 (2013).
\newblock \doi{10.1051/0004-6361/201321906}

\bibitem{bersten14}
M.C. {Bersten}, O.G. {Benvenuto}, G.~{Folatelli}, K.~{Nomoto},
  H.~{Kuncarayakti}, S.~{Srivastav}, G.C. {Anupama}, R.~{Quimby}, D.K. {Sahu},
  \aj \textbf{148}, 68 (2014).
\newblock \doi{10.1088/0004-6256/148/4/68}

\bibitem{eldridge15}
J.J. {Eldridge}, M.~{Fraser}, J.R. {Maund}, S.J. {Smartt}, \mnras \textbf{446},
  2689 (2015).
\newblock \doi{10.1093/mnras/stu2197}

\bibitem{eldridge16}
J.J. {Eldridge}, J.R. {Maund}, \mnras \textbf{461}, L117 (2016).
\newblock \doi{10.1093/mnrasl/slw099}

\bibitem{galyam09}
A.~{Gal-Yam}, D.C. {Leonard}, \nat \textbf{458}, 865 (2009).
\newblock \doi{10.1038/nature07934}

\bibitem{smith11}
N.~{Smith}, W.~{Li}, A.A. {Miller}, J.M. {Silverman}, A.V. {Filippenko}, J.C.
  {Cuillandre}, M.C. {Cooper}, T.~{Matheson}, S.D. {Van Dyk}, \apj
  \textbf{732}, 63 (2011).
\newblock \doi{10.1088/0004-637X/732/2/63}

\bibitem{pastorello07}
A.~{Pastorello}, S.J. {Smartt}, S.~{Mattila}, J.J. {Eldridge}, D.~{Young},
  K.~{Itagaki}, H.~{Yamaoka}, H.~{Navasardyan}, S.~{Valenti}, F.~{Patat},
  I.~{Agnoletto}, T.~{Augusteijn}, S.~{Benetti}, E.~{Cappellaro}, T.~{Boles},
  J.M. {Bonnet-Bidaud}, M.T. {Botticella}, F.~{Bufano}, C.~{Cao}, J.~{Deng},
  M.~{Dennefeld}, N.~{Elias-Rosa}, A.~{Harutyunyan}, F.P. {Keenan},
  T.~{Iijima}, V.~{Lorenzi}, P.A. {Mazzali}, X.~{Meng}, S.~{Nakano}, T.B.
  {Nielsen}, J.V. {Smoker}, V.~{Stanishev}, M.~{Turatto}, D.~{Xu},
  L.~{Zampieri}, \nat \textbf{447}, 829 (2007).
\newblock \doi{10.1038/nature05825}

\bibitem{ofek14}
E.O. {Ofek}, M.~{Sullivan}, N.J. {Shaviv}, A.~{Steinbok}, I.~{Arcavi},
  A.~{Gal-Yam}, D.~{Tal}, S.R. {Kulkarni}, P.E. {Nugent}, S.~{Ben-Ami}, M.M.
  {Kasliwal}, S.B. {Cenko}, R.~{Laher}, J.~{Surace}, J.S. {Bloom}, A.V.
  {Filippenko}, J.M. {Silverman}, O.~{Yaron}, \apj \textbf{789}, 104 (2014).
\newblock \doi{10.1088/0004-637X/789/2/104}

\bibitem{pastorello13}
A.~{Pastorello}, E.~{Cappellaro}, C.~{Inserra}, S.J. {Smartt}, G.~{Pignata},
  S.~{Benetti}, S.~{Valenti}, M.~{Fraser}, K.~{Tak{\'a}ts}, S.~{Benitez}, M.T.
  {Botticella}, J.~{Brimacombe}, F.~{Bufano}, F.~{Cellier-Holzem}, M.T.
  {Costado}, G.~{Cupani}, I.~{Curtis}, N.~{Elias-Rosa}, M.~{Ergon}, J.P.U.
  {Fynbo}, F.J. {Hambsch}, M.~{Hamuy}, A.~{Harutyunyan}, K.M. {Ivarson},
  E.~{Kankare}, J.C. {Martin}, R.~{Kotak}, A.P. {LaCluyze}, K.~{Maguire},
  S.~{Mattila}, J.~{Maza}, M.~{McCrum}, M.~{Miluzio}, H.U. {Norgaard-Nielsen},
  M.C. {Nysewander}, P.~{Ochner}, Y.C. {Pan}, M.L. {Pumo}, D.E. {Reichart},
  T.G. {Tan}, S.~{Taubenberger}, L.~{Tomasella}, M.~{Turatto}, D.~{Wright},
  \apj \textbf{767}, 1 (2013).
\newblock \doi{10.1088/0004-637X/767/1/1}

\bibitem{fraser13}
M.~{Fraser}, C.~{Inserra}, A.~{Jerkstrand}, R.~{Kotak}, G.~{Pignata},
  S.~{Benetti}, M.T. {Botticella}, F.~{Bufano}, M.~{Childress}, S.~{Mattila},
  A.~{Pastorello}, S.J. {Smartt}, M.~{Turatto}, F.~{Yuan}, J.P. {Anderson},
  D.D.R. {Bayliss}, F.E. {Bauer}, T.W. {Chen}, F.~{F{\"o}rster Bur{\'o}n},
  A.~{Gal-Yam}, J.B. {Haislip}, C.~{Knapic}, L.~{Le Guillou}, S.~{Marchi},
  P.~{Mazzali}, M.~{Molinaro}, J.P. {Moore}, D.~{Reichart}, R.~{Smareglia},
  K.W. {Smith}, A.~{Sternberg}, M.~{Sullivan}, K.~{Tak{\'a}ts}, B.E. {Tucker},
  S.~{Valenti}, O.~{Yaron}, D.R. {Young}, G.~{Zhou}, \mnras \textbf{433}, 1312
  (2013).
\newblock \doi{10.1093/mnras/stt813}

\bibitem{mauerhan13a}
J.C. {Mauerhan}, N.~{Smith}, A.V. {Filippenko}, K.B. {Blanchard}, P.K.
  {Blanchard}, C.F.E. {Casper}, S.B. {Cenko}, K.I. {Clubb}, D.P. {Cohen}, K.L.
  {Fuller}, G.Z. {Li}, J.M. {Silverman}, \mnras \textbf{430}, 1801 (2013).
\newblock \doi{10.1093/mnras/stt009}

\bibitem{margutti14}
R.~{Margutti}, D.~{Milisavljevic}, A.M. {Soderberg}, R.~{Chornock}, B.A.
  {Zauderer}, K.~{Murase}, C.~{Guidorzi}, N.E. {Sanders}, P.~{Kuin},
  C.~{Fransson}, E.M. {Levesque}, P.~{Chandra}, E.~{Berger}, F.B. {Bianco},
  P.J. {Brown}, P.~{Challis}, E.~{Chatzopoulos}, C.C. {Cheung}, C.~{Choi},
  L.~{Chomiuk}, N.~{Chugai}, C.~{Contreras}, M.R. {Drout}, R.~{Fesen}, R.J.
  {Foley}, W.~{Fong}, A.S. {Friedman}, C.~{Gall}, N.~{Gehrels}, J.~{Hjorth},
  E.~{Hsiao}, R.~{Kirshner}, M.~{Im}, G.~{Leloudas}, R.~{Lunnan}, G.H.
  {Marion}, J.~{Martin}, N.~{Morrell}, K.F. {Neugent}, N.~{Omodei}, M.M.
  {Phillips}, A.~{Rest}, J.M. {Silverman}, J.~{Strader}, M.D. {Stritzinger},
  T.~{Szalai}, N.B. {Utterback}, J.~{Vinko}, J.C. {Wheeler}, D.~{Arnett},
  S.~{Campana}, R.~{Chevalier}, A.~{Ginsburg}, A.~{Kamble}, P.W.A. {Roming},
  T.~{Pritchard}, G.~{Stringfellow}, \apj \textbf{780}, 21 (2014).
\newblock \doi{10.1088/0004-637X/780/1/21}

\bibitem{smith10}
N.~{Smith}, A.~{Miller}, W.~{Li}, A.V. {Filippenko}, J.M. {Silverman}, A.W.
  {Howard}, P.~{Nugent}, G.W. {Marcy}, J.S. {Bloom}, A.M. {Ghez}, J.~{Lu},
  S.~{Yelda}, R.A. {Bernstein}, J.E. {Colucci}, \aj \textbf{139}, 1451 (2010).
\newblock \doi{10.1088/0004-6256/139/4/1451}

\bibitem{foley11}
R.J. {Foley}, E.~{Berger}, O.~{Fox}, E.M. {Levesque}, P.J. {Challis}, I.I.
  {Ivans}, J.E. {Rhoads}, A.M. {Soderberg}, \apj \textbf{732}, 32 (2011).
\newblock \doi{10.1088/0004-637X/732/1/32}

\bibitem{soker13}
N.~{Soker}, A.~{Kashi}, \apjl \textbf{764}, L6 (2013).
\newblock \doi{10.1088/2041-8205/764/1/L6}

\bibitem{mackey14}
J.~{Mackey}, S.~{Mohamed}, V.V. {Gvaramadze}, R.~{Kotak}, N.~{Langer}, D.M.A.
  {Meyer}, T.J. {Moriya}, H.R. {Neilson}, \nat \textbf{512}, 282 (2014).
\newblock \doi{10.1038/nature13522}

\bibitem{eldridge04}
J.J. {Eldridge}, C.A. {Tout}, \mnras \textbf{353}, 87 (2004).
\newblock \doi{10.1111/j.1365-2966.2004.08041.x}

\bibitem{Chandrasekhar:1938}
S.~Chandrasekhar, \emph{Stellar Structure} (Dover, New York, 1938)

\bibitem{Sekiguchi:2011}
Y.~{Sekiguchi}, M.~{Shibata}, \apj \textbf{737}, 6 (2011).
\newblock \doi{10.1088/0004-637X/737/1/6}

\bibitem{Janka:2008}
H.T. {Janka}, B.~{M{\"u}ller}, F.S. {Kitaura}, R.~{Buras}, \aap \textbf{485},
  199 (2008).
\newblock \doi{10.1051/0004-6361:20079334}

\bibitem{Buras:2006}
R.~{Buras}, M.~{Rampp}, H.T. {Janka}, K.~{Kifonidis}, \aap \textbf{447}, 1049
  (2006).
\newblock \doi{10.1051/0004-6361:20053783}

\bibitem{Marek:2009}
A.~{Marek}, H.T. {Janka}, E.~{M{\"u}ller}, \aap \textbf{496}, 475 (2009).
\newblock \doi{10.1051/0004-6361/200810883}

\bibitem{Blondin:2003}
J.M. {Blondin}, A.~{Mezzacappa}, C.~{DeMarino}, \apj \textbf{584}, 971 (2003).
\newblock \doi{10.1086/345812}

\bibitem{Foglizzo:2000}
T.~{Foglizzo}, M.~{Tagger}, \aap \textbf{363}, 174 (2000)

\bibitem{Yamasaki:2007}
T.~{Yamasaki}, S.~{Yamada}, \apj \textbf{656}, 1019 (2007).
\newblock \doi{10.1086/510505}

\bibitem{Yamada:1999}
S.~{Yamada}, H.T. {Janka}, H.~{Suzuki}, \aap \textbf{344}, 533 (1999)

\bibitem{Liebendoerfer:2004}
M.~{Liebend{\"o}rfer}, O.E.B. {Messer}, A.~{Mezzacappa}, S.W. {Bruenn}, C.Y.
  {Cardall}, F.K. {Thielemann}, \apjs \textbf{150}, 263 (2004).
\newblock \doi{10.1086/380191}

\bibitem{Burrows:2000}
A.~{Burrows}, T.~{Young}, P.~{Pinto}, R.~{Eastman}, T.A. {Thompson}, \apj
  \textbf{539}, 865 (2000).
\newblock \doi{10.1086/309244}

\bibitem{Mueller:2010}
B.~{M{\"u}ller}, H.T. {Janka}, H.~{Dimmelmeier}, \apjs \textbf{189}, 104
  (2010).
\newblock \doi{10.1088/0067-0049/189/1/104}

\bibitem{Rampp:2002}
M.~{Rampp}, H.T. {Janka}, \aap \textbf{396}, 361 (2002).
\newblock \doi{10.1051/0004-6361:20021398}

\bibitem{Burrows:2006}
A.~{Burrows}, E.~{Livne}, L.~{Dessart}, C.D. {Ott}, J.~{Murphy}, \apj
  \textbf{640}, 878 (2006).
\newblock \doi{10.1086/500174}

\bibitem{Burrows:2007a}
A.~{Burrows}, E.~{Livne}, L.~{Dessart}, C.D. {Ott}, J.~{Murphy}, \apj
  \textbf{655}, 416 (2007).
\newblock \doi{10.1086/509773}

\bibitem{Burrows:2007b}
A.~{Burrows}, L.~{Dessart}, E.~{Livne}, C.D. {Ott}, J.~{Murphy}, \apj
  \textbf{664}, 416 (2007).
\newblock \doi{10.1086/519161}

\bibitem{Swesty:2009}
F.D. {Swesty}, E.S. {Myra}, \apjs \textbf{181}, 1 (2009).
\newblock \doi{10.1088/0067-0049/181/1/1}

\bibitem{Ott:2008}
C.D. {Ott}, A.~{Burrows}, L.~{Dessart}, E.~{Livne}, \apj \textbf{685},
  1069-1088 (2008).
\newblock \doi{10.1086/591440}

\bibitem{Brandt:2011}
T.D. {Brandt}, A.~{Burrows}, C.D. {Ott}, E.~{Livne}, \apj \textbf{728}, 8
  (2011).
\newblock \doi{10.1088/0004-637X/728/1/8}

\bibitem{Obergaulinger:2014}
M.~{Obergaulinger}, O.~{Just}, H.T. {Janka}, M.A. {Aloy}, C.~{Aloy}, in
  \emph{8th International Conference of Numerical Modeling of Space Plasma
  Flows (ASTRONUM 2013)}, \emph{Astronomical Society of the Pacific Conference
  Series}, vol. 488, ed. by N.V. {Pogorelov}, E.~{Audit}, G.P. {Zank} (2014),
  \emph{Astronomical Society of the Pacific Conference Series}, vol. 488, p.
  255

\bibitem{Liebendoerfer:2009}
M.~{Liebend{\"o}rfer}, S.C. {Whitehouse}, T.~{Fischer}, \apj \textbf{698}, 1174
  (2009).
\newblock \doi{10.1088/0004-637X/698/2/1174}

\bibitem{Mueller:2015a}
B.~{M{\"u}ller}, H.T. {Janka}, \mnras \textbf{448}, 2141 (2015).
\newblock \doi{10.1093/mnras/stv101}

\bibitem{Mueller:2015b}
B.~{M{\"u}ller}, \mnras \textbf{453}, 287 (2015).
\newblock \doi{10.1093/mnras/stv1611}

\bibitem{Oconnor:2010}
E.~{O'Connor}, C.D. {Ott}, Classical and Quantum Gravity \textbf{27}(11),
  114103 (2010).
\newblock \doi{10.1088/0264-9381/27/11/114103}

\bibitem{Ott:2012}
C.D. {Ott}, E.~{Abdikamalov}, E.~{O'Connor}, C.~{Reisswig}, R.~{Haas},
  P.~{Kalmus}, S.~{Drasco}, A.~{Burrows}, E.~{Schnetter}, \prd \textbf{86}(2),
  024026 (2012).
\newblock \doi{10.1103/PhysRevD.86.024026}

\bibitem{Perego:2016}
A.~{Perego}, R.M. {Cabez{\'o}n}, R.~{K{\"a}ppeli}, \apjs \textbf{223}, 22
  (2016).
\newblock \doi{10.3847/0067-0049/223/2/22}

\bibitem{Lentz:2012}
E.J. {Lentz}, A.~{Mezzacappa}, O.E.B. {Messer}, M.~{Liebend{\"o}rfer}, W.R.
  {Hix}, S.W. {Bruenn}, \apj \textbf{747}, 73 (2012).
\newblock \doi{10.1088/0004-637X/747/1/73}

\bibitem{Burrows:1998}
A.~{Burrows}, R.F. {Sawyer}, \prc \textbf{58}, 554 (1998).
\newblock \doi{10.1103/PhysRevC.58.554}

\bibitem{Langanke:2003}
K.~{Langanke}, G.~{Mart{\'{\i}}nez-Pinedo}, J.M. {Sampaio}, D.J. {Dean}, W.R.
  {Hix}, O.E. {Messer}, A.~{Mezzacappa}, M.~{Liebend{\"o}rfer}, H.T. {Janka},
  M.~{Rampp}, Physical Review Letters \textbf{90}(24), 241102 (2003).
\newblock \doi{10.1103/PhysRevLett.90.241102}

\bibitem{Horowitz:1997}
C.J. {Horowitz}, \prd \textbf{55}, 4577 (1997).
\newblock \doi{10.1103/PhysRevD.55.4577}

\bibitem{Mezzacappa:1993}
A.~{Mezzacappa}, S.W. {Bruenn}, \apj \textbf{405}, 669 (1993).
\newblock \doi{10.1086/172395}

\bibitem{Hannestad:1998}
S.~{Hannestad}, G.~{Raffelt}, \apj \textbf{507}, 339 (1998).
\newblock \doi{10.1086/306303}

\bibitem{Bruenn:1985}
S.W. {Bruenn}, \apjs \textbf{58}, 771 (1985).
\newblock \doi{10.1086/191056}

\bibitem{Pons:1998}
J.A. {Pons}, J.A. {Miralles}, J.M.A. {Ibanez}, \aaps \textbf{129}, 343 (1998).
\newblock \doi{10.1051/aas:1998189}

\bibitem{Buras:2003}
R.~{Buras}, H.T. {Janka}, M.T. {Keil}, G.G. {Raffelt}, M.~{Rampp}, \apj
  \textbf{587}, 320 (2003).
\newblock \doi{10.1086/368015}

\bibitem{Rampp:2000}
M.~Rampp, Radiation hydrodynamics with neutrinos: Stellar core collapse and the
  explosion mechanism of type ii supernovae.
\newblock Dissertation, Technische Universität München, München (2000)

\bibitem{Burrows:2016}
A.~{Burrows}, D.~{Vartanyan}, J.C. {Dolence}, M.A. {Skinner}, D.~{Radice},
  ArXiv e-prints  (2016)

\bibitem{Mueller:2012}
B.~{M{\"u}ller}, H.T. {Janka}, A.~{Marek}, \apj \textbf{756}, 84 (2012).
\newblock \doi{10.1088/0004-637X/756/1/84}

\bibitem{Ott:2007a}
C.D. {Ott}, H.~{Dimmelmeier}, A.~{Marek}, H.T. {Janka}, I.~{Hawke}, B.~{Zink},
  E.~{Schnetter}, Physical Review Letters \textbf{98}(26), 261101 (2007).
\newblock \doi{10.1103/PhysRevLett.98.261101}

\bibitem{Ott:2013}
C.D. {Ott}, E.~{Abdikamalov}, P.~{M{\"o}sta}, R.~{Haas}, S.~{Drasco}, E.P.
  {O'Connor}, C.~{Reisswig}, C.A. {Meakin}, E.~{Schnetter}, \apj \textbf{768},
  115 (2013).
\newblock \doi{10.1088/0004-637X/768/2/115}

\bibitem{Moesta:2014}
P.~{M{\"o}sta}, S.~{Richers}, C.D. {Ott}, R.~{Haas}, A.L. {Piro},
  K.~{Boydstun}, E.~{Abdikamalov}, C.~{Reisswig}, E.~{Schnetter}, \apjl
  \textbf{785}, L29 (2014).
\newblock \doi{10.1088/2041-8205/785/2/L29}

\bibitem{Abdikamalov:2015}
E.~{Abdikamalov}, C.D. {Ott}, D.~{Radice}, L.F. {Roberts}, R.~{Haas},
  C.~{Reisswig}, P.~{M{\"o}sta}, H.~{Klion}, E.~{Schnetter}, \apj \textbf{808},
  70 (2015).
\newblock \doi{10.1088/0004-637X/808/1/70}

\bibitem{Shibata:1995}
M.~{Shibata}, T.~{Nakamura}, \prd \textbf{52}, 5428 (1995).
\newblock \doi{10.1103/PhysRevD.52.5428}

\bibitem{Baumgarte:1999}
T.W. {Baumgarte}, S.L. {Shapiro}, \prd \textbf{59}(2), 024007 (1999).
\newblock \doi{10.1103/PhysRevD.59.024007}

\bibitem{Isenberg:2008}
J.A. {Isenberg}, International Journal of Modern Physics D \textbf{17}, 265
  (2008).
\newblock \doi{10.1142/S0218271808011997}.
\newblock This article appeared for the first time in 1978 as a University of
  Maryland preprint.

\bibitem{Wilson:1996}
J.R. {Wilson}, G.J. {Mathews}, P.~{Marronetti}, \prd \textbf{54}, 1317 (1996).
\newblock \doi{10.1103/PhysRevD.54.1317}

\bibitem{Dimmelmeier:2002}
H.~Dimmelmeier, J.A. Font, E.~M{\"u}ller, Astronomy \& Astrophysics
  \textbf{388}(3), 917 (2002)

\bibitem{Cerda-Duran:2013}
P.~{Cerd{\'a}-Dur{\'a}n}, N.~{DeBrye}, M.A. {Aloy}, J.A. {Font},
  M.~{Obergaulinger}, \apjl \textbf{779}, L18 (2013).
\newblock \doi{10.1088/2041-8205/779/2/L18}

\bibitem{Shibata:2004}
M.~{Shibata}, Y.I. {Sekiguchi}, \prd \textbf{69}(8), 084024 (2004).
\newblock \doi{10.1103/PhysRevD.69.084024}

\bibitem{Ott:2007b}
C.D. {Ott}, H.~{Dimmelmeier}, A.~{Marek}, H.T. {Janka}, B.~{Zink}, I.~{Hawke},
  E.~{Schnetter}, Classical and Quantum Gravity \textbf{24}, S139 (2007).
\newblock \doi{10.1088/0264-9381/24/12/S10}

\bibitem{Cordero-Carrion:2009}
I.~{Cordero-Carri{\'o}n}, P.~{Cerd{\'a}-Dur{\'a}n}, H.~{Dimmelmeier}, J.L.
  {Jaramillo}, J.~{Novak}, E.~{Gourgoulhon}, \prd \textbf{79}(2), 024017
  (2009).
\newblock \doi{10.1103/PhysRevD.79.024017}

\bibitem{Cordero-Carrion:2014}
I.~{Cordero-Carri{\'o}n}, N.~{Vasset}, J.~{Novak}, J.L. {Jaramillo}, \prd
  \textbf{90}(4), 044062 (2014).
\newblock \doi{10.1103/PhysRevD.90.044062}

\bibitem{Cerda-Duran:2005}
P.~{Cerd{\'a}-Dur{\'a}n}, G.~{Faye}, H.~{Dimmelmeier}, J.A. {Font}, J.M.
  {Ib{\'a}{\~n}ez}, E.~{M{\"u}ller}, G.~{Sch{\"a}fer}, \aap \textbf{439}, 1033
  (2005).
\newblock \doi{10.1051/0004-6361:20042602}

\bibitem{Marek:2006}
A.~{Marek}, H.~{Dimmelmeier}, H.T. {Janka}, E.~{M{\"u}ller}, R.~{Buras}, \aap
  \textbf{445}, 273 (2006).
\newblock \doi{10.1051/0004-6361:20052840}

\bibitem{Kitaura:2006}
F.S. {Kitaura}, H.T. {Janka}, W.~{Hillebrandt}, \aap \textbf{450}, 345 (2006).
\newblock \doi{10.1051/0004-6361:20054703}

\bibitem{Obergaulinger:2006}
M.~{Obergaulinger}, M.A. {Aloy}, H.~{Dimmelmeier}, E.~{M{\"u}ller}, \aap
  \textbf{457}, 209 (2006).
\newblock \doi{10.1051/0004-6361:20064982}

\bibitem{Scheidegger:2008}
S.~{Scheidegger}, T.~{Fischer}, S.C. {Whitehouse}, M.~{Liebend{\"o}rfer}, \aap
  \textbf{490}, 231 (2008).
\newblock \doi{10.1051/0004-6361:20078577}

\bibitem{Wongwathanarat:2013}
A.~{Wongwathanarat}, H.T. {Janka}, E.~{M{\"u}ller}, \aap \textbf{552}, A126
  (2013).
\newblock \doi{10.1051/0004-6361/201220636}

\bibitem{Hanke:2013}
F.~{Hanke}, B.~{M{\"u}ller}, A.~{Wongwathanarat}, A.~{Marek}, H.T. {Janka},
  \apj \textbf{770}, 66 (2013).
\newblock \doi{10.1088/0004-637X/770/1/66}

\bibitem{Bruenn:2013}
S.W. {Bruenn}, A.~{Mezzacappa}, W.R. {Hix}, E.J. {Lentz}, O.E.B. {Messer}, E.J.
  {Lingerfelt}, J.M. {Blondin}, E.~{Endeve}, P.~{Marronetti}, K.N. {Yakunin},
  \apjl \textbf{767}, L6 (2013).
\newblock \doi{10.1088/2041-8205/767/1/L6}

\bibitem{Oconnor:2015}
E.~{O'Connor}, S.~{Couch}, ArXiv e-prints  (2015)

\bibitem{Bruenn:2016}
S.W. {Bruenn}, E.J. {Lentz}, W.R. {Hix}, A.~{Mezzacappa}, J.A. {Harris}, O.E.B.
  {Messer}, E.~{Endeve}, J.M. {Blondin}, M.A. {Chertkow}, E.J. {Lingerfelt},
  P.~{Marronetti}, K.N. {Yakunin}, \apj \textbf{818}, 123 (2016).
\newblock \doi{10.3847/0004-637X/818/2/123}

\bibitem{Summa:2016}
A.~{Summa}, F.~{Hanke}, H.T. {Janka}, T.~{Melson}, A.~{Marek}, B.~{M{\"u}ller},
  \apj \textbf{825}, 6 (2016).
\newblock \doi{10.3847/0004-637X/825/1/6}

\bibitem{Reisswig:2011}
C.~{Reisswig}, C.D. {Ott}, U.~{Sperhake}, E.~{Schnetter}, \prd \textbf{83}(6),
  064008 (2011).
\newblock \doi{10.1103/PhysRevD.83.064008}

\bibitem{Lattimer:1991}
J.M. {Lattimer}, F.~{Douglas Swesty}, Nuclear Physics A \textbf{535}, 331
  (1991).
\newblock \doi{10.1016/0375-9474(91)90452-C}

\bibitem{HShen:1998a}
H.~{Shen}, H.~{Toki}, K.~{Oyamatsu}, K.~{Sumiyoshi}, Nuclear Physics A
  \textbf{637}, 435 (1998).
\newblock \doi{10.1016/S0375-9474(98)00236-X}

\bibitem{HShen:1998b}
H.~{Shen}, H.~{Toki}, K.~{Oyamatsu}, K.~{Sumiyoshi}, Progress of Theoretical
  Physics \textbf{100}, 1013 (1998).
\newblock \doi{10.1143/PTP.100.1013}

\bibitem{HShen:2011}
H.~{Shen}, H.~{Toki}, K.~{Oyamatsu}, K.~{Sumiyoshi}, \apjs \textbf{197}, 20
  (2011).
\newblock \doi{10.1088/0067-0049/197/2/20}

\bibitem{Furusawa:2011}
S.~{Furusawa}, S.~{Yamada}, K.~{Sumiyoshi}, H.~{Suzuki}, \apj \textbf{738}, 178
  (2011).
\newblock \doi{10.1088/0004-637X/738/2/178}

\bibitem{Furusawa:2013}
S.~{Furusawa}, K.~{Sumiyoshi}, S.~{Yamada}, H.~{Suzuki}, \apj \textbf{772}, 95
  (2013).
\newblock \doi{10.1088/0004-637X/772/2/95}

\bibitem{Hempel:2010}
M.~{Hempel}, J.~{Schaffner-Bielich}, Nuclear Physics A \textbf{837}, 210
  (2010).
\newblock \doi{10.1016/j.nuclphysa.2010.02.010}

\bibitem{Steiner:2013}
A.W. {Steiner}, M.~{Hempel}, T.~{Fischer}, \apj \textbf{774}, 17 (2013).
\newblock \doi{10.1088/0004-637X/774/1/17}

\bibitem{GShen:2011a}
G.~{Shen}, C.J. {Horowitz}, S.~{Teige}, \prc \textbf{83}(3), 035802 (2011).
\newblock \doi{10.1103/PhysRevC.83.035802}

\bibitem{GShen:2011b}
G.~{Shen}, C.J. {Horowitz}, E.~{O'Connor}, \prc \textbf{83}(6), 065808 (2011).
\newblock \doi{10.1103/PhysRevC.83.065808}

\bibitem{Oertel:2017}
M.~{Oertel}, M.~{Hempel}, T.~{Kl{\"a}hn}, S.~{Typel}, Reviews of Modern Physics
  \textbf{89}(1), 015007 (2017).
\newblock \doi{10.1103/RevModPhys.89.015007}

\bibitem{Lattimer:2000}
J.M. {Lattimer}, M.~{Prakash}, \physrep \textbf{333}, 121 (2000).
\newblock \doi{10.1016/S0370-1573(00)00019-3}

\bibitem{Sumiyoshi:2005}
K.~{Sumiyoshi}, S.~{Yamada}, H.~{Suzuki}, H.~{Shen}, S.~{Chiba}, H.~{Toki},
  \apj \textbf{629}, 922 (2005).
\newblock \doi{10.1086/431788}

\bibitem{Suwa:2013}
Y.~{Suwa}, T.~{Takiwaki}, K.~{Kotake}, T.~{Fischer}, M.~{Liebend{\"o}rfer},
  K.~{Sato}, \apj \textbf{764}, 99 (2013).
\newblock \doi{10.1088/0004-637X/764/1/99}

\bibitem{Hempel:2012}
M.~{Hempel}, T.~{Fischer}, J.~{Schaffner-Bielich}, M.~{Liebend{\"o}rfer}, \apj
  \textbf{748}, 70 (2012).
\newblock \doi{10.1088/0004-637X/748/1/70}

\bibitem{Timmes_Arnett:1999}
F.X. {Timmes}, D.~{Arnett}, \apjs \textbf{125}, 277 (1999).
\newblock \doi{10.1086/313271}

\bibitem{Timmes:2000}
F.X. {Timmes}, F.D. {Swesty}, \apjs \textbf{126}, 501 (2000).
\newblock \doi{10.1086/313304}

\bibitem{Timmes:1999}
F.X. {Timmes}, \apjs \textbf{124}, 241 (1999).
\newblock \doi{10.1086/313257}

\bibitem{Mueller:2014}
B.~{M{\"u}ller}, H.T. {Janka}, \apj \textbf{788}, 82 (2014).
\newblock \doi{10.1088/0004-637X/788/1/82}

\bibitem{Lentz:2015}
E.J. {Lentz}, S.W. {Bruenn}, W.R. {Hix}, A.~{Mezzacappa}, O.E.B. {Messer},
  E.~{Endeve}, J.M. {Blondin}, J.A. {Harris}, P.~{Marronetti}, K.N. {Yakunin},
  \apjl \textbf{807}, L31 (2015).
\newblock \doi{10.1088/2041-8205/807/2/L31}

\bibitem{Melson:2015b}
T.~{Melson}, H.T. {Janka}, R.~{Bollig}, F.~{Hanke}, A.~{Marek},
  B.~{M{\"u}ller}, \apjl \textbf{808}, L42 (2015).
\newblock \doi{10.1088/2041-8205/808/2/L42}

\bibitem{Takiwaki:2016}
T.~{Takiwaki}, K.~{Kotake}, Y.~{Suwa}, \mnras \textbf{461}, L112 (2016).
\newblock \doi{10.1093/mnrasl/slw105}

\bibitem{Roberts:2016}
L.F. {Roberts}, C.D. {Ott}, R.~{Haas}, E.P. {O'Connor}, P.~{Diener},
  E.~{Schnetter}, \apj \textbf{831}, 98 (2016).
\newblock \doi{10.3847/0004-637X/831/1/98}

\bibitem{Dolence:2015}
J.C. {Dolence}, A.~{Burrows}, W.~{Zhang}, \apj \textbf{800}, 10 (2015).
\newblock \doi{10.1088/0004-637X/800/1/10}

\bibitem{Nagakura:2017}
H.~{Nagakura}, W.~{Iwakami}, S.~{Furusawa}, H.~{Okawa}, A.~{Harada},
  K.~{Sumiyoshi}, S.~{Yamada}, H.~{Matsufuru}, A.~{Imakura}, ArXiv e-prints
  (2017)

\bibitem{Suwa:2016}
Y.~{Suwa}, E.~{M{\"u}ller}, \mnras \textbf{460}, 2664 (2016).
\newblock \doi{10.1093/mnras/stw1150}

\bibitem{Pan:2016}
K.C. {Pan}, M.~{Liebend{\"o}rfer}, M.~{Hempel}, F.K. {Thielemann}, \apj
  \textbf{817}, 72 (2016).
\newblock \doi{10.3847/0004-637X/817/1/72}

\bibitem{Melson:2015a}
T.~{Melson}, H.T. {Janka}, A.~{Marek}, \apjl \textbf{801}, L24 (2015).
\newblock \doi{10.1088/2041-8205/801/2/L24}

\bibitem{Couch:2015}
S.M. {Couch}, C.D. {Ott}, \apj \textbf{799}, 5 (2015).
\newblock \doi{10.1088/0004-637X/799/1/5}

\bibitem{Nomoto:1984}
K.~{Nomoto}, \apj \textbf{277}, 791 (1984).
\newblock \doi{10.1086/161749}

\bibitem{Nomoto:1987}
K.~{Nomoto}, \apj \textbf{322}, 206 (1987).
\newblock \doi{10.1086/165716}

\bibitem{Fischer:2010}
T.~{Fischer}, S.C. {Whitehouse}, A.~{Mezzacappa}, F.K. {Thielemann},
  M.~{Liebend{\"o}rfer}, \aap \textbf{517}, A80 (2010).
\newblock \doi{10.1051/0004-6361/200913106}

\bibitem{Smartt:2009}
S.J. {Smartt}, \araa \textbf{47}, 63 (2009).
\newblock \doi{10.1146/annurev-astro-082708-101737}

\bibitem{botticella09}
M.T. {Botticella}, A.~{Pastorello}, S.J. {Smartt}, W.P.S. {Meikle},
  S.~{Benetti}, R.~{Kotak}, E.~{Cappellaro}, R.M. {Crockett}, S.~{Mattila},
  M.~{Sereno}, F.~{Patat}, D.~{Tsvetkov}, J.T. {van Loon}, D.~{Abraham},
  I.~{Agnoletto}, R.~{Arbour}, C.~{Benn}, G.~{di Rico}, N.~{Elias-Rosa}, D.L.
  {Gorshanov}, A.~{Harutyunyan}, D.~{Hunter}, V.~{Lorenzi}, F.P. {Keenan},
  K.~{Maguire}, J.~{Mendez}, M.~{Mobberley}, H.~{Navasardyan}, C.~{Ries},
  V.~{Stanishev}, S.~{Taubenberger}, C.~{Trundle}, M.~{Turatto}, I.M. {Volkov},
  \mnras \textbf{398}, 1041 (2009).
\newblock \doi{10.1111/j.1365-2966.2009.15082.x}

\bibitem{Hillebrandt:1982}
W.~{Hillebrandt}, \aap \textbf{110}, L3 (1982)

\bibitem{Nomoto:1982}
K.~{Nomoto}, D.~{Sugimoto}, W.M. {Sparks}, R.A. {Fesen}, T.R. {Gull},
  S.~{Miyaji}, \nat \textbf{299}, 803 (1982).
\newblock \doi{10.1038/299803a0}

\bibitem{Jones:2013}
S.~{Jones}, R.~{Hirschi}, K.~{Nomoto}, T.~{Fischer}, F.X. {Timmes},
  F.~{Herwig}, B.~{Paxton}, H.~{Toki}, T.~{Suzuki},
  G.~{Mart{\'{\i}}nez-Pinedo}, Y.H. {Lam}, M.G. {Bertolli}, \apj \textbf{772},
  150 (2013).
\newblock \doi{10.1088/0004-637X/772/2/150}

\bibitem{Schwab:2010}
J.~{Schwab}, P.~{Podsiadlowski}, S.~{Rappaport}, \apj \textbf{719}, 722 (2010).
\newblock \doi{10.1088/0004-637X/719/1/722}

\bibitem{Poelarends:2008}
A.J.T. {Poelarends}, F.~{Herwig}, N.~{Langer}, A.~{Heger}, \apj \textbf{675},
  614-625 (2008).
\newblock \doi{10.1086/520872}

\bibitem{Podsiadlowski:2004}
P.~{Podsiadlowski}, N.~{Langer}, A.J.T. {Poelarends}, S.~{Rappaport},
  A.~{Heger}, E.~{Pfahl}, \apj \textbf{612}, 1044 (2004).
\newblock \doi{10.1086/421713}

\bibitem{Pumo:2009}
M.L. {Pumo}, M.~{Turatto}, M.T. {Botticella}, A.~{Pastorello}, S.~{Valenti},
  L.~{Zampieri}, S.~{Benetti}, E.~{Cappellaro}, F.~{Patat}, \apjl \textbf{705},
  L138 (2009).
\newblock \doi{10.1088/0004-637X/705/2/L138}

\bibitem{Wanajo:2011}
S.~{Wanajo}, H.T. {Janka}, B.~{M{\"u}ller}, \apjl \textbf{726}, L15 (2011).
\newblock \doi{10.1088/2041-8205/726/2/L15}

\bibitem{Pastorello:2004}
A.~{Pastorello}, L.~{Zampieri}, M.~{Turatto}, E.~{Cappellaro}, W.P.S. {Meikle},
  S.~{Benetti}, D.~{Branch}, E.~{Baron}, F.~{Patat}, M.~{Armstrong},
  G.~{Altavilla}, M.~{Salvo}, M.~{Riello}, \mnras \textbf{347}, 74 (2004).
\newblock \doi{10.1111/j.1365-2966.2004.07173.x}

\bibitem{vandenHeuvel:2004}
E.P.J. {van den Heuvel}, in \emph{5th INTEGRAL Workshop on the INTEGRAL
  Universe}, \emph{ESA Special Publication}, vol. 552, ed. by
  V.~{Schoenfelder}, G.~{Lichti}, C.~{Winkler} (2004), \emph{ESA Special
  Publication}, vol. 552, p. 185

\bibitem{Oezel:2012}
F.~{{\"O}zel}, D.~{Psaltis}, R.~{Narayan}, A.~{Santos Villarreal}, \apj
  \textbf{757}, 55 (2012).
\newblock \doi{10.1088/0004-637X/757/1/55}

\bibitem{Andrews:2015}
J.J. {Andrews}, W.M. {Farr}, V.~{Kalogera}, B.~{Willems}, \apj \textbf{801}, 32
  (2015).
\newblock \doi{10.1088/0004-637X/801/1/32}

\bibitem{Oezel:2016}
F.~{{\"O}zel}, P.~{Freire}, \araa \textbf{54}, 401 (2016).
\newblock \doi{10.1146/annurev-astro-081915-023322}

\bibitem{Limongi:2006}
M.~{Limongi}, A.~{Chieffi}, \apj \textbf{647}, 483 (2006).
\newblock \doi{10.1086/505164}

\bibitem{Nomoto:2006}
K.~{Nomoto}, N.~{Tominaga}, H.~{Umeda}, C.~{Kobayashi}, K.~{Maeda}, Nuclear
  Physics A \textbf{777}, 424 (2006).
\newblock \doi{10.1016/j.nuclphysa.2006.05.008}

\bibitem{Woosley:2007}
S.E. {Woosley}, A.~{Heger}, \physrep \textbf{442}, 269 (2007).
\newblock \doi{10.1016/j.physrep.2007.02.009}

\bibitem{Chieffi:2013}
A.~{Chieffi}, M.~{Limongi}, \apj \textbf{764}, 21 (2013).
\newblock \doi{10.1088/0004-637X/764/1/21}

\bibitem{Sukhbold:2014}
T.~{Sukhbold}, S.E. {Woosley}, \apj \textbf{783}, 10 (2014).
\newblock \doi{10.1088/0004-637X/783/1/10}

\bibitem{Woosley:2015}
S.E. {Woosley}, A.~{Heger}, \apj \textbf{810}, 34 (2015).
\newblock \doi{10.1088/0004-637X/810/1/34}

\bibitem{Sukhbold:2016}
T.~{Sukhbold}, T.~{Ertl}, S.E. {Woosley}, J.M. {Brown}, H.T. {Janka}, \apj
  \textbf{821}, 38 (2016).
\newblock \doi{10.3847/0004-637X/821/1/38}

\bibitem{Mueller:2012b}
B.~{M{\"u}ller}, H.T. {Janka}, A.~{Heger}, \apj \textbf{761}, 72 (2012).
\newblock \doi{10.1088/0004-637X/761/1/72}

\bibitem{Mueller:2013}
B.~{M{\"u}ller}, H.T. {Janka}, A.~{Marek}, \apj \textbf{766}, 43 (2013).
\newblock \doi{10.1088/0004-637X/766/1/43}

\bibitem{Shigeyama:1990}
T.~{Shigeyama}, K.~{Nomoto}, \apj \textbf{360}, 242 (1990).
\newblock \doi{10.1086/169114}

\bibitem{Woosley:1988}
S.E. {Woosley}, P.A. {Pinto}, L.~{Ensman}, \apj \textbf{324}, 466 (1988).
\newblock \doi{10.1086/165908}

\bibitem{OConnor:2011}
E.~{O'Connor}, C.D. {Ott}, \apj \textbf{730}, 70 (2011).
\newblock \doi{10.1088/0004-637X/730/2/70}

\bibitem{Zhang:2008}
W.~{Zhang}, S.E. {Woosley}, A.~{Heger}, \apj \textbf{679}, 639-654 (2008).
\newblock \doi{10.1086/526404}

\bibitem{Ugliano:2012}
M.~{Ugliano}, H.T. {Janka}, A.~{Marek}, A.~{Arcones}, \apj \textbf{757}, 69
  (2012).
\newblock \doi{10.1088/0004-637X/757/1/69}

\bibitem{Pejcha:2015}
O.~{Pejcha}, T.A. {Thompson}, \apj \textbf{801}, 90 (2015).
\newblock \doi{10.1088/0004-637X/801/2/90}

\bibitem{Ertl:2016}
T.~{Ertl}, H.T. {Janka}, S.E. {Woosley}, T.~{Sukhbold}, M.~{Ugliano}, \apj
  \textbf{818}, 124 (2016).
\newblock \doi{10.3847/0004-637X/818/2/124}

\bibitem{Horiuchi:2011}
S.~{Horiuchi}, J.F. {Beacom}, C.S. {Kochanek}, J.L. {Prieto}, K.Z. {Stanek},
  T.A. {Thompson}, \apj \textbf{738}, 154 (2011).
\newblock \doi{10.1088/0004-637X/738/2/154}

\bibitem{Nadezhin:1980}
D.K. {Nadezhin}, \apss \textbf{69}, 115 (1980).
\newblock \doi{10.1007/BF00638971}

\bibitem{Lovegrove:2013}
E.~{Lovegrove}, S.E. {Woosley}, \apj \textbf{769}, 109 (2013).
\newblock \doi{10.1088/0004-637X/769/2/109}

\bibitem{Piro:2013}
A.L. {Piro}, \apjl \textbf{768}, L14 (2013).
\newblock \doi{10.1088/2041-8205/768/1/L14}

\bibitem{Colgate:1971}
S.A. {Colgate}, \apj \textbf{163}, 221 (1971).
\newblock \doi{10.1086/150760}

\bibitem{Chevalier:1989}
R.A. {Chevalier}, \apj \textbf{346}, 847 (1989).
\newblock \doi{10.1086/168066}

\bibitem{Kifonidis:2003}
K.~{Kifonidis}, T.~{Plewa}, H.T. {Janka}, E.~{M{\"u}ller}, \aap \textbf{408},
  621 (2003).
\newblock \doi{10.1051/0004-6361:20030863}

\bibitem{Scheck:2006}
L.~{Scheck}, K.~{Kifonidis}, H.T. {Janka}, E.~{M{\"u}ller}, \aap \textbf{457},
  963 (2006).
\newblock \doi{10.1051/0004-6361:20064855}

\bibitem{Hammer:2010}
N.J. {Hammer}, H.T. {Janka}, E.~{M{\"u}ller}, \apj \textbf{714}, 1371 (2010).
\newblock \doi{10.1088/0004-637X/714/2/1371}

\bibitem{Joggerst:2010}
C.C. {Joggerst}, A.~{Almgren}, S.E. {Woosley}, \apj \textbf{723}, 353 (2010).
\newblock \doi{10.1088/0004-637X/723/1/353}

\bibitem{Wongwathanarat:2015}
A.~{Wongwathanarat}, E.~{M{\"u}ller}, H.T. {Janka}, \aap \textbf{577}, A48
  (2015).
\newblock \doi{10.1051/0004-6361/201425025}

\bibitem{Moriya:2010}
T.~{Moriya}, N.~{Tominaga}, M.~{Tanaka}, K.~{Nomoto}, D.N. {Sauer}, P.A.
  {Mazzali}, K.~{Maeda}, T.~{Suzuki}, \apj \textbf{719}, 1445 (2010).
\newblock \doi{10.1088/0004-637X/719/2/1445}

\bibitem{Young:1995}
E.J. {Young}, G.~{Chanmugam}, \apjl \textbf{442}, L53 (1995).
\newblock \doi{10.1086/187814}

\bibitem{Muslimov:1995}
A.~{Muslimov}, D.~{Page}, \apjl \textbf{440}, L77 (1995).
\newblock \doi{10.1086/187765}

\bibitem{Geppert:1999}
U.~{Geppert}, D.~{Page}, T.~{Zannias}, \aap \textbf{345}, 847 (1999)

\bibitem{Shabaltas:2012}
N.~{Shabaltas}, D.~{Lai}, \apj \textbf{748}, 148 (2012).
\newblock \doi{10.1088/0004-637X/748/2/148}

\bibitem{Payne:2004}
D.J.B. {Payne}, A.~{Melatos}, \mnras \textbf{351}, 569 (2004).
\newblock \doi{10.1111/j.1365-2966.2004.07798.x}

\bibitem{Payne:2007}
D.J.B. {Payne}, A.~{Melatos}, \mnras \textbf{376}, 609 (2007).
\newblock \doi{10.1111/j.1365-2966.2007.11451.x}

\bibitem{Bernal:2010}
C.G. {Bernal}, W.H. {Lee}, D.~{Page}, \rmxaa \textbf{46}, 309 (2010)

\bibitem{Mukherjee:2013a}
D.~{Mukherjee}, D.~{Bhattacharya}, A.~{Mignone}, \mnras \textbf{430}, 1976
  (2013).
\newblock \doi{10.1093/mnras/stt020}

\bibitem{Bernal:2013}
C.G. {Bernal}, D.~{Page}, W.H. {Lee}, \apj \textbf{770}, 106 (2013).
\newblock \doi{10.1088/0004-637X/770/2/106}

\bibitem{Mukherjee:2013b}
D.~{Mukherjee}, D.~{Bhattacharya}, A.~{Mignone}, \mnras \textbf{435}, 718
  (2013).
\newblock \doi{10.1093/mnras/stt1344}

\bibitem{Torres-Forne:2016}
A.~{Torres-Forn{\'e}}, P.~{Cerd{\'a}-Dur{\'a}n}, J.A. {Pons}, J.A. {Font},
  \mnras \textbf{456}, 3813 (2016).
\newblock \doi{10.1093/mnras/stv2926}

\bibitem{Sana:2012}
H.~{Sana}, S.E. {de Mink}, A.~{de Koter}, N.~{Langer}, C.J. {Evans},
  M.~{Gieles}, E.~{Gosset}, R.G. {Izzard}, J.B. {Le Bouquin}, F.R.N.
  {Schneider}, Science \textbf{337}, 444 (2012).
\newblock \doi{10.1126/science.1223344}

\bibitem{Hobbs:2005}
G.~{Hobbs}, D.R. {Lorimer}, A.G. {Lyne}, M.~{Kramer}, \mnras \textbf{360}, 974
  (2005).
\newblock \doi{10.1111/j.1365-2966.2005.09087.x}

\bibitem{Arzoumanian:2002}
Z.~{Arzoumanian}, D.F. {Chernoff}, J.M. {Cordes}, \apj \textbf{568}, 289
  (2002).
\newblock \doi{10.1086/338805}

\bibitem{Janka:1994}
H.T. {Janka}, E.~{Mueller}, \aap \textbf{290}, 496 (1994)

\bibitem{Fryer:2007}
C.L. {Fryer}, P.A. {Young}, \apj \textbf{659}, 1438 (2007).
\newblock \doi{10.1086/513003}

\bibitem{Nordhaus:2010}
J.~{Nordhaus}, T.D. {Brandt}, A.~{Burrows}, E.~{Livne}, C.D. {Ott}, \prd
  \textbf{82}(10), 103016 (2010).
\newblock \doi{10.1103/PhysRevD.82.103016}

\bibitem{Nordhaus:2012}
J.~{Nordhaus}, T.D. {Brandt}, A.~{Burrows}, A.~{Almgren}, \mnras \textbf{423},
  1805 (2012).
\newblock \doi{10.1111/j.1365-2966.2012.21002.x}

\bibitem{Scheck:2004}
L.~{Scheck}, T.~{Plewa}, H.T. {Janka}, K.~{Kifonidis}, E.~{M{\"u}ller},
  Physical Review Letters \textbf{92}(1), 011103 (2004).
\newblock \doi{10.1103/PhysRevLett.92.011103}

\bibitem{Wongwathanarat:2010}
A.~{Wongwathanarat}, H.T. {Janka}, E.~{M{\"u}ller}, \apjl \textbf{725}, L106
  (2010).
\newblock \doi{10.1088/2041-8205/725/1/L106}

\bibitem{Spruit:1998}
H.~{Spruit}, E.S. {Phinney}, \nat \textbf{393}, 139 (1998).
\newblock \doi{10.1038/30168}

\bibitem{Blaauw:1961}
A.~{Blaauw}, \bain \textbf{15}, 265 (1961)

\bibitem{Boersma:1961}
J.~{Boersma}, \bain \textbf{15}, 291 (1961)

\bibitem{Postnov:2014}
K.A. Postnov, L.R. Yungelson, Living Reviews in Relativity \textbf{17}(1), 3
  (2014).
\newblock \doi{10.12942/lrr-2014-3}.
\newblock \urlprefix\url{http://dx.doi.org/10.12942/lrr-2014-3}

\bibitem{Woosley:1995}
S.E. {Woosley}, T.A. {Weaver}, \apjs \textbf{101}, 181 (1995).
\newblock \doi{10.1086/192237}

\bibitem{Marigo:2002}
P.~{Marigo}, L.~{Girardi}, C.~{Chiosi}, P.R. {Wood}, \aap \textbf{371}, 152
  (2001).
\newblock \doi{10.1051/0004-6361:20010309}

\bibitem{Heger:2002}
A.~{Heger}, S.E. {Woosley}, \apj \textbf{567}, 532 (2002).
\newblock \doi{10.1086/338487}

\bibitem{Umeda:2005}
H.~{Umeda}, K.~{Nomoto}, \apj \textbf{619}, 427 (2005).
\newblock \doi{10.1086/426097}

\bibitem{Hirschi:2006}
R.~{Hirschi}, {et al.}, in \emph{Reviews in Modern Astronomy}, \emph{Reviews in
  Modern Astronomy}, vol.~19, ed. by S.~{Roeser} (2006), \emph{Reviews in
  Modern Astronomy}, vol.~19, p. 101.
\newblock \doi{10.1002/9783527619030.ch5}

\bibitem{Tominaga:2007}
N.~{Tominaga}, H.~{Umeda}, K.~{Nomoto}, \apj \textbf{660}, 516 (2007).
\newblock \doi{10.1086/513063}

\bibitem{Limongi:2012}
M.~{Limongi}, A.~{Chieffi}, \apjs \textbf{199}, 38 (2012).
\newblock \doi{10.1088/0067-0049/199/2/38}

\bibitem{Fukuda:1982}
I.~{Fukuda}, \pasp \textbf{94}, 271 (1982).
\newblock \doi{10.1086/130977}

\bibitem{Langer:1998}
N.~{Langer}, \aap \textbf{329}, 551 (1998)

\bibitem{Heger:2000}
A.~{Heger}, N.~{Langer}, S.E. {Woosley}, \apj \textbf{528}, 368 (2000).
\newblock \doi{10.1086/308158}

\bibitem{Maeder:2000a}
A.~{Maeder}, G.~{Meynet}, \araa \textbf{38}, 143 (2000).
\newblock \doi{10.1146/annurev.astro.38.1.143}

\bibitem{Maeder:2000b}
A.~{Maeder}, G.~{Meynet}, \aap \textbf{361}, 159 (2000)

\bibitem{Hirschi:2004}
R.~{Hirschi}, G.~{Meynet}, A.~{Maeder}, \aap \textbf{425}, 649 (2004).
\newblock \doi{10.1051/0004-6361:20041095}

\bibitem{Spruit:1999}
H.C. {Spruit}, \aap \textbf{349}, 189 (1999)

\bibitem{Spruit:2002}
H.C. {Spruit}, \aap \textbf{381}, 923 (2002).
\newblock \doi{10.1051/0004-6361:20011465}

\bibitem{Charbonneau:2013}
P.~{Charbonneau}, Solar and Stellar Dynamos: Saas-Fee Advanced Course 39 Swiss
  Society for Astrophysics and Astronomy, Saas-Fee Advanced Courses, Volume
  39.~ISBN 978-3-642-32092-7.~Springer-Verlag Berlin Heidelberg, 2013
  \textbf{39} (2013).
\newblock \doi{10.1007/978-3-642-32093-4}

\bibitem{Heger:2005}
A.~{Heger}, S.E. {Woosley}, H.C. {Spruit}, \apj \textbf{626}, 350 (2005).
\newblock \doi{10.1086/429868}

\bibitem{Maeder:1987}
A.~{Maeder}, \aap \textbf{178}, 159 (1987)

\bibitem{Woosley:2006}
S.E. {Woosley}, A.~{Heger}, \apj \textbf{637}, 914 (2006).
\newblock \doi{10.1086/498500}

\bibitem{Yoon:2005}
S.C. {Yoon}, N.~{Langer}, \aap \textbf{443}, 643 (2005).
\newblock \doi{10.1051/0004-6361:20054030}

\bibitem{MacFadyen:1999}
A.I. {MacFadyen}, S.E. {Woosley}, \apj \textbf{524}, 262 (1999).
\newblock \doi{10.1086/307790}

\bibitem{Usov:1994}
V.V. {Usov}, \mnras \textbf{267}, 1035 (1994).
\newblock \doi{10.1093/mnras/267.4.1035}

\bibitem{Wheeler:2000}
J.C. {Wheeler}, I.~{Yi}, P.~{H{\"o}flich}, L.~{Wang}, \apj \textbf{537}, 810
  (2000).
\newblock \doi{10.1086/309055}

\bibitem{nicholl13}
M.~{Nicholl}, S.J. {Smartt}, A.~{Jerkstrand}, C.~{Inserra}, M.~{McCrum},
  R.~{Kotak}, M.~{Fraser}, D.~{Wright}, T.W. {Chen}, K.~{Smith}, D.R. {Young},
  S.A. {Sim}, S.~{Valenti}, D.A. {Howell}, F.~{Bresolin}, R.P. {Kudritzki},
  J.L. {Tonry}, M.E. {Huber}, A.~{Rest}, A.~{Pastorello}, L.~{Tomasella},
  E.~{Cappellaro}, S.~{Benetti}, S.~{Mattila}, E.~{Kankare}, T.~{Kangas},
  G.~{Leloudas}, J.~{Sollerman}, F.~{Taddia}, E.~{Berger}, R.~{Chornock},
  G.~{Narayan}, C.W. {Stubbs}, R.J. {Foley}, R.~{Lunnan}, A.~{Soderberg},
  N.~{Sanders}, D.~{Milisavljevic}, R.~{Margutti}, R.P. {Kirshner},
  N.~{Elias-Rosa}, A.~{Morales-Garoffolo}, S.~{Taubenberger}, M.T.
  {Botticella}, S.~{Gezari}, Y.~{Urata}, S.~{Rodney}, A.G. {Riess},
  D.~{Scolnic}, W.M. {Wood-Vasey}, W.S. {Burgett}, K.~{Chambers}, H.A.
  {Flewelling}, E.A. {Magnier}, N.~{Kaiser}, N.~{Metcalfe}, J.~{Morgan}, P.A.
  {Price}, W.~{Sweeney}, C.~{Waters}, \nat \textbf{502}, 346 (2013).
\newblock \doi{10.1038/nature12569}

\bibitem{Shibata:2006}
M.~{Shibata}, Y.T. {Liu}, S.L. {Shapiro}, B.C. {Stephens}, \prd
  \textbf{74}(10), 104026 (2006).
\newblock \doi{10.1103/PhysRevD.74.104026}

\bibitem{Ott:2006}
C.D. {Ott}, A.~{Burrows}, T.A. {Thompson}, E.~{Livne}, R.~{Walder}, \apjs
  \textbf{164}, 130 (2006).
\newblock \doi{10.1086/500832}

\bibitem{Burrows:2007}
A.~{Burrows}, L.~{Dessart}, E.~{Livne}, C.D. {Ott}, J.~{Murphy}, \apj
  \textbf{664}, 416 (2007).
\newblock \doi{10.1086/519161}

\bibitem{CerdaDuran:2008}
P.~{Cerd{\'a}-Dur{\'a}n}, J.A. {Font}, L.~{Ant{\'o}n}, E.~{M{\"u}ller}, \aap
  \textbf{492}, 937 (2008).
\newblock \doi{10.1051/0004-6361:200810086}

\bibitem{Takiwaki:2011}
T.~{Takiwaki}, K.~{Kotake}, \apj \textbf{743}, 30 (2011).
\newblock \doi{10.1088/0004-637X/743/1/30}

\bibitem{Sawai:2013}
H.~{Sawai}, S.~{Yamada}, H.~{Suzuki}, \apjl \textbf{770}, L19 (2013).
\newblock \doi{10.1088/2041-8205/770/2/L19}

\bibitem{Sawai:2016}
H.~{Sawai}, S.~{Yamada}, \apj \textbf{817}, 153 (2016).
\newblock \doi{10.3847/0004-637X/817/2/153}

\bibitem{Obergaulinger:2017}
M.~{Obergaulinger}, M.~{{\'A}ngel Aloy}, ArXiv e-prints  (2017)

\bibitem{Mikami:2008}
H.~{Mikami}, Y.~{Sato}, T.~{Matsumoto}, T.~{Hanawa}, \apj \textbf{683}, 357-374
  (2008).
\newblock \doi{10.1086/589759}

\bibitem{Kuroda:2010}
T.~{Kuroda}, H.~{Umeda}, \apjs \textbf{191}, 439 (2010).
\newblock \doi{10.1088/0067-0049/191/2/439}

\bibitem{Scheidegger:2010}
S.~{Scheidegger}, R.~{K{\"a}ppeli}, S.C. {Whitehouse}, T.~{Fischer},
  M.~{Liebend{\"o}rfer}, \aap \textbf{514}, A51 (2010).
\newblock \doi{10.1051/0004-6361/200913220}

\bibitem{Moesta:2015}
P.~{M{\"o}sta}, C.D. {Ott}, D.~{Radice}, L.F. {Roberts}, E.~{Schnetter},
  R.~{Haas}, \nat \textbf{528}, 376 (2015).
\newblock \doi{10.1038/nature15755}

\bibitem{Winteler:2012}
C.~{Winteler}, R.~{K{\"a}ppeli}, A.~{Perego}, A.~{Arcones}, N.~{Vasset},
  N.~{Nishimura}, M.~{Liebend{\"o}rfer}, F.K. {Thielemann}, \apjl \textbf{750},
  L22 (2012).
\newblock \doi{10.1088/2041-8205/750/1/L22}

\bibitem{Akiyama:2003}
S.~{Akiyama}, J.C. {Wheeler}, D.L. {Meier}, I.~{Lichtenstadt}, \apj
  \textbf{584}, 954 (2003).
\newblock \doi{10.1086/344135}

\bibitem{Velikhov:1959}
V.~E., Sov. Phys. JETP \textbf{36} (1959)

\bibitem{Chandrasekhar:1960}
S.~{Chandrasekhar}, Proceedings of the National Academy of Science \textbf{46},
  253 (1960).
\newblock \doi{10.1073/pnas.46.2.253}

\bibitem{Balbus:1991}
S.A. {Balbus}, J.F. {Hawley}, \apj \textbf{376}, 214 (1991).
\newblock \doi{10.1086/170270}

\bibitem{Rembiasz:2016b}
T.~{Rembiasz}, J.~{Guilet}, M.~{Obergaulinger}, P.~{Cerd{\'a}-Dur{\'a}n}, M.A.
  {Aloy}, E.~{M{\"u}ller}, \mnras \textbf{460}, 3316 (2016).
\newblock \doi{10.1093/mnras/stw1201}

\bibitem{Obergaulinger:2009}
M.~{Obergaulinger}, P.~{Cerd{\'a}-Dur{\'a}n}, E.~{M{\"u}ller}, M.A. {Aloy},
  \aap \textbf{498}, 241 (2009).
\newblock \doi{10.1051/0004-6361/200811323}

\bibitem{Guilet:2015a}
J.~{Guilet}, E.~{M{\"u}ller}, \mnras \textbf{450}, 2153 (2015).
\newblock \doi{10.1093/mnras/stv727}

\bibitem{Rembiasz:2016a}
T.~{Rembiasz}, M.~{Obergaulinger}, P.~{Cerd{\'a}-Dur{\'a}n}, E.~{M{\"u}ller},
  M.A. {Aloy}, \mnras \textbf{456}, 3782 (2016).
\newblock \doi{10.1093/mnras/stv2917}

\bibitem{Guilet:2015b}
J.~{Guilet}, E.~{M{\"u}ller}, H.T. {Janka}, \mnras \textbf{447}, 3992 (2015).
\newblock \doi{10.1093/mnras/stu2550}

\bibitem{Brandenburg:2005}
A.~{Brandenburg}, K.~{Subramanian}, \physrep \textbf{417}, 1 (2005).
\newblock \doi{10.1016/j.physrep.2005.06.005}

\bibitem{quimby11}
R.M. {Quimby}, S.R. {Kulkarni}, M.M. {Kasliwal}, A.~{Gal-Yam}, I.~{Arcavi},
  M.~{Sullivan}, P.~{Nugent}, R.~{Thomas}, D.A. {Howell}, E.~{Nakar},
  L.~{Bildsten}, C.~{Theissen}, N.M. {Law}, R.~{Dekany}, G.~{Rahmer},
  D.~{Hale}, R.~{Smith}, E.O. {Ofek}, J.~{Zolkower}, V.~{Velur}, R.~{Walters},
  J.~{Henning}, K.~{Bui}, D.~{McKenna}, D.~{Poznanski}, S.B. {Cenko},
  D.~{Levitan}, \nat \textbf{474}, 487 (2011).
\newblock \doi{10.1038/nature10095}

\bibitem{pastorello10}
A.~{Pastorello}, M.T. {Botticella}, C.~{Trundle}, S.~{Taubenberger},
  S.~{Mattila}, E.~{Kankare}, N.~{Elias-Rosa}, S.~{Benetti}, G.~{Duszanowicz},
  L.~{Hermansson}, J.E. {Beckman}, F.~{Bufano}, M.~{Fraser}, A.~{Harutyunyan},
  H.~{Navasardyan}, S.J. {Smartt}, S.D. {van Dyk}, J.S. {Vink}, R.M. {Wagner},
  \mnras \textbf{408}, 181 (2010).
\newblock \doi{10.1111/j.1365-2966.2010.17142.x}

\bibitem{benetti14}
S.~{Benetti}, M.~{Nicholl}, E.~{Cappellaro}, A.~{Pastorello}, S.J. {Smartt},
  N.~{Elias-Rosa}, A.J. {Drake}, L.~{Tomasella}, M.~{Turatto},
  A.~{Harutyunyan}, S.~{Taubenberger}, S.~{Hachinger}, A.~{Morales-Garoffolo},
  T.W. {Chen}, S.G. {Djorgovski}, M.~{Fraser}, A.~{Gal-Yam}, C.~{Inserra},
  P.~{Mazzali}, M.L. {Pumo}, J.~{Sollerman}, S.~{Valenti}, D.R. {Young},
  M.~{Dennefeld}, L.~{Le Guillou}, M.~{Fleury}, P.F. {L{\'e}get}, \mnras
  \textbf{441}, 289 (2014).
\newblock \doi{10.1093/mnras/stu538}

\bibitem{inserra14}
C.~{Inserra}, S.J. {Smartt}, \apj \textbf{796}, 87 (2014).
\newblock \doi{10.1088/0004-637X/796/2/87}

\bibitem{pastorello08}
A.~{Pastorello}, S.~{Mattila}, L.~{Zampieri}, M.~{Della Valle}, S.J. {Smartt},
  S.~{Valenti}, I.~{Agnoletto}, S.~{Benetti}, C.R. {Benn}, D.~{Branch},
  E.~{Cappellaro}, M.~{Dennefeld}, J.J. {Eldridge}, A.~{Gal-Yam},
  A.~{Harutyunyan}, I.~{Hunter}, H.~{Kjeldsen}, Y.~{Lipkin}, P.A. {Mazzali},
  P.~{Milne}, H.~{Navasardyan}, E.O. {Ofek}, E.~{Pian}, O.~{Shemmer},
  S.~{Spiro}, R.A. {Stathakis}, S.~{Taubenberger}, M.~{Turatto}, H.~{Yamaoka},
  \mnras \textbf{389}, 113 (2008).
\newblock \doi{10.1111/j.1365-2966.2008.13602.x}

\bibitem{pastorello16}
A.~{Pastorello}, X.F. {Wang}, F.~{Ciabattari}, D.~{Bersier}, P.A. {Mazzali},
  X.~{Gao}, Z.~{Xu}, J.J. {Zhang}, S.~{Tokuoka}, S.~{Benetti}, E.~{Cappellaro},
  N.~{Elias-Rosa}, A.~{Harutyunyan}, F.~{Huang}, M.~{Miluzio}, J.~{Mo},
  P.~{Ochner}, L.~{Tartaglia}, G.~{Terreran}, L.~{Tomasella}, M.~{Turatto},
  \mnras \textbf{456}, 853 (2016).
\newblock \doi{10.1093/mnras/stv2634}

\bibitem{foley07}
R.J. {Foley}, N.~{Smith}, M.~{Ganeshalingam}, W.~{Li}, R.~{Chornock}, A.V.
  {Filippenko}, \apjl \textbf{657}, L105 (2007).
\newblock \doi{10.1086/513145}

\bibitem{valenti09}
S.~{Valenti}, A.~{Pastorello}, E.~{Cappellaro}, S.~{Benetti}, P.A. {Mazzali},
  J.~{Manteca}, S.~{Taubenberger}, N.~{Elias-Rosa}, R.~{Ferrando},
  A.~{Harutyunyan}, V.P. {Hentunen}, M.~{Nissinen}, E.~{Pian}, M.~{Turatto},
  L.~{Zampieri}, S.J. {Smartt}, \nat \textbf{459}, 674 (2009).
\newblock \doi{10.1038/nature08023}

\bibitem{perets10}
H.B. {Perets}, A.~{Gal-Yam}, P.A. {Mazzali}, D.~{Arnett}, D.~{Kagan}, A.V.
  {Filippenko}, W.~{Li}, I.~{Arcavi}, S.B. {Cenko}, D.B. {Fox}, D.C. {Leonard},
  D.S. {Moon}, D.J. {Sand}, A.M. {Soderberg}, J.P. {Anderson}, P.A. {James},
  R.J. {Foley}, M.~{Ganeshalingam}, E.O. {Ofek}, L.~{Bildsten}, G.~{Nelemans},
  K.J. {Shen}, N.N. {Weinberg}, B.D. {Metzger}, A.L. {Piro}, E.~{Quataert},
  M.~{Kiewe}, D.~{Poznanski}, \nat \textbf{465}, 322 (2010).
\newblock \doi{10.1038/nature09056}

\bibitem{pumo10}
M.L. {Pumo}, Memorie della Societa Astronomica Italiana Supplementi
  \textbf{14}, 115 (2010)

\bibitem{vandyk_rev12}
S.D. {Van Dyk}, T.~{Matheson}, in \emph{Eta Carinae and the Supernova
  Impostors}, \emph{Astrophysics and Space Science Library}, vol. 384, ed. by
  K.~{Davidson}, R.M. {Humphreys} (2012), \emph{Astrophysics and Space Science
  Library}, vol. 384, p. 249.
\newblock \doi{10.1007/978-1-4614-2275-4_11}

\bibitem{tartaglia16}
L.~{Tartaglia}, A.~{Pastorello}, M.~{Sullivan}, C.~{Baltay}, D.~{Rabinowitz},
  P.~{Nugent}, A.J. {Drake}, S.G. {Djorgovski}, A.~{Gal-Yam}, S.~{Fabrika},
  E.A. {Barsukova}, V.P. {Goranskij}, A.F. {Valeev}, T.~{Fatkhullin},
  S.~{Schulze}, A.~{Mehner}, F.E. {Bauer}, S.~{Taubenberger}, J.~{Nordin},
  S.~{Valenti}, D.A. {Howell}, S.~{Benetti}, E.~{Cappellaro}, G.~{Fasano},
  N.~{Elias-Rosa}, M.~{Barbieri}, D.~{Bettoni}, A.~{Harutyunyan}, T.~{Kangas},
  E.~{Kankare}, J.C. {Martin}, S.~{Mattila}, A.~{Morales-Garoffolo},
  P.~{Ochner}, U.D. {Rebbapragada}, G.~{Terreran}, L.~{Tomasella},
  M.~{Turatto}, E.~{Verroi}, P.R. {Wo{\'z}niak}, \mnras \textbf{459}, 1039
  (2016).
\newblock \doi{10.1093/mnras/stw675}

\bibitem{eliasrosa16}
N.~{Elias-Rosa}, A.~{Pastorello}, S.~{Benetti}, E.~{Cappellaro},
  S.~{Taubenberger}, G.~{Terreran}, M.~{Fraser}, P.J. {Brown}, L.~{Tartaglia},
  A.~{Morales-Garoffolo}, J.~{Harmanen}, N.D. {Richardson}, {\'E}.~{Artigau},
  L.~{Tomasella}, R.~{Margutti}, S.J. {Smartt}, M.~{Dennefeld}, M.~{Turatto},
  G.C. {Anupama}, R.~{Arbour}, M.~{Berton}, K.S. {Bjorkman}, T.~{Boles},
  F.~{Briganti}, R.~{Chornock}, F.~{Ciabattari}, G.~{Cortini}, A.~{Dimai}, C.J.
  {Gerhartz}, K.~{Itagaki}, R.~{Kotak}, R.~{Mancini}, F.~{Martinelli},
  D.~{Milisavljevic}, K.~{Misra}, P.~{Ochner}, D.~{Patnaude}, J.~{Polshaw},
  D.K. {Sahu}, S.~{Zaggia}, \mnras \textbf{463}, 3894 (2016).
\newblock \doi{10.1093/mnras/stw2253}

\bibitem{tartaglia15}
L.~{Tartaglia}, A.~{Pastorello}, S.~{Taubenberger}, E.~{Cappellaro}, J.R.
  {Maund}, S.~{Benetti}, T.~{Boles}, F.~{Bufano}, G.~{Duszanowicz},
  N.~{Elias-Rosa}, A.~{Harutyunyan}, L.~{Hermansson}, P.~{H{\"o}flich},
  K.~{Maguire}, H.~{Navasardyan}, S.J. {Smartt}, F.~{Taddia}, M.~{Turatto},
  \mnras \textbf{447}, 117 (2015).
\newblock \doi{10.1093/mnras/stu2384}

\bibitem{valenti12}
S.~{Valenti}, S.~{Taubenberger}, A.~{Pastorello}, L.~{Aramyan}, M.T.
  {Botticella}, M.~{Fraser}, S.~{Benetti}, S.J. {Smartt}, E.~{Cappellaro},
  N.~{Elias-Rosa}, M.~{Ergon}, L.~{Magill}, E.~{Magnier}, R.~{Kotak}, P.A.
  {Price}, J.~{Sollerman}, L.~{Tomasella}, M.~{Turatto}, D.E. {Wright}, \apjl
  \textbf{749}, L28 (2012).
\newblock \doi{10.1088/2041-8205/749/2/L28}

\bibitem{drout13}
M.R. {Drout}, A.M. {Soderberg}, P.A. {Mazzali}, J.T. {Parrent}, R.~{Margutti},
  D.~{Milisavljevic}, N.E. {Sanders}, R.~{Chornock}, R.J. {Foley}, R.P.
  {Kirshner}, A.V. {Filippenko}, W.~{Li}, P.J. {Brown}, S.B. {Cenko},
  S.~{Chakraborti}, P.~{Challis}, A.~{Friedman}, M.~{Ganeshalingam},
  M.~{Hicken}, C.~{Jensen}, M.~{Modjaz}, H.B. {Perets}, J.M. {Silverman}, D.S.
  {Wong}, \apj \textbf{774}, 58 (2013).
\newblock \doi{10.1088/0004-637X/774/1/58}

\bibitem{spiro14}
S.~{Spiro}, A.~{Pastorello}, M.L. {Pumo}, L.~{Zampieri}, M.~{Turatto}, S.J.
  {Smartt}, S.~{Benetti}, E.~{Cappellaro}, S.~{Valenti}, I.~{Agnoletto},
  G.~{Altavilla}, T.~{Aoki}, E.~{Brocato}, E.M. {Corsini}, A.~{Di Cianno},
  N.~{Elias-Rosa}, M.~{Hamuy}, K.~{Enya}, M.~{Fiaschi}, G.~{Folatelli},
  S.~{Desidera}, A.~{Harutyunyan}, D.A. {Howell}, A.~{Kawka}, Y.~{Kobayashi},
  B.~{Leibundgut}, T.~{Minezaki}, H.~{Navasardyan}, K.~{Nomoto}, S.~{Mattila},
  A.~{Pietrinferni}, G.~{Pignata}, G.~{Raimondo}, M.~{Salvo}, B.P. {Schmidt},
  J.~{Sollerman}, J.~{Spyromilio}, S.~{Taubenberger}, G.~{Valentini},
  S.~{Vennes}, Y.~{Yoshii}, \mnras \textbf{439}, 2873 (2014).
\newblock \doi{10.1093/mnras/stu156}

\bibitem{inserra13}
C.~{Inserra}, A.~{Pastorello}, M.~{Turatto}, M.L. {Pumo}, S.~{Benetti},
  E.~{Cappellaro}, M.T. {Botticella}, F.~{Bufano}, N.~{Elias-Rosa},
  A.~{Harutyunyan}, S.~{Taubenberger}, S.~{Valenti}, L.~{Zampieri}, \aap
  \textbf{555}, A142 (2013).
\newblock \doi{10.1051/0004-6361/201220496}

\bibitem{turatto98}
M.~{Turatto}, P.A. {Mazzali}, T.R. {Young}, K.~{Nomoto}, K.~{Iwamoto},
  S.~{Benetti}, E.~{Cappellaro}, I.J. {Danziger}, D.F. {de Mello}, M.M.
  {Phillips}, N.B. {Suntzeff}, A.~{Clocchiatti}, A.~{Piemonte},
  B.~{Leibundgut}, R.~{Covarrubias}, J.~{Maza}, J.~{Sollerman}, \apjl
  \textbf{498}, L129 (1998).
\newblock \doi{10.1086/311324}

\bibitem{atwood09}
W.B. {Atwood}, A.A. {Abdo}, M.~{Ackermann}, W.~{Althouse}, B.~{Anderson},
  M.~{Axelsson}, L.~{Baldini}, J.~{Ballet}, D.L. {Band}, G.~{Barbiellini},
  et~al., \apj \textbf{697}, 1071 (2009).
\newblock \doi{10.1088/0004-637X/697/2/1071}

\bibitem{SuperKamiokande:2007}
M.~{Ikeda}, A.~{Takeda}, Y.~{Fukuda}, M.R. {Vagins}, K.~{Abe}, T.~{Iida},
  K.~{Ishihara}, J.~{Kameda}, Y.~{Koshio}, A.~{Minamino}, C.~{Mitsuda},
  M.~{Miura}, S.~{Moriyama}, M.~{Nakahata}, Y.~{Obayashi}, H.~{Ogawa},
  H.~{Sekiya}, M.~{Shiozawa}, Y.~{Suzuki}, Y.~{Takeuchi}, K.~{Ueshima},
  H.~{Watanabe}, S.~{Yamada}, I.~{Higuchi}, C.~{Ishihara}, M.~{Ishitsuka},
  T.~{Kajita}, K.~{Kaneyuki}, G.~{Mitsuka}, S.~{Nakayama}, H.~{Nishino},
  K.~{Okumura}, C.~{Saji}, Y.~{Takenaga}, S.~{Clark}, S.~{Desai}, F.~{Dufour},
  E.~{Kearns}, S.~{Likhoded}, M.~{Litos}, J.L. {Raaf}, J.L. {Stone}, L.R.
  {Sulak}, W.~{Wang}, M.~{Goldhaber}, D.~{Casper}, J.P. {Cravens},
  J.~{Dunmore}, W.R. {Kropp}, D.W. {Liu}, S.~{Mine}, C.~{Regis}, M.B. {Smy},
  H.W. {Sobel}, K.S. {Ganezer}, J.~{Hill}, W.E. {Keig}, J.S. {Jang}, J.Y.
  {Kim}, I.T. {Lim}, K.~{Scholberg}, N.~{Tanimoto}, C.W. {Walter},
  R.~{Wendell}, R.W. {Ellsworth}, S.~{Tasaka}, G.~{Guillian}, J.G. {Learned},
  S.~{Matsuno}, M.D. {Messier}, Y.~{Hayato}, A.K. {Ichikawa}, T.~{Ishida},
  T.~{Ishii}, T.~{Iwashita}, T.~{Kobayashi}, T.~{Nakadaira}, K.~{Nakamura},
  K.~{Nitta}, Y.~{Oyama}, Y.~{Totsuka}, A.T. {Suzuki}, M.~{Hasegawa},
  K.~{Hiraide}, H.~{Maesaka}, T.~{Nakaya}, K.~{Nishikawa}, T.~{Sasaki},
  S.~{Yamamoto}, M.~{Yokoyama}, T.J. {Haines}, S.~{Dazeley}, S.~{Hatakeyama},
  R.~{Svoboda}, G.W. {Sullivan}, D.~{Turcan}, A.~{Habig}, T.~{Sato}, Y.~{Itow},
  T.~{Koike}, T.~{Tanaka}, C.K. {Jung}, T.~{Kato}, K.~{Kobayashi}, M.~{Malek},
  C.~{McGrew}, A.~{Sarrat}, R.~{Terri}, C.~{Yanagisawa}, N.~{Tamura},
  Y.~{Idehara}, M.~{Sakuda}, M.~{Sugihara}, Y.~{Kuno}, M.~{Yoshida}, S.B.
  {Kim}, B.S. {Yang}, J.~{Yoo}, T.~{Ishizuka}, H.~{Okazawa}, Y.~{Choi}, H.K.
  {Seo}, Y.~{Gando}, T.~{Hasegawa}, K.~{Inoue}, Y.~{Furuse}, H.~{Ishii},
  K.~{Nishijima}, H.~{Ishino}, Y.~{Watanabe}, M.~{Koshiba}, S.~{Chen},
  Z.~{Deng}, Y.~{Liu}, D.~{Kielczewska}, J.~{Zalipska}, H.~{Berns}, R.~{Gran},
  K.K. {Shiraishi}, A.~{Stachyra}, E.~{Thrane}, K.~{Washburn}, R.J. {Wilkes},
  {Super-KAMIOKANDE Collaboration}, \apj \textbf{669}, 519 (2007).
\newblock \doi{10.1086/521547}

\bibitem{IceCube:2009}
A.~{Karle}, {IceCube Collaboration}, Nuclear Instruments and Methods in Physics
  Research A \textbf{604}, S46 (2009).
\newblock \doi{10.1016/j.nima.2009.03.180}

\bibitem{Antares:2012}
M.~{Ageron}, J.A. {Aguilar}, I.~{Al Samarai}, A.~{Albert}, M.~{Andr{\'e}},
  M.~{Anghinolfi}, G.~{Anton}, S.~{Anvar}, M.~{Ardid}, A.C. {Assis Jesus},
  T.~{Astraatmadja}, J.J. {Aubert}, B.~{Baret}, S.~{Basa}, V.~{Bertin},
  S.~{Biagi}, A.~{Bigi}, C.~{Bigongiari}, C.~{Bogazzi}, M.~{Bou-Cabo},
  B.~{Bouhou}, M.C. {Bouwhuis}, J.~{Brunner}, J.~{Busto}, F.~{Camarena},
  A.~{Capone}, C.~{C{\^a}rloganu}, G.~{Carminati}, J.~{Carr}, S.~{Cecchini},
  Z.~{Charif}, P.~{Charvis}, T.~{Chiarusi}, M.~{Circella}, R.~{Coniglione},
  H.~{Costantini}, P.~{Coyle}, C.~{Curtil}, M.P. {Decowski}, I.~{Dekeyser},
  A.~{Deschamps}, C.~{Distefano}, C.~{Donzaud}, D.~{Dornic}, Q.~{Dorosti},
  D.~{Drouhin}, T.~{Eberl}, U.~{Emanuele}, A.~{Enzenh{\"o}fer}, J.P.
  {Ernenwein}, S.~{Escoffier}, P.~{Fermani}, M.~{Ferri}, V.~{Flaminio},
  F.~{Folger}, U.~{Fritsch}, J.L. {Fuda}, S.~{Galat{\`a}}, P.~{Gay},
  G.~{Giacomelli}, V.~{Giordano}, J.P. {G{\'o}mez-Gonz{\'a}lez}, K.~{Graf},
  G.~{Guillard}, G.~{Halladjian}, G.~{Hallewell}, H.~{van Haren}, J.~{Hartman},
  A.J. {Heijboer}, Y.~{Hello}, J.J. {Hern{\'a}ndez-Rey}, B.~{Herold},
  J.~{H{\"o}{\ss}l}, C.C. {Hsu}, M.~{de Jong}, M.~{Kadler}, O.~{Kalekin},
  A.~{Kappes}, U.~{Katz}, O.~{Kavatsyuk}, P.~{Kooijman}, C.~{Kopper},
  A.~{Kouchner}, I.~{Kreykenbohm}, V.~{Kulikovskiy}, R.~{Lahmann}, P.~{Lamare},
  G.~{Larosa}, D.~{Lattuada}, D.~{Lef{\`e}vre}, G.~{Lim}, D.~{Lo Presti},
  H.~{Loehner}, S.~{Loucatos}, S.~{Mangano}, M.~{Marcelin}, A.~{Margiotta},
  J.A. {Mart{\'{\i}}nez-Mora}, A.~{Meli}, T.~{Montaruli}, L.~{Moscoso},
  H.~{Motz}, M.~{Neff}, E.~{Nezri}, D.~{Palioselitis}, G.E. {P{\u a}v{\u
  a}la{\c s}}, K.~{Payet}, P.~{Payre}, J.~{Petrovic}, P.~{Piattelli},
  N.~{Picot-Clemente}, V.~{Popa}, T.~{Pradier}, E.~{Presani}, C.~{Racca},
  C.~{Reed}, C.~{Richardt}, R.~{Richter}, C.~{Rivi{\`e}re}, A.~{Robert},
  K.~{Roensch}, A.~{Rostovtsev}, J.~{Ruiz-Rivas}, M.~{Rujoiu}, G.V. {Russo},
  F.~{Salesa}, P.~{Sapienza}, F.~{Sch{\"o}ck}, J.P. {Schuller},
  F.~{Sch{\"u}ssler}, R.~{Shanidze}, F.~{Simeone}, A.~{Spies}, M.~{Spurio},
  J.J.M. {Steijger}, T.~{Stolarczyk}, A.~{S{\'a}nchez-Losa}, M.~{Taiuti},
  C.~{Tamburini}, S.~{Toscano}, B.~{Vallage}, V.~{van Elewyck}, G.~{Vannoni},
  M.~{Vecchi}, P.~{Vernin}, G.~{Wijnker}, J.~{Wilms}, E.~{de Wolf}, H.~{Yepes},
  D.~{Zaborov}, J.D. {Zornoza}, J.~{Z{\'u}{\~n}iga}, Astroparticle Physics
  \textbf{35}, 530 (2012).
\newblock \doi{10.1016/j.astropartphys.2011.11.011}

\bibitem{abbott16a}
B.P. {Abbott}, R.~{Abbott}, T.D. {Abbott}, M.R. {Abernathy}, F.~{Acernese},
  K.~{Ackley}, C.~{Adams}, T.~{Adams}, P.~{Addesso}, R.X. {Adhikari}, et~al.,
  Physical Review Letters \textbf{116}(6), 061102 (2016).
\newblock \doi{10.1103/PhysRevLett.116.061102}

\bibitem{Cutler:2002}
C.~{Cutler}, K.S. {Thorne}, ArXiv General Relativity and Quantum Cosmology
  e-prints  (2002)

\bibitem{aLIGO:2015}
{LIGO Scientific Collaboration}, J.~{Aasi}, B.P. {Abbott}, R.~{Abbott},
  T.~{Abbott}, M.R. {Abernathy}, K.~{Ackley}, C.~{Adams}, T.~{Adams},
  P.~{Addesso}, et~al., Classical and Quantum Gravity \textbf{32}(7), 074001
  (2015).
\newblock \doi{10.1088/0264-9381/32/7/074001}

\bibitem{aVirgo:2015}
F.~{Acernese}, M.~{Agathos}, K.~{Agatsuma}, D.~{Aisa}, N.~{Allemandou},
  A.~{Allocca}, J.~{Amarni}, P.~{Astone}, G.~{Balestri}, G.~{Ballardin},
  et~al., Classical and Quantum Gravity \textbf{32}(2), 024001 (2015).
\newblock \doi{10.1088/0264-9381/32/2/024001}

\bibitem{KAGRA:2013}
Y.~{Aso}, Y.~{Michimura}, K.~{Somiya}, M.~{Ando}, O.~{Miyakawa},
  T.~{Sekiguchi}, D.~{Tatsumi}, H.~{Yamamoto}, \prd \textbf{88}(4), 043007
  (2013).
\newblock \doi{10.1103/PhysRevD.88.043007}

\bibitem{LISA:2017}
P.~{Amaro-Seoane}, H.~{Audley}, S.~{Babak}, J.~{Baker}, E.~{Barausse},
  P.~{Bender}, E.~{Berti}, P.~{Binetruy}, M.~{Born}, D.~{Bortoluzzi},
  J.~{Camp}, C.~{Caprini}, V.~{Cardoso}, M.~{Colpi}, J.~{Conklin},
  N.~{Cornish}, C.~{Cutler}, K.~{Danzmann}, R.~{Dolesi}, L.~{Ferraioli},
  V.~{Ferroni}, E.~{Fitzsimons}, J.~{Gair}, L.~{Gesa Bote}, D.~{Giardini},
  F.~{Gibert}, C.~{Grimani}, H.~{Halloin}, G.~{Heinzel}, T.~{Hertog},
  M.~{Hewitson}, K.~{Holley-Bockelmann}, D.~{Hollington}, M.~{Hueller},
  H.~{Inchauspe}, P.~{Jetzer}, N.~{Karnesis}, C.~{Killow}, A.~{Klein},
  B.~{Klipstein}, N.~{Korsakova}, S.L. {Larson}, J.~{Livas}, I.~{Lloro},
  N.~{Man}, D.~{Mance}, J.~{Martino}, I.~{Mateos}, K.~{McKenzie}, S.T.
  {McWilliams}, C.~{Miller}, G.~{Mueller}, G.~{Nardini}, G.~{Nelemans},
  M.~{Nofrarias}, A.~{Petiteau}, P.~{Pivato}, E.~{Plagnol}, E.~{Porter},
  J.~{Reiche}, D.~{Robertson}, N.~{Robertson}, E.~{Rossi}, G.~{Russano},
  B.~{Schutz}, A.~{Sesana}, D.~{Shoemaker}, J.~{Slutsky}, C.F. {Sopuerta},
  T.~{Sumner}, N.~{Tamanini}, I.~{Thorpe}, M.~{Troebs}, M.~{Vallisneri},
  A.~{Vecchio}, D.~{Vetrugno}, S.~{Vitale}, M.~{Volonteri}, G.~{Wanner},
  H.~{Ward}, P.~{Wass}, W.~{Weber}, J.~{Ziemer}, P.~{Zweifel}, ArXiv e-prints
  (2017)

\bibitem{PTA:2010}
G.~{Hobbs}, A.~{Archibald}, Z.~{Arzoumanian}, D.~{Backer}, M.~{Bailes}, N.D.R.
  {Bhat}, M.~{Burgay}, S.~{Burke-Spolaor}, D.~{Champion}, I.~{Cognard},
  W.~{Coles}, J.~{Cordes}, P.~{Demorest}, G.~{Desvignes}, R.D. {Ferdman},
  L.~{Finn}, P.~{Freire}, M.~{Gonzalez}, J.~{Hessels}, A.~{Hotan},
  G.~{Janssen}, F.~{Jenet}, A.~{Jessner}, C.~{Jordan}, V.~{Kaspi}, M.~{Kramer},
  V.~{Kondratiev}, J.~{Lazio}, K.~{Lazaridis}, K.J. {Lee}, Y.~{Levin},
  A.~{Lommen}, D.~{Lorimer}, R.~{Lynch}, A.~{Lyne}, R.~{Manchester},
  M.~{McLaughlin}, D.~{Nice}, S.~{Oslowski}, M.~{Pilia}, A.~{Possenti},
  M.~{Purver}, S.~{Ransom}, J.~{Reynolds}, S.~{Sanidas}, J.~{Sarkissian},
  A.~{Sesana}, R.~{Shannon}, X.~{Siemens}, I.~{Stairs}, B.~{Stappers},
  D.~{Stinebring}, G.~{Theureau}, R.~{van Haasteren}, W.~{van Straten}, J.P.W.
  {Verbiest}, D.R.B. {Yardley}, X.P. {You}, Classical and Quantum Gravity
  \textbf{27}(8), 084013 (2010).
\newblock \doi{10.1088/0264-9381/27/8/084013}

\bibitem{Dimmelmeier:2002b}
H.~{Dimmelmeier}, J.A. {Font}, E.~{M{\"u}ller}, \aap \textbf{393}, 523 (2002).
\newblock \doi{10.1051/0004-6361:20021053}

\bibitem{Dimmelmeier:2008}
H.~{Dimmelmeier}, C.D. {Ott}, A.~{Marek}, H.T. {Janka}, \prd \textbf{78}(6),
  064056 (2008).
\newblock \doi{10.1103/PhysRevD.78.064056}

\bibitem{Abdikamalov:2014}
E.~{Abdikamalov}, S.~{Gossan}, A.M. {DeMaio}, C.D. {Ott}, \prd \textbf{90}(4),
  044001 (2014).
\newblock \doi{10.1103/PhysRevD.90.044001}

\bibitem{Richers:2017}
S.~{Richers}, C.D. {Ott}, E.~{Abdikamalov}, E.~{O'Connor}, C.~{Sullivan}, ArXiv
  e-prints  (2017)

\bibitem{Murphy:2009}
J.W. {Murphy}, C.D. {Ott}, A.~{Burrows}, \apj \textbf{707}, 1173 (2009).
\newblock \doi{10.1088/0004-637X/707/2/1173}

\bibitem{Kuroda:2016}
T.~{Kuroda}, K.~{Kotake}, T.~{Takiwaki}, \apjl \textbf{829}, L14 (2016).
\newblock \doi{10.3847/2041-8205/829/1/L14}

\bibitem{Andresen:2017}
H.~{Andresen}, B.~{M{\"u}ller}, E.~{M{\"u}ller}, H.T. {Janka}, \mnras
  \textbf{468}, 2032 (2017).
\newblock \doi{10.1093/mnras/stx618}

\bibitem{Sotani:2016}
H.~{Sotani}, T.~{Takiwaki}, \prd \textbf{94}(4), 044043 (2016).
\newblock \doi{10.1103/PhysRevD.94.044043}

\bibitem{Abbot:2016c}
B.P. {Abbott}, R.~{Abbott}, T.D. {Abbott}, M.R. {Abernathy}, F.~{Acernese},
  K.~{Ackley}, C.~{Adams}, T.~{Adams}, P.~{Addesso}, R.X. {Adhikari}, et~al.,
  \prd \textbf{94}(10), 102001 (2016).
\newblock \doi{10.1103/PhysRevD.94.102001}

\bibitem{ott09}
C.D. {Ott}, Classical and Quantum Gravity \textbf{26}(6), 063001 (2009).
\newblock \doi{10.1088/0264-9381/26/6/063001}

\bibitem{SkyLoc:2016}
B.P. {Abbott}, R.~{Abbott}, T.D. {Abbott}, M.R. {Abernathy}, F.~{Acernese},
  K.~{Ackley}, C.~{Adams}, T.~{Adams}, P.~{Addesso}, R.X. {Adhikari}, et~al.,
  Living Reviews in Relativity \textbf{19}, 1 (2016).
\newblock \doi{10.1007/lrr-2016-1}

\bibitem{levesque09}
E.M. {Levesque}, P.~{Massey}, B.~{Plez}, K.A.G. {Olsen}, \aj \textbf{137}, 4744
  (2009).
\newblock \doi{10.1088/0004-6256/137/6/4744}

\bibitem{adams16}
S.M. {Adams}, C.S. {Kochanek}, J.R. {Gerke}, K.Z. {Stanek}, X.~{Dai}, ArXiv
  e-prints  (2016)

\bibitem{galyam14}
A.~{Gal-Yam}, in \emph{American Astronomical Society Meeting Abstracts \#223},
  \emph{American Astronomical Society Meeting Abstracts}, vol. 223 (2014),
  \emph{American Astronomical Society Meeting Abstracts}, vol. 223, p. 235.02

\bibitem{terreran16}
G.~{Terreran}, A.~{Jerkstrand}, S.~{Benetti}, S.J. {Smartt}, P.~{Ochner},
  L.~{Tomasella}, D.A. {Howell}, A.~{Morales-Garoffolo}, A.~{Harutyunyan},
  E.~{Kankare}, I.~{Arcavi}, E.~{Cappellaro}, N.~{Elias-Rosa},
  G.~{Hosseinzadeh}, T.~{Kangas}, A.~{Pastorello}, L.~{Tartaglia},
  M.~{Turatto}, S.~{Valenti}, P.~{Wiggins}, F.~{Yuan}, \mnras \textbf{462}, 137
  (2016).
\newblock \doi{10.1093/mnras/stw1591}

\bibitem{smartt15_pessto}
S.J. {Smartt}, S.~{Valenti}, M.~{Fraser}, C.~{Inserra}, D.R. {Young},
  M.~{Sullivan}, A.~{Pastorello}, S.~{Benetti}, A.~{Gal-Yam}, C.~{Knapic},
  M.~{Molinaro}, R.~{Smareglia}, K.W. {Smith}, S.~{Taubenberger}, O.~{Yaron},
  J.P. {Anderson}, C.~{Ashall}, C.~{Balland}, C.~{Baltay}, C.~{Barbarino}, F.E.
  {Bauer}, S.~{Baumont}, D.~{Bersier}, N.~{Blagorodnova}, S.~{Bongard}, M.T.
  {Botticella}, F.~{Bufano}, M.~{Bulla}, E.~{Cappellaro}, H.~{Campbell},
  F.~{Cellier-Holzem}, T.W. {Chen}, M.J. {Childress}, A.~{Clocchiatti},
  C.~{Contreras}, M.~{Dall'Ora}, J.~{Danziger}, T.~{de Jaeger}, A.~{De Cia},
  M.~{Della Valle}, M.~{Dennefeld}, N.~{Elias-Rosa}, N.~{Elman}, U.~{Feindt},
  M.~{Fleury}, E.~{Gall}, S.~{Gonzalez-Gaitan}, L.~{Galbany}, A.~{Morales
  Garoffolo}, L.~{Greggio}, L.L. {Guillou}, S.~{Hachinger}, E.~{Hadjiyska},
  P.E. {Hage}, W.~{Hillebrandt}, S.~{Hodgkin}, E.Y. {Hsiao}, P.A. {James},
  A.~{Jerkstrand}, T.~{Kangas}, E.~{Kankare}, R.~{Kotak}, M.~{Kromer},
  H.~{Kuncarayakti}, G.~{Leloudas}, P.~{Lundqvist}, J.D. {Lyman}, I.M. {Hook},
  K.~{Maguire}, I.~{Manulis}, S.J. {Margheim}, S.~{Mattila}, J.R. {Maund}, P.A.
  {Mazzali}, M.~{McCrum}, R.~{McKinnon}, M.E. {Moreno-Raya}, M.~{Nicholl},
  P.~{Nugent}, R.~{Pain}, G.~{Pignata}, M.M. {Phillips}, J.~{Polshaw}, M.L.
  {Pumo}, D.~{Rabinowitz}, E.~{Reilly}, C.~{Romero-Ca{\~n}izales}, R.~{Scalzo},
  B.~{Schmidt}, S.~{Schulze}, S.~{Sim}, J.~{Sollerman}, F.~{Taddia},
  L.~{Tartaglia}, G.~{Terreran}, L.~{Tomasella}, M.~{Turatto}, E.~{Walker},
  N.A. {Walton}, L.~{Wyrzykowski}, F.~{Yuan}, L.~{Zampieri}, \aap \textbf{579},
  A40 (2015).
\newblock \doi{10.1051/0004-6361/201425237}

\bibitem{Thompson:1993}
C.~{Thompson}, R.C. {Duncan}, \apj \textbf{408}, 194 (1993).
\newblock \doi{10.1086/172580}

\bibitem{Shakura:1973}
N.I. {Shakura}, R.A. {Sunyaev}, \aap \textbf{24}, 337 (1973)

\bibitem{Moffatt:1978}
H.K. {Moffatt}, \emph{{Magnetic field generation in electrically conducting
  fluids}} (Cambridge, England, Cambridge University Press, 1978.~353 p., 1978)

\bibitem{Bugli:2014}
M.~{Bugli}, L.~{Del Zanna}, N.~{Bucciantini}, \mnras \textbf{440}, L41 (2014).
\newblock \doi{10.1093/mnrasl/slu017}

\bibitem{Sadowski:2015}
A.~{S{\c a}dowski}, R.~{Narayan}, A.~{Tchekhovskoy}, D.~{Abarca}, Y.~{Zhu},
  J.C. {McKinney}, \mnras \textbf{447}, 49 (2015).
\newblock \doi{10.1093/mnras/stu2387}

\bibitem{Giacomazzo:2015}
B.~{Giacomazzo}, J.~{Zrake}, P.C. {Duffell}, A.I. {MacFadyen}, R.~{Perna}, \apj
  \textbf{809}, 39 (2015).
\newblock \doi{10.1088/0004-637X/809/1/39}

\bibitem{Shibata:2017}
M.~{Shibata}, K.~{Kiuchi}, Y.i. {Sekiguchi}, \prd \textbf{95}(8), 083005
  (2017).
\newblock \doi{10.1103/PhysRevD.95.083005}

\bibitem{Baraffe:2011}
M.~{Viallet}, I.~{Baraffe}, R.~{Walder}, \aap \textbf{531}, A86 (2011).
\newblock \doi{10.1051/0004-6361/201016374}

\bibitem{Mueller:2016b}
B.~{M{\"u}ller}, M.~{Viallet}, A.~{Heger}, H.T. {Janka}, \apj \textbf{833}, 124
  (2016).
\newblock \doi{10.3847/1538-4357/833/1/124}

\bibitem{Couch:2013}
S.M. {Couch}, C.D. {Ott}, \apjl \textbf{778}, L7 (2013).
\newblock \doi{10.1088/2041-8205/778/1/L7}

\bibitem{Couch:2015b}
S.M. {Couch}, E.~{Chatzopoulos}, W.D. {Arnett}, F.X. {Timmes}, \apjl
  \textbf{808}, L21 (2015).
\newblock \doi{10.1088/2041-8205/808/1/L21}

\bibitem{Mueller:2015}
B.~{M{\"u}ller}, H.T. {Janka}, \mnras \textbf{448}, 2141 (2015).
\newblock \doi{10.1093/mnras/stv101}

\bibitem{Motl:2002}
P.M. {Motl}, J.E. {Tohline}, J.~{Frank}, \apjs \textbf{138}, 121 (2002).
\newblock \doi{10.1086/324159}

\bibitem{Lajoie:2011}
C.P. {Lajoie}, A.~{Sills}, \apj \textbf{726}, 66 (2011).
\newblock \doi{10.1088/0004-637X/726/2/66}

\bibitem{Lombardi:2011}
J.C. {Lombardi}, Jr., W.~{Holtzman}, K.L. {Dooley}, K.~{Gearity},
  V.~{Kalogera}, F.A. {Rasio}, \apj \textbf{737}, 49 (2011).
\newblock \doi{10.1088/0004-637X/737/2/49}

\end{thebibliography}

\end{document}